\documentclass[10pt,DIV=10]{scrartcl}

\usepackage[utf8]{inputenc}
\usepackage{tikzHA}
\usepackage{amsmath,amsfonts,amssymb}
\usepackage{colortbl}
\usepackage{enumitem}
\usepackage{graphicx}
\graphicspath{{./figures/}}
\usepackage{subfigure}
\usepackage{floatflt}
\usepackage{wrapfig}
\usepackage{booktabs}
\usepackage[breaklinks,hidelinks]{hyperref} 

\setenumerate[1]{label={(\arabic*)},leftmargin=*,widest=9}

\newcommand{\ignore}[1]{}

\usepackage{etoolbox}
\newcommand{\memberfield}[3]{{%
   \ifstrequal{.}{#1}{%
     {\sf \dot{#3}}{\ifx&#2&{}\else (#2)\fi}%
     }{%
     {\sf #3 #1}{\ifx&#2&{}\else (#2)\fi}%
     }
   }}

\usepackage{listings}
\definecolor{gray}{rgb}{0.4,0.4,0.4}
\definecolor{identifiercolor}{rgb}{0.0,0.0,0.5}
\definecolor{keywordcolor}{rgb}{0.0,0.3,0.8}
\lstdefinelanguage{XML}
{
    morestring=[b]",
    morestring=[s]{>}{<},
    morecomment=[s]{<?}{?>},
    stringstyle=\color{black},
    identifierstyle=\color{identifiercolor},
    keywordstyle=\color{keywordcolor},
    morekeywords={flow,invariant,mode,jump,guard,reset}
}
\newcommand{\hpilot}{H-PILoT}
\newcommand{\pointer}[2][]{{\memberfield{#1}{#2}{index}}}
\newcommand{\variable}[2][]{{\memberfield{#1}{#2}{x}}}
\newcommand{\pos}[2][]{{\memberfield{#1}{#2}{pos}}}

\newcommand{\lane}[2][]{{\memberfield{#1}{#2}{lane}}}
\newcommand{\back}[2][]{{\memberfield{#1}{#2}{back}}}
\newcommand{\front}[2][]{{\memberfield{#1}{#2}{front}}}

\newcommand{\sideback}[2][]{{\memberfield{#1}{#2}{sideback}}}
\newcommand{\sidefront}[2][]{{\memberfield{#1}{#2}{sidefront}}}

\newcommand{\nil}{{\sf nil}}

\newcommand{\environment}{\text{\sf Top}}

\newtheorem{thm}{Theorem}
\newtheorem{lem}[thm]{Lemma}
\newtheorem{cor}[thm]{Corollary}
\newtheorem{defi}[thm]{Definition}
\newtheorem{rem}[thm]{Remark}
\newtheorem{ex}[thm]{Example}

\newcommand{\flow}{{\sf flow}}
\newcommand{\invariant}{{\sf Inv}}
\newcommand{\initialStates}{{\sf Init}}
\newcommand{\guard}{{\sf guard}}
\newcommand{\jump}{{\sf jump}}

\begin{document}

\title{
  Decidability of Verification of Safety Properties
  of Spatial Families of Linear Hybrid Automata}
\author{
 {\small  Werner Damm$^1$, 
  Matthias Horbach$^{2,3}$ and
  Viorica Sofronie-Stokkermans$^{2,3}$}\\
{\small $^1$  Carl von Ossietzky University, Oldenburg, Germany and} \\[-1ex]
{\small $^2$  University Koblenz-Landau, Koblenz, Germany and } \\[-1ex]
{\small$^2$  Max-Planck-Institut f{\"u}r Informatik, Saarbr{\"u}cken, Germany}}

\date{}

\maketitle

\begin{abstract}
We consider systems composed of an unbounded number of
uniformly designed linear hybrid automata, whose dynamic
behavior is determined by their relation to neighboring systems.
We present a class of such systems and a 
class of safety properties  whose verification 
can be reduced to the 
verification of (small) families of ``neighboring'' 
systems of bounded size, and identify situations in which such 
verification problems are decidable, resp.\ fixed parameter tractable.  
We illustrate the approach with an example from coordinated vehicle
guidance, 
and describe an implementation which allows us to perform such
verification tasks automatically. 
\end{abstract}

\section{Introduction}
Verification of families of interacting
systems is very important nowadays. 
Next generations cars will perform 
cooperative maneuvers for collision 
avoidance, lane changing, overtaking, and passing intersections. 
They will rely on an internal digital representation of the
environment -- capturing relative distance and speed 
of surrounding vehicles 
through on board sensors, sensor fusion, and 
vehicle2vehicle communication 
in determining which coalition of vehicles will follow 
what dynamics to achieve e.g. collision freedom. 
While prototype realizations of such highly automated driving
functions have been demonstrated (cf.\ e.g.\ HAVEit project~\cite{HAVEit}), 
the challenge in deploying such solutions rests in proving their
safety.

In this paper, we propose a general mathematical model capturing the essence
of such interacting systems 
as {\em spatial families of  hybrid automata} and provide
efficient  verification methods for proving safety when abstracting 
the dynamics to linear hybrid automata. It thus provides efficient 
verification methods for systems composed of an unbounded 
dynamically communicating parallel composition of uniformly 
defined linear hybrid automata.

The main contributions can be 
summarized as follows: 
\begin{itemize}
\vspace{-1mm}
\item We identify a class of 
systems composed of dynamically communicating 
uniformly defined linear hybrid automata 
and a class of safety properties (with exhaustive entry conditions) 
for which the verification of the 
whole system can be reduced to the 
verification of subsystems of bounded size of 
``neighboring'' components. 

\vspace{-1mm}
\item We identify situations when verification is 
decidable and fixed parameter tractable. 

\vspace{-1mm}
\item We identify situations when checking whether the 
safety property has ``exhaustive entry conditions'' is 
decidable resp.\ fixed parameter tractable. 

\vspace{-1mm}
\item We analyze the complexity 
of parametric verification resp. synthesis.

\vspace{-1mm}
\item We illustrate all concepts we introduce and all 
steps of our method on a running example from coordinated 
vehicle guidance. 

\vspace{-1mm}
\item We implemented these ideas in the tool HAHA (Hierarchical
  Analysis of Hybrid Automata), which employs \hpilot{} for the
  reasoning tests. We present several tests and 
comparisons. 
\end{itemize}

\subsection{Related work} 
A considerable amount of work has been dedicated to identifying 
classes of hybrid automata for which checking 
safety is decidable. Reachability and safety in linear hybrid automata are 
in general undecidable, while invariant checking 
and bounded reachability are decidable.
There are various approaches to the parametric verification of 
individual  hybrid automata \cite{AlurHH96}, 
the development of a dynamic 
hybrid logic \cite{Platzer08}, and of tools (cf.\ e.g.\ 
\cite{FrehseJK08,FribourgK13}). 
A survey of existing decidability 
and undecidability results for individual  hybrid automata
can be found in 
\cite{Sofronie-Stokkermans10,DammIS11}, which gives 
an overview of papers in which 
classes of hybrid automata resp.\ classes of verification problems for 
which decidability results can be established. 

\medskip 
In this paper we analyze {\em systems} of hybrid automata. 
In recent years, systems of systems have been studied in various
papers. 

Small model or cutoff properties for the verification of families of 
systems have been studied, but only for systems of discrete 
(or even finite state) systems.  
In \cite{EmersonSrinivasan90} an indexed temporal logic
 is introduced that can be
used to specify programs with arbitrarily many similar processes.
It is shown that the problems of checking 
``almost always satisfiability'' and 
``almost always unsatisfiability'' are decidable, and a small model
property is given. In \cite{AbdullaHH13}, a framework for the automatic
verification of systems with a parametric number of communicating 
processes (organized in various topologies such as words, multisets,
rings, or trees) is proposed; a method for the verification of such
systems is given which needs to inspect only a small number of
processes in order to show correctness of the whole system (the method
relies on an abstraction function that views the system from the
perspective of a fixed number of processes).
In \cite{KaiserKW10}, the class of finite-state programs executed
 by an unbounded number of replicated threads communicating via 
shared variables is studied. The thread-state reachability problem 
for this class is decidable via Petri net coverability analysis, but
as techniques solely based on coverability are inefficient,
\cite{KaiserKW10}
presents an alternative method based on a thread-state cutoff.  
Modularity results (and similar cutoff results) are presented for 
the special case of systems of trains on a complex track topology 
in \cite{Sofronie-GETCO09} and \cite{FaberIJS10}. 
In \cite{JacobsB14} a cutoff property is used for parameterized synthesis 
in token ring networks (the synthesis problem is reduced to
distributed synthesis in a network consisting of a few copies of a single process). 
Our work generalizes previous results on verification of classes of 
systems such as
\cite{EmersonSrinivasan90,AbdullaHH13,KaiserKW10,FaberIJS10,DammPRW13,JacobsB14} 
in supporting the much richer system model of linear hybrid automata.
The temporal logic we use for specifying the safety properties we
consider is similar to that introduced in \cite{EmersonSrinivasan90}. 
\medskip

Among the existing work in which the safety of 
cooperative driver assistance systems (modeling 
autonomous cars on highways performing lane-change maneuvers) 
we mention the results in \cite{FreseB10}, \cite{HilscherLOR11} and 
\cite{DammPRW13}. 

\cite{DammPRW13} proposes a design and verification methodology 
for cooperative driver assistance systems (with focus on applications 
where drivers are supported in complex driving tasks by safe 
strategies involving the coordinated movements of multiple vehicles 
to complete the driving task successfully). A ``divide and conquer'' 
approach for formally verifying timed probabilistic requirements on 
successful completion of the driving task and collision freedom is 
proposed. Our method is different, mainly because it relies on 
locality properties of the logical theories used for modeling the 
problems. 
In \cite{HilscherLOR11}, an alternative approach to prove safety (collision freedom) of
multi-lane motorway traffic with lane-change maneuvers is proposed,
based on a new spatial interval logic based on the view of each car.
The compositional approach \cite{HilscherLOR11} addresses an 
application class that is related to our running example, but does not
use hybrid automata to model the systems and does not provide
decidability or complexity results.
\cite{FreseB10} searches for strategies controlling 
all vehicles, and employs heuristic methods 
to determine strategies for coordinated 
vehicle movements.
An excellent survey of alternative methods for controlling all 
vehicles to perform collision-free driving tasks is given in \cite{Frese10}.
Both methods share the restriction of the analysis to a small 
number of vehicles, whereas we consider 
an unbounded number of systems. 
\medskip

\cite{HenzingerMineaPrabhu} analyzes the interplay of fixed combinations of
hybrid systems using assume-guarantee reasoning. 
In \cite{JohnsonMitra12b,JohnsonMitra12} 
a small  model theorem for {\em finite} families of 
automata with constant derivatives,  
with a parametric bound on the number of  components, 
is established; 
the discrete transitions describe changes in exactly one system
(thus no global updates of sensors can be modeled).  
Our approach allows
us to consider families with an {\em unbounded} or {\em infinite} number of  
components which are {\em parametric linear hybrid automata}. 
We moreover allow for parallel mode switches and global topology updates. 
In \cite{MickelinOM14}, robust finite abstractions 
with bounded estimation errors are provided for reducing the synthesis of winning 
strategies for LTL objectives to finite state synthesis; 
the approach is used for an aerospace control application. 
\cite{Platzer2010} proposes a quantified differential dynamic logic for specifying and verifying 
distributed hybrid systems but the focus is not
  on providing decidability results or small model property results. 
\medskip

Our current work stands in the tradition of 
\cite{Sofronie-Stokkermans10,DammIS11,sofronie-cade24}, where we 
studied linear hybrid systems in which both mode changes and the 
dynamics can be parametrized. We presented first results on the
verification of families of LHA in \cite{DammHS15}.  This paper 
considerably extends the results presented in \cite{DammHS15}. 
In particular, compared to 
 \cite{DammHS15}, the theoretical results are extended and 
the experimental 
results reported in Section~\ref{experiments} are an order of magnitude
faster than the ones reported in~\cite{DammHS15}; we also 
explain how to use our system and our theory prover \hpilot{} for 
generating (and visualizing) counterexamples to safety. 

\subsection{Paper Structure} 
In Section~\ref{sec:system model} we 
present our model of
spatial families of hybrid automata with its semantics. In
Section~\ref{verif} we introduce the verification properties we consider.
The notions are illustrated on a running
example of cars on a highway. In Section~\ref{Sec:HierarchicalReasoning} we present 
classes of decidable and tractable logical theories, which we use 
in Section~\ref{inv-bmc} for solving the verification tasks and 
proving modularity and
complexity results. In Section~\ref{sec:consequences of locality} we
summarize the main results in the form of a small model property, 
as well as a discussion of the decidability
and complexity of the verification problems we consider. We identify
situations in which the problems are fixed parameter tractable; and
give decidability and complexity results also for 
parametric verification and parameter synthesis.
In Section~\ref{experiments} we discuss our tests
with our systems H-PILoT and HAHA. 
In Section~\ref{sec:conclusions} we present a summary of the results we obtained, followed by
plans for future work. 

\newpage
{\small 
\section*{Contents}
\vspace{-1mm}
\contentsline {section}{\numberline {1}Introduction}{1}{section.1}
\contentsline {subsection}{\numberline {1.1}Related work}{2}{subsection.1.1}
\contentsline {subsection}{\numberline {1.2}Paper Structure}{3}{subsection.1.2}
\vspace{-1mm}
\contentsline {section}{\numberline {2}Spatial Families of Hybrid Automata}{5}{section.2}
\contentsline {subsection}{\numberline {2.1}The language.}{6}{subsection.2.1}
\contentsline {subsection}{\numberline {2.2}Component systems.}{6}{subsection.2.2}
\contentsline {subsection}{\numberline {2.3}Topology}{8}{subsection.2.3}
\contentsline {subsubsection}{\numberline {2.3.1}Topology automata}{8}{subsubsection.2.3.1}
\contentsline {subsubsection}{\numberline {2.3.2}Timed topology automata}{10}{subsubsection.2.3.2}
\contentsline {subsection}{\numberline {2.4}Spatial family of hybrid automata}{10}{subsection.2.4}
\vspace{-1mm}
\contentsline {section}{\numberline {3}Verification Tasks}{12}{section.3}
\contentsline {subsection}{\numberline {3.1}Safety properties}{12}{subsection.3.1}
\contentsline {subsubsection}{\numberline {3.1.1}Safety properties with exhaustive entry conditions}{13}{subsubsection.3.1.1}
\contentsline {subsubsection}{\numberline {3.1.2}Reduction to GMR invariant checking}{14}{subsubsection.3.1.2}
\contentsline {subsubsection}{\numberline {3.1.3}Safety properties with GMR-exhaustive entry conditions}{15}{subsubsection.3.1.3}
\contentsline {subsection}{\numberline {3.2}Reducing verification tasks to satisfiability checking}{16}{subsection.3.2}
\contentsline {subsubsection}{\numberline {3.2.1}Sequentializing parallel jumps}{16}{subsubsection.3.2.1}
\contentsline {subsubsection}{\numberline {3.2.2}Verification of safety properties and satisfiability checking}{17}{subsubsection.3.2.2}
\contentsline {subsubsection}{\numberline {3.2.3}Checking exhaustive entry conditions}{18}{subsubsection.3.2.3}
\vspace{-1mm}
\contentsline {section}{\numberline {4}Automated Reasoning}{20}{section.4}
\contentsline {subsection}{\numberline {4.1}Local theory extensions}{20}{subsection.4.1}
\contentsline {subsection}{\numberline {4.2}Hierarchical reasoning in local theory extensions}{20}{subsection.4.2}
\contentsline {subsection}{\numberline {4.3}Examples of local theories and theory extensions}{21}{subsection.4.3}
\contentsline {subsubsection}{\numberline {4.3.1}Update rules}{21}{subsubsection.4.3.1}
\contentsline {subsubsection}{\numberline {4.3.2}A theory of pointers}{21}{subsubsection.4.3.2}
\contentsline {subsection}{\numberline {4.4}Chains of local theory extensions}{22}{subsection.4.4}
\vspace{-1mm}
\contentsline {section}{\numberline {5}Verification: Decidability and Complexity}{23}{section.5}
\contentsline {subsection}{\numberline {5.1}Verification tasks: Chains of local theory extensions}{25}{subsection.5.1}
\contentsline {subsection}{\numberline {5.2}Verification of safety properties.}{26}{subsection.5.2}
\contentsline {subsubsection}{\numberline {5.2.1}Entry conditions}{26}{subsubsection.5.2.1}
\contentsline {subsubsection}{\numberline {5.2.2}Flows}{29}{subsubsection.5.2.2}
\contentsline {subsubsection}{\numberline {5.2.3}Jumps}{33}{subsubsection.5.2.3}
\contentsline {subsubsection}{\numberline {5.2.4}Topology updates}{36}{subsubsection.5.2.4}
\contentsline {subsection}{\numberline {5.3}Checking exhaustive entry conditions}{38}{subsection.5.3}
\vspace{-1mm}
\contentsline {section}{\numberline {6}Consequences of Locality}{40}{section.6}
\contentsline {subsection}{\numberline {6.1}A small model property}{40}{subsection.6.1}
\contentsline {subsection}{\numberline {6.2}Decidability, Complexity}{41}{subsection.6.2}
\vspace{-1mm}
\contentsline {section}{\numberline {7}Tool Support}{42}{section.7}
\contentsline {subsection}{\numberline {7.1}Input syntax}{42}{subsection.7.1}
\contentsline {subsection}{\numberline {7.2}System architecture}{43}{subsection.7.2}
\contentsline {subsection}{\numberline {7.3}Experiments}{44}{subsection.7.3}
\contentsline {subsubsection}{\numberline {7.3.1}Decision Problems}{44}{subsubsection.7.3.1}
\contentsline {subsubsection}{\numberline {7.3.2}Model generation}{45}{subsubsection.7.3.2}
\contentsline {subsubsection}{\numberline {7.3.3}Complexity}{45}{subsubsection.7.3.3}
\vspace{-1mm}
\contentsline {section}{\numberline {8}Conclusions}{46}{section.8}
\contentsline {subsection}{\numberline {8.1}Summary of results}{46}{subsection.8.1}
\contentsline {subsection}{\numberline {8.2}Plans for further work}{47}{subsection.8.2}
}
\newpage

\section{Spatial Families of Hybrid Automata}
\label{sec:system model}
We study families $\{ S(i) \mid i \in I \}$ consisting of 
an unbounded number of similar 
systems. To describe them, we have to specify 
the properties of the component 
systems  
and the way they obtain information about neighboring systems: 
\begin{itemize}
\item We model the systems $S(i)$ using  
hybrid automata. 
\item For describing the information about neighboring or other
  observed systems we use structures  
$(I, \{ p : I \rightarrow I \}_{p \in P})$, where $I$ is a 
countably infinite set and $P = P_S 
\cup P_N$ is a finite set of unary 
 function symbols 
 which model the way the systems perceive  
other systems using 
 sensors in $P_S$, or by neighborhood connections 
(e.g. established by communication channels) in $P_N$. 
\end{itemize}
We use highway control as a running example. 

\begin{figure}[t]%
\centering 
\includegraphics[width=8cm]{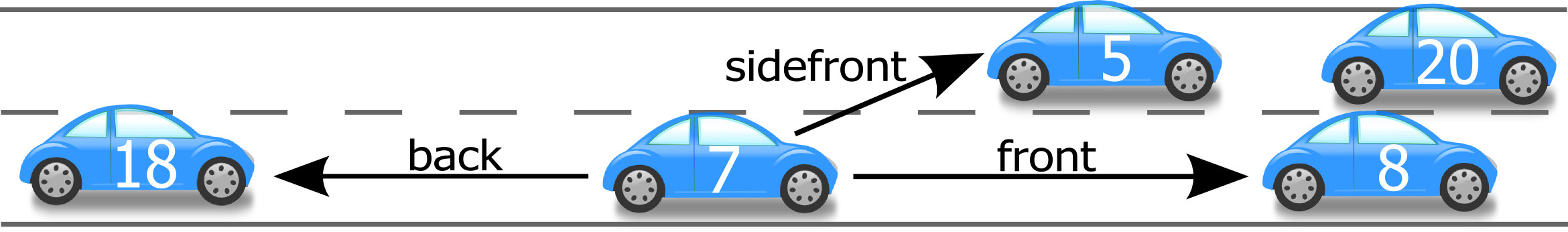}%
\caption{Traffic situation on a highway}%
\label{fig:snapshot}%
\end{figure}  

\begin{ex}
\label{example}
{\em 
Let $I$ be a set of car identities, including the special constant \nil. 

\begin{enumerate}
\item
A car can observe other cars through sensors; these 
are modeled by a finite application-dependent set $P_S$ of 
functions $p: I \to I$, where $p(i)=j$ 
represents the fact that $i$'s $p$-sensor 
observes car $j$. 
We choose $P_S$ to include 
$\back{}$, $\front{}$, $\sidefront{}$, $\sideback{}$,
which indicate the closest car in the respective directions: In Figure~\ref{fig:snapshot}, 
we have $\sidefront{7}=5$, $\back{7}=18$, $\front{7}=8$. 
If sensor $p \in P$ of car $i$ sees no car then $p(i) = \nil$.
We will make these notions more precise in Examples~\ref{run-ex} and~\ref{ex-top-updates}.

\item \label{ex:platoon} 
Car platoons of length at most $n$ can be modeled e.g.\ by choosing a set of 
neighborhood connections $P_N$ including ${\sf leader},$ ${\sf follower}_1,
\dots {\sf follower}_n,$ ${\sf next},$  ${\sf prev}$. 
Car $i$ is leader if ${\sf leader}(i) = i$; 
if ${\sf leader}(j) = i \neq j$, then
$j = {\sf follower}_k(i)$ for some $k \leq n$. 
\end{enumerate}
}\end{ex}

\begin{defi}[Hybrid automata, linear hybrid automata \cite{AlurHH96}] 
\label{def-lha}
A \emph{hybrid automaton} (HA) is a tuple
\[S=(X, Q, \initialStates, \flow, \invariant, E, \guard, \jump)\]
consisting of:  
\begin{enumerate}
\item finite sets $X = \{ x_1, \dots, x_n \}$ (\emph{real-valued variables}) and 
$Q$ (\emph{control modes}); a finite multiset $E$ 
with elements in $Q \times Q$ (\emph{control switches}); 
\item families $\initialStates = \{ \initialStates_q \mid q \in Q \}$
  and $\invariant = \{ \invariant_q \mid q \in Q \}$ of predicates
  over $X$, 
defining the \emph{initial states} and \emph{invariant conditions} for each control
mode,  
and 
$\flow = \{ {\sf flow}_q \mid q \in Q \}$ of predicates over $X\cup
{\dot X}$ specifying the dynamics in each control mode, 
where ${\dot X} =\{ {\dot x_1}, \dots, {\dot x_n} \}$ 
(${\dot x_i}$ is the derivative of $x_i$); 
\item families $\{ \guard_e \mid e \in E \}$ of predicates over $X$ (\emph{guards}) 
and $\{ \jump_e \mid e \in E \}$ of predicates over
$X \cup X'$ (\emph{jump conditions}) for the control
switches, 
where $X' =\{ x'_1, \dots, x'_n \}$ is a copy of $X$. 
\end{enumerate}
A {\em linear hybrid automaton} (LHA) is a HA in which 
for every $q \in Q, e \in E$:
\begin{enumerate}[label=(\roman*)]
\item $\invariant_q$, $\initialStates_q$, $\jump_e$ and 
$\guard_e$ are convex linear predicates\footnote{A convex 
	linear predicate is a finite conjunction of linear inequalities over 
	${\mathbb R}$.} 
and
\item $\flow_q$ is a convex linear predicate (with only  {\em
	non-strict} inequalities) over ${\dot X}$. 
\end{enumerate}
\end{defi}

A {\em state} of $S$ is a pair $(q, a)$, where $q \in Q$ and 
$a {=} (a_1, \dots, a_n)$, where $a_i {\in} {\mathbb R}$ is a value for $x_i {\in} X$.
A state $s = (q, a)$ is {\em admissible} (resp.\ {\em initial}) 
if ${\sf Inv}_q$ (resp.\ ${\sf Init}_q$) is true when each $x_i$ is replaced by $a_i$.
A state can change by a jump (instantaneous transition that changes
the control mode and the values of the variables according to the
jump conditions), or 
by a flow (evolution in a mode $q$ where the values of the variables 
change according to the ${\sf flow}_q$).
 
\subsection{The language.} 
To describe the families $\{ S(i) \mid i \in I \}$, 
the topology $(I, \{ p : I \rightarrow I \}_{p \in P})$ and its 
updates, and the safety properties we are interested
in, we use a two-sorted first-order language 
${\cal L}_{{\sf index},{\sf num}}$ 
of a theory of pointers with two sorts, ${\sf index}$ and ${\sf
  num}$. Sort ${\sf index}$ is used for representing the indices
and sort ${\sf num}$ is used for numerical values. 
The signature of the theory contains  a constant $\nil$ of sort ${\sf index}$, 
unary function symbols in $P$ (sort ${\sf index}
\rightarrow {\sf index}$) for modeling pointer fields, and a 
set $X$ (sort  ${\sf index} \rightarrow {\sf num}$) for modeling 
the scalar (numeric) information associated with the indices
(values of the continuous variables of the systems). 
A theory ${\cal T}_{\sf num}$ (sort ${\sf num}$) is used 
for describing properties of the values of the continuous variables
of the systems (e.g.\ the 
theory ${\mathbb R}$ of real numbers, or 
linear real arithmetic $LI({\mathbb R})$). 
We consider first-order formulae in the language ${\cal L}_{{\sf index},{\sf num}}$. 
Variables of sort {\sf index}  are denoted with indexed versions of $i, j,
k$; variables of sort ${\sf num}$ are denoted $x_1, \dots, x_n$. 

\subsection{Component systems.} 
The component systems are similar\footnote{The results can be adapted to 
the situation when a finite number of types of systems are given 
and the description of each $S(i)$ is of one of these types.} 
hybrid automata $\{ S(i) \mid i \in I \}$, with:
\begin{itemize}
\item the same set of control modes $Q$ and the same 
mode switches $E \subseteq Q \times Q$, 
\item real valued variables $X_{S(i)}$, partitioned into a set 
$X(i) = \{ x(i) \mid x \in X \}$ of variables describing the
states of the system $S(i)$ and  a set $X_P(i) = \{ x_p(i) \mid x \in
X, p \in P \}$ describing the state of the neighbors $\{ p(i)  \mid p
\in P \}$  of $i$,  
where $X = \{ x_1, \dots, x_n \}$. 
\end{itemize}
We consider two possibilities for $x_p(i)$:
\label{a-b}
\begin{enumerate}[label=(\alph*)]
\item{} \label{variant:continuous sensors} {\em Continuous sensors:} $x_p(i)$ is at any moment the value of $x(p(i))$, the value 
  of variable $x$ for the system $S(p(i))$ and is controlled by suitable flow/jump
  conditions of $S(p(i))$;
\item{} \label{variant:intermittent sensors} {\em Intermittent sensors:} $x_p(i)$ is the value of $x(p(i))$ which was sensed by the 
  sensor in the last measurement, and does not change between 
measurements. 
\end{enumerate}
We assume that all sets $X(i), i \in I$ are disjoint. 
Every component system $S(i)$ has the form: 
\newcommand{\variablesofautomaton}{X(i) \cup X_P(i)}
\newcommand{\derivativesofautomaton}{\{\variable[.]{i},
  \variable[.]{\pointer{i}} \mid \variable{}\in X, \pointer{} \in P\}}
\[ S(i) = (\variablesofautomaton, Q, \flow(i), \invariant(i), \initialStates(i), E, \guard(i), \jump(i))\]
where -- 
with the notations in Definition~\ref{def-lha}: 
\begin{itemize}
\item for every $q \in Q$ 
${\sf Inv}_q(i)$, ${\sf Init}_q(i)$
is a conjunction of formulae of the form ${\cal E} \vee C$, where $C$ is a predicate over 
$X_{S(i)}$ 
and ${\cal E}$ is a disjunction 
of equalities of the form $i = \nil$ and $p(i) = \nil$ if 
$x_p(i)$ occurs in $C$. We will in general assume that ${\sf Init}_q$
includes ${\sf Inv}_q$ as a conjunct. 
\item 
for every $q \in Q$, ${\sf flow}_q(i)$ is a  conjunction of formulae
of the form ${\cal E} \vee C$, where $C$ is a predicate 
over $X_{S(i)} \cup \dot{X}_{S(i)}$ and ${\cal E}$ is a disjunction 
of equalities of the form $i = \nil$ and $p(i) = \nil$ if 
$x_p(i)$ occurs in $C$. 
\item  for every $e \in E$,  ${\sf guard}_e(i)$ is a conjunction of
  formulae of 
  the form $\neg ({\cal E} \vee C)$, where $C$ is a predicate 
over $X_{S(i)}$ and ${\cal E}$ is a disjunction 
of equalities of the form $i = \nil$ and $p(i) = \nil$ if 
$x_p(i)$ occurs in $C$. 

\item  for every $e \in E$, $ {\sf jump}_e(i)$ is a conjunction of
  formulae of the form ${\cal E} \vee C$, where $C$ is a predicate over 
$X_{S(i)} \cup X'(i)$ and ${\cal E}$ is a disjunction 
of equalities of the form $i = \nil$ and $p(i) = \nil$ if 
$x_p(i)$ occurs in $C$. 
\end{itemize}
All these formulae can also be regarded as 
${\cal L}_{\sf  index, num}$-formulae; 
for all $i \in I$ 
they differ only in the variable index.

The component 
$S(i)$ is {\em linear} if 
\begin{enumerate}[label=(\roman*)]
\item for every $q \in Q$, $\flow_q(i)$ contains only variables in
$\dot{X}_{S(i)}$ and 
\item for every $q \in Q$ and $e \in E$, $\flow_q(i),$ $\invariant_q(i),$ $\initialStates_q(i),$
$\guard_e(i)$, $\jump_e(i)$ are conjunctions of formulae ${\cal E} \vee C$,
as above, where $C$ is a linear inequality (non-strict for flows).  
\end{enumerate}
 We also consider systems of {\em parametric} LHA, in which
some coefficients in the linear inequalities 
(and also bounds for invariants, guards or jumps) 
are parameters in a set ${\sf Par}$.

\begin{ex}
\label{run-ex}
{\em 
Consider the following model of a system of cars, which is also depicted in Figure~\ref{fig:cars}:
\begin{figure}[t]%
\begin{center}
\hspace{-5em}
\arrayrulecolor{blue!40!black!60}
\scalebox{0.8}{
\begin{tikzpicture}
  \HATikzStyle
  \node (accbot) [rectangle split, line width=1.2pt]
             {\textsf{Appr} \nodepart{second} 
    \multimath{{\sf Inv}_{\textsf{Appr}}{:}\\\ \\{\sf flow}_{\textsf{Appr}}{:}\ {}\\\ }%
		\multimath{1\leq \lane{i}\leq 2\\
    \front{i}=\nil\vee {\sf pos}_{\sf front}(i)-\pos{i}\geq d\\\hline
    \lane[.]{i}=0\\
		\front{i}=\nil\vee {\dot {\sf pos}}_{\sf front}(i) \leq {\pos[.]{i}}\leq 100
		}};
  \node (decbot) [rectangle split, below=3.8cm of accbot, line width=1.2pt]
             {\textsf{Rec} \nodepart{second} 
    \multimath{{\sf Inv}_{\textsf{Rec}}{:}\\\ \\{\sf flow}_{\textsf{Rec}}{:}\ {}\\\ \\\ }%
    \multimath{1\leq \lane{i}\leq 2\\
		\front{i}=\nil\vee {\sf pos}_{\sf front}(i)-\pos{i}\leq D\\\hline
    \lane[.]{i}=0\\
		0\leq{\dot {\sf pos}}_{\sf front}(i)\\
    \front{i}=\nil\vee {\dot {\sf pos}}_{\sf front}(i)\leq{\pos[.]{\front{i}}}
		}};
  \tikzstyle{every node}=[font=\scriptsize,fill=none];
  \path[jump,line width=1.2pt]
      (decbot) edge[bend right,transform canvas={xshift=4.5em}]
		          node [right,pos=.5]{
          \multimath{{\sf guard}{:}\ {}\\[4.2em]{\sf jump}{:}}%
					\multimath{
          \front{i}\not=\nil\\ {\sf pos}_{\sf front}(i)-\pos{i}\leq D'
          \\\back{i}=\nil \vee \pos{i}-{\sf pos}_{\sf back}(i)\geq d'
          \\\sideback{i}=\nil \vee \pos{i}-{\sf pos}_{\sf sideback}(i)\geq d'
          \\\sidefront{i}=\nil \vee {\sf pos}_{\sf sidefront}(i)-\pos{i}\geq d'
          \\\hline\lane[']{i}= 3-\lane{i}}}
					(accbot)
      (accbot) edge[loop right, looseness=4]
		          node [right,pos=.5] {
          \multimath{{\sf guard}{:}\ {}\\[4.2em]{\sf jump}{:}}%
					\multimath{
          \front{i}\not=\nil\\ {\sf pos}_{\sf front}(i)-\pos{i}\leq D'
          \\\back{i}=\nil \vee \pos{i}-{\sf pos}_{\sf back}(i)\geq d'
          \\\sideback{i}=\nil \vee \pos{i}-{\sf pos}_{\sf sideback}(i)\geq d'
          \\\sidefront{i}=\nil \vee {\sf pos}_{\sf sidefront}(i)-\pos{i}\geq d'
          \\\hline\lane[']{i}= 3-\lane{i}}}
					(accbot)
     (accbot) edge[transform canvas={xshift=-1em}]
		          node[rotate=-90,anchor=south] {
							                               \multimath{{\sf guard}{:}\ {}\\[1em]}%
							                               \multimath{\front{i}\not=\nil\\ {\sf pos}_{\sf front}(i)-\pos{i}\leq D'
																						}} (decbot)
     (decbot) edge[transform canvas={xshift=-3em}]
		          node[rotate=90,anchor=south] {
							                              \multimath{{\sf guard}{:}\ {}\\[1em]}%
							                              \multimath{\front{i}\not=\nil\\ {\sf pos}_{\sf front}(i)-\pos{i}\geq d'
																						}} (accbot)
        ;
\end{tikzpicture}}
\hspace{-4em}
\vspace{-1.5em}
\end{center}
\caption{Hybrid automaton modeling the behavior of a car on a two-lane highway}%
\label{fig:cars}%
\end{figure}
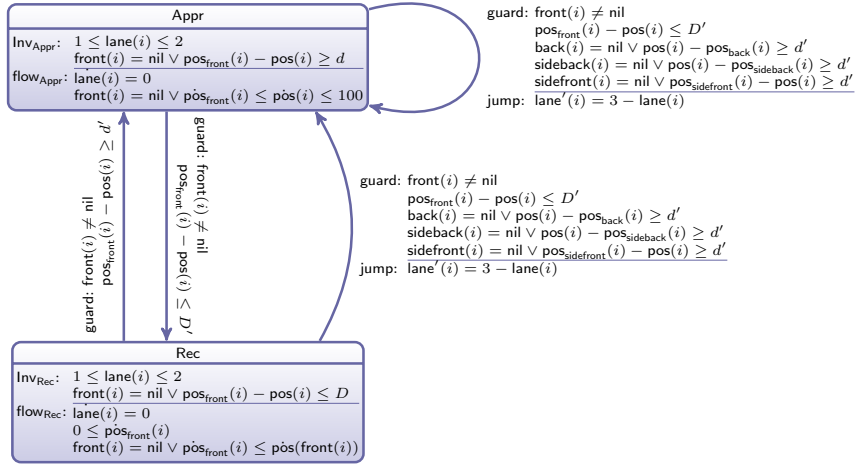 
The controlled variables are the position and the lane of the car, so  $X = \{ {\sf pos}, {\sf lane} \}$. 
The car can drive on either lane 1 or lane 2. 
Its sensors provide information about the car in 
front and back on the same lane ($\front{}, \back{}$) 
and about the closest cars on the 
other lane ($\sidefront{}, \sideback{}$). 
Thus the set of sensors is \[P=\{\back{},\front{},\sideback{},\sidefront{}\}\ .\] 
Each car is modeled by a hybrid 
automaton with set of continuous variables 
\[X=\{ {\sf pos}(i), {\sf lane}(i) \} \cup  \{ {\sf pos}_p(i), {\sf
  lane}_p(i) \mid p \in P \}\]
and modes 
\[Q = \{ {\sf Appr}, {\sf Rec} \}\ .\]
%
We assume that $x_p(i) = x(p(i))$ (continuous sensors, variant~\ref{variant:continuous sensors} above)
and use parameters ${\sf Par} = \{ d, d', D, D' \}$.

\begin{description}
\item[Initial states:] As initial states, we allow all states where
  ${\sf pos}_{\sf front}(i)-\pos{i}\geq d'$ if $\front{i}\not=\nil$,
  and where the respective mode invariant is satisfied: 
\begin{itemize}
\item ${\sf Init}_{\sf Appr}$ and ${\sf Init}_{\sf Rec}$ are $(i = \nil
  \vee {\sf front}(i) = \nil \vee {\sf pos}_{\sf front}(i)-\pos{i}\geq
  d')$. 
\end{itemize}
\item[Invariants; flow conditions:] ~~\\
{\bf Mode {\sf Appr}:} car $i$  keeps its velocity high
enough to approach the car ahead. 
\begin{itemize}
\item ${\sf Inv}_{\sf Appr}$ is 
$(i = \nil \vee 1 {\leq} \lane{i} \leq 2)\wedge(i = \nil \vee
\front{i} = \nil\vee {\sf pos}_{\sf front}(i) - \pos{i} \geq 
d)$; 
\item ${\sf flow}_{\sf Appr}$ is 
$(i = \nil \vee {\sf la\dot ne}(i) = 0)\wedge(i = \nil \vee \front{i}
= \nil\vee {\dot {\sf pos}}_{\sf
  front}(i) \leq {\pos[.]{i}})$ \\
$~~~~~~~~~~~~~~~~~~~~~~~~~~~~~~~~~~~~~~~~~~\,\wedge (i = \nil \vee 0 {\leq} {\pos[.]{i}} {\leq} 100)$.
\end{itemize}

{\bf Mode {\sf Rec}:} car $i$ maintains a lower  
velocity to fall back.  
\begin{itemize}
\item ${\sf Inv}_{\sf Rec}$ is  
$(i = \nil \vee 1 {\leq} \lane{i} \leq 2)\wedge(i = \nil \vee \front{i}{=}\nil\vee {\sf pos}_{\sf front}(i){-}\pos{i}{\leq}
D)$; 

\item ${\sf flow}_{\sf Rec}$ is $((i = \nil \vee {\sf la\dot ne}(i)=0)
  \wedge
(i = \nil \vee 0\leq{\pos[.]{i}}) $ \\
$~~~~~~~~~~~~~~~~~~~~~~~~~~~~~~~~~~~~~~~~~~\,\wedge (i = \nil \vee \front{i}=\nil\vee {\pos[.]{i}}\leq {\dot {\sf pos}}_{\sf
  front}(i) ))$. 
\end{itemize}

\item[Mode switches:]  ~~ \\
A mode switch (without resets) can happen if $i \neq \nil$, $\front{i}\not=\nil$ (there is a car ahead)
and the distance to that car leaves a predefined
range, i.e.\
\begin{itemize}
\item ${\sf pos}_{\sf front}(i)-\pos{i}\leq D'$ (switch from {\sf Appr} to
{\sf Rec}) or 
\item ${\sf pos}_{\sf front}(i)-\pos{i}\geq d'$ (switch from {\sf Rec} to
{\sf Appr}). 
\end{itemize}
Another mode switch to mode {\sf Appr}, 
which changes between lanes 1 and 2 with reset $\lane[']{i}{=} 3{-}\lane{i}$,
can happen when $i \neq \nil$ and:  
\begin{itemize}
\item the car in front is too close ($\front{i}\not=\nil\wedge {\sf pos}_{\sf front}(i)
{-}\pos{i}\leq D'$) and 
\item there is space to start the
maneuver: $\back{i} {=} \nil \vee \pos{i}{-} {\sf pos}_{\sf back}(i){\geq} d'$.
Similarly for $\sideback{i}$ and $\sidefront{i}$.
\end{itemize}
\end{description}
}
\end{ex}

\subsection{Topology}

We now present a possibility of modeling the topology of the family of
systems using a one-state automaton, where the transitions are
labeled with updates of the values of the pointers
(Section~\ref{top-automata}), 
and a refinement of this model in which clocks are additionally used
(Section~\ref{timed-top-automata}). 

\subsubsection{Topology automata} 
\label{top-automata}

We model the topology of the family of systems and its updates using an automaton 
$\environment$ with one mode, having 
as read-only-variables all  variables in $\{ x(i) \mid x \in X,  i
\in I \}$ and as write variables  $\{ x_p(i) \mid p \in P,  i\in I \}$, where $P
= P_S \cup P_N$. In addition, $\environment$ updates the 
functions $p : I \rightarrow I$, where $P
= P_S \cup P_N$.
 
The  initial states ${\sf Init}$ are described using ${\cal L}_{{\sf index}, {\sf num}}$-formulae.  
The jumps can represent updates of the sensor values $p(i), p \in P_S$, for a single system $S(i)$, 
but also synchronized 
global updates of the sensors $p \in P_S$ or neighborhood connections $p \in P_N$ for subsets 
of systems with a certain property (described by a 
formula). This can be useful when modeling 
systems of systems with an external controller (e.g.\ systems of car
platoons)  and entails 
a simultaneous update of an unbounded set of variables.\footnote{
Our choice allows us
  to uniformly represent various types of topology updates, from 
purely local ones to global updates, without loss of generality.} 
Therefore, the description of the mode switches (topology updates) 
in $\environment$ is of a global nature and is done using 
${\cal L}_{{\sf index}, {\sf num}}$-formulae. 

The update rules for $p \in P$, which we denote as ${\sf Update}(p, p')$, are conjunctions of
implications of the form 
\begin{eqnarray}
\forall i (i \neq \nil \wedge \phi^p_k(i) \rightarrow F^p_k(p'(i), i)),
\quad \quad k \in \{ 1, \dots, m \},
\end{eqnarray} 
which describe how the values of the pointer $p$ change depending on a
set of mutually exclusive conditions $\{ \phi^p_1(i),
\dots,\phi^p_m(i) \}$ such that: 
\begin{itemize}
\item $\phi^p_k(i)$ and $F^p_k(j, i)$ are formulae over the 2-sorted language ${\cal L}_{{\sf index}, {\sf num}}$
without any occurrence of unary functions in $P'$;
\item if $p \in P_S$ ($p$ represents a sensor), the formulae
  $\phi^p_k(i)$ and $F^p_k(j, i)$ 
also do not contain functions in $P$;  
\item under the condition $\phi^p_k(i)$, the existence 
of a value for $p'(i)$ such that $F^p_k(p'(i), i)$ holds must be
guaranteed, i.e.\ 
\[ \models \phi^p_k(i) \rightarrow \exists j \, F^p_k(j, i); \]

\item The variables 
$\{ x(i) \mid x {\in} X, i {\in} I \}$ can be used in the guards of ${\sf Update}(p, p')$, 
but cannot be updated by $\environment$. 

\item If $x_p(i)$ stores the value of $x(p(i))$ at the update of $p$ (variant~\ref{variant:intermittent sensors} on page~\pageref{a-b}),  
then the update rules also change $x_p(i)$, so  $F^p_k(p'(i), i)$ 
must contain $x'_p(i) = x(p'(i))$ as a conjunct. 
\end{itemize}

\begin{ex}
\label{ex-top-updates}
{\em 
We present possible update rules for the topology and initial states
for the model of cars in Example~\ref{run-ex}. 
Consider the following formulae: 
\begin{itemize}
\item  
$\textsf{ASL}(j,i){:\ }j \neq \nil \wedge \lane{j} = \lane{i} \wedge
\pos{j} >\pos{i}$, which expresses the fact that $j$ is ahead of $i$ on the same 
lane, and 
\item ${\sf Closest_f}(j,i){:\ }\textsf{ASL}(j, i) \wedge \forall k (\textsf{ASL}(k, i) 
{\to} \pos{k} \geq \pos{j})$, which  
expresses the fact that $j$ 
is ahead of $i$ on the same 
lane and there is no car between them. 
\end{itemize}
{\em Update rules.} The rule for updating the front
sensor of all cars with a given property expressed by a formula ${\sf
  Prop}$ and of no other car 
is described by ${\sf Update}({\sf front}, {\sf front}')$: 
 \begin{align*}
  \forall i\big(i \neq \nil \wedge {\sf Prop}(i) \wedge \neg \exists j (\textsf{ASL}(j,i)) &\to \front[']{i} = \nil\big)\\
  \forall i\big(i \neq \nil \wedge {\sf Prop}(i) \wedge \phantom{\neg}\exists j
  (\textsf{ASL}(j,i)) &\to {\sf Closest_f}(\front[']{i}, i)\big) \\
\forall i\big(i \neq \nil \wedge \neg {\sf Prop}(i) & \to \front[']{i} = \front[]{i} \big)
\end{align*}
Below are three examples of formulae which can describe a property ${\sf
  Prop}$: 
\begin{itemize}
\item[(1)] If ${\sf Prop}(i) = (i = i_0)$, only the front sensor of
car $i_0$ is updated. 
\item[(2)] For car platoons, ${\sf Prop}(i)$ can be ${\sf leader}(i) = i_0$; we then
obtain a coordinated update for all platoon members. 
\item[(3)] If ${\sf Prop}(i) =$ true,  ${\sf Update}({\sf front}, {\sf
  front}')$ describes a global update. 
\end{itemize}
{\em Initial states.} The initial states can e.g.\ be the states in which 
all sensor pointers have the correct value, as if they had 
just been updated. For ${\sf front}$ this can be expressed by
the following set of formulae: 

\medskip

\noindent $\begin{array}{@{}rl} 
\forall i (i \neq \nil \wedge {\sf front}(i) = \nil & \to \forall k (k \neq \nil \wedge k \neq i \wedge  {\sf pos}(k) \geq {\sf
  pos}(i) \rightarrow {\sf lane}(k) \neq {\sf lane}(i))) \\
\forall i (i \neq \nil \wedge {\sf front}(i) \neq \nil & \to {\sf
  pos}_{\sf front}(i) > {\sf pos}(i)  \wedge {\sf lane}_{\sf
  front}(i) = {\sf lane}(i) \wedge \\
& ~~~\forall k (k \neq \nil \wedge k \neq i \wedge  {\sf pos}(k) \geq {\sf
  pos}(i) \wedge {\sf lane}(k) = {\sf lane}(i) \\
& ~~~~~~~\to {\sf pos}(k) \geq {\sf
  pos}_{\sf front}(i)) \wedge \\
& ~~~{\sf pos}({\sf front}(i)) = {\sf pos}_{\sf front}(i) \wedge {\sf
  lane}({\sf front}(i)) = {\sf lane}_{\sf front}(i)). 
\end{array}$ 

\medskip
\noindent Alternatively, we can express this using formulae similar to the update rules: 
\begin{align*}
  \forall i\big(i \neq \nil \wedge {\sf Prop}(i) \wedge \neg \exists j (\textsf{ASL}(j,i)) &\to \front[]{i} = \nil\big)\\
  \forall i\big(i \neq \nil \wedge {\sf Prop}(i) \wedge \phantom{\neg}\exists j
  (\textsf{ASL}(j,i)) &\to {\sf Closest_f}({\sf front}(i), i)\big) 
\end{align*} 
}
\end{ex}
\begin{ex}
{\em 
Consider a car platoon as in Example~\ref{example}~\ref{ex:platoon}. 
The situation when a car $i_0$ (who is not a leader) leaves
the platoon can e.g.\ be described by:
\begin{align*}
 {\sf leader}'(i_0) = i_0  \quad \quad  & {\sf next}'(i_0) = \nil  \quad\quad  {\sf prev}'(i_0) = \nil\\
{\sf prev}(i_0) \neq \nil \to{} & {\sf next}'({\sf prev}(i_0)) = {\sf
  next}(i_0)\\
{\sf next}(i_0) \neq \nil \to{} & {\sf prev}'({\sf next}(i_0)) = {\sf
  prev}(i_0)\\
\forall i (i \neq i_0 \wedge i \neq {\sf prev}(i_0) \to{} & {\sf next}'(i) = {\sf
    next}(i))\\
\forall i (i \neq i_0 \wedge i \neq {\sf next}(i_0) \to{} & {\sf prev}'(i) = {\sf
    prev}(i))\\
\end{align*}}
\end{ex}

\subsubsection{Timed topology automata} 
\label{timed-top-automata}

If we want to ensure that the component systems 
update the information about their neighbors sufficiently often, 
we can use additional clock variables $\{ c_p(i) \mid i
\in I, p \in P \}$, satisfying flow conditions of the form $\dot{c}_p(i) = 1$. 
Every topology update involving a set of systems and pointer field $p$
has the effect that the clocks $c_p(i)$ for all systems $i$ in that set are set to 
$0$ (added to the conclusion of the topology updates). 

\begin{ex}
{\em In Example~\ref{ex-top-updates} 
the consequence of the update rules ${\sf Update}({\sf front}, {\sf
  front}')$ for ${\sf front}$ 
would contain as a conjunct the formula $c'_{\sf front}(i) = 0$.
} 
\end{ex}
In addition, we can require that 
for every system $i$ the interval between two updates of $p \in P$ is 
at most $\Delta t(i)$.  
Then $\initialStates_{\environment}$ 
contains $\forall i \, c_p(i) = 0$ as a conjunct; 
the invariant of the mode of ${\sf Top}$ 
contains $\forall i \, 0 \leq c_p(i) \leq \Delta t (i)$; 
and if $c_p(i) = \Delta t(i)$ a topology update for system $i$
must take place. 

\subsection{Spatial family of hybrid automata}

\begin{defi}[Spatial Family of Hybrid Automata]
A \emph{spatial family of hybrid automata (SFHA)} 
is a family of the form
\[S = (\environment, \{S(i) \mid i\in I\}),\]
where $\{ S(i) \mid i \in I \}$ is a system of similar hybrid automata
and $\environment$ is a topology automaton. 
If for every $i \in I$, $S(i)$ is a linear hybrid automaton, we talk
about a \emph{spatial family of linear hybrid automata (SFLHA)}. 
If the topology automaton is timed, we speak of a {\em  spatial family of
  timed (linear) hybrid automata (SFT(L)HA)}.
\end{defi}

\begin{defi}[Decoupling]
An SFLHA $S$ is {\em decoupled} if 
the real-valued variables in the guard of a
mode switch of $S(i)$ can only be reset in a jump by $S(i)$ or by
$\environment$.
\end{defi}

\noindent
{\bf Remark:} In the variant with continuous sensors (variant~\ref{variant:continuous sensors} on page~\pageref{a-b}), we have $x_p(i) = x(p(i))$ for every $i\in I$.
If $x_p(i)$ is used in the guard of a mode switch of $S(i)$,
then in order to ensure that $S$ is decoupled, no jump of $S(p(i))$ should reset $x(p(i))$. 

In the variant with intermittent sensors (variant \ref{variant:intermittent sensors}),
$x_p(i)$ is the value sensed by the sensor $p$  
in the last measurement
and so $S$ is always decoupled.

\begin{ex} 
{\em 
In our running highway example (Example~\ref{run-ex},~\ref{ex-top-updates}) 
only the variables $\pos{i}$, ${\sf pos}_{\sf front}(i)$, ${\sf pos}_{\sf back}(i)$, ${\sf pos}_{\sf sidefront}(i)$, and ${\sf pos}_{\sf back}(i)$ are used in jump guards. Since no jump of a car resets its position, the system is decoupled. Note that if ${\sf lane}_{\sf front}(i)$ were used in any jump guard, the system would not be decoupled in variant~\ref{variant:continuous sensors}, because ${\sf front}(i)$ can reset its lane during a jump.
}
\end{ex} 

\begin{defi}[States and Runs] Let $S = ({\sf Top}, \{ S(i) \mid i \in
I \})$ be a spatial family of hybrid automata. 
\begin{itemize}
\item A state $s = (q, a)$ of $S$ consists of a tuple $q = (q_i)_{i \in 
  I} \in Q^I$ of modes of the component automata and a tuple $a$ of 
values of the variables of all components. 
A state $(q, a)$ is admissible if the values in $a$ satisfy the
invariants of ${\sf Top}$ and the restriction to the variables of
  $S(i)$ satisfies ${\sf Inv}_{q_i}(i)$, for all $i \in I$.

\item Initial states of $S$ are the initial states of $\environment$ whose 
restriction to the variables of $S(i)$ are initial states of 
$S(i)$, for all $i \in I$.
\item A state change $(s, s')$ is a flow of length $t$ if 
its restriction 
to the variables of $S(i)$ is a flow of length $t$, for all $i \in I$.

\item A state change $(s, s')$ is a jump if its restriction 
to the variables of $S(i)$ is a jump or else a flow of length 0, for
all $i \in I$. 

\item A run of $S$ is a sequence 
$s_0,s_1,\ldots$ of admissible states where: 
\begin{enumerate}[label=(\roman*)]
\item $s_0$ is an initial state of $S$, 
\item each pair $(s_j, s_{j+1})$ is a jump, a flow or a topology
update, and 
\item each flow is followed by a jump or a topology update.
\end{enumerate}
\end{itemize}
\end{defi}

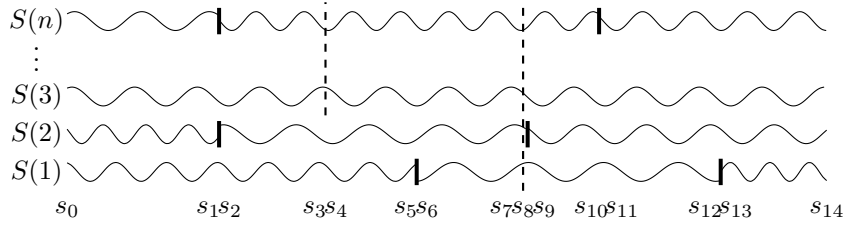
\begin{figure}[t]%
  \pgfmathsetseed{1138}
  \newcommand{\flowline}[3]{
    \pgfmathsetmacro{\secondx}{#2+.0249}
    \pgfmathsetmacro{\lastx}{#3+.001} 
    \pgfmathparse{rnd}
    \draw[smooth, variable=\x, samples at={#2,\secondx,...,\lastx}]
      plot (\x, { #1+sin((3.14*\pgfmathresult+(10.0+15*\pgfmathresult)*\x) r)/4});
  }
  \newcommand{\jumpmark}[2]{
    \draw[line width=1.5pt] (#2,#1+.35)--++(0,-.7);
  }
  \newcommand{\flowjump}[3]{
    \flowline{#1}{#2}{#3}
    \jumpmark{#1}{#3}
  }
  \centering
  \begin{tikzpicture}[yscale=.5,xscale=2]
    \foreach \i in {1,...,3} {\node at (-.2,\i) {$S(\i)$};}
      \node at (-.2,4.2) {$\vdots$};
      \node at (-.2,5) {$S(n)$};      
    \foreach \i [count=\ci] in {.93,1.07,1.63,1.77,2.23,2.37,2.86,3,3.14,3.45,3.65,4.2,4.4,5} {\node at (\i,0) {$s_{\ci}$};}
      \node at (0,0) {$s_{0}$};
    \flowjump 501 \flowjump 51{3.5} \flowline 5{3.5}5
    \flowline 305
    \flowjump 20{1} \flowjump 2{1}{3.03} \flowline 2{3.03}5
    \flowjump 10{2.3} \flowjump 1{2.3}{4.3} \flowline 1{4.3}5
    \draw[dashed,thick] (1.7,2.5)--++(0,3);
    \draw[dashed,thick] (3,0.5)--++(0,5);
  \end{tikzpicture}
  \caption{Visualization of a run of an SFLHA with components $S(1),\ldots,S(n)$.
    Time passes towards the right.
    Jumps in a system $S(i)$ are marked by a short vertical line, and local or global updates by a
    dashed line.
    The curve corresponding to $S(i)$ between such lines represents a
    flow in system $S(i)$.}%
  \label{fig:run}%
\end{figure}

A visualization of a run of an SFLHA is depicted in Figure~\ref{fig:run}.
(Note that property (iii) of runs does not restrict the set of states that are
reachable in a run.)

\section{Verification Tasks}
\label{verif}

\noindent 
The properties of 
SFLHA we consider are specified in 
a logic which combines first-order logic over the language
${\cal L}_{{\sf index},{\sf num}}$  and temporal logic: 
Formulae are constructed inductively 
from atoms using temporal operators and quantification over 
variables of  sort {\sf index}. 
Since runs of the system define valuations of variables 
for each point in time,  
the semantics of such formulae 
is defined canonically, see e.g.~\cite{HungarGD95}.
We consider {\em safety properties} of the form: 

\[ \Phi_{\sf entry}
  \rightarrow \Box \Phi_{\sf safe}, \] 

\noindent which state that for every run of the composed system,  
if $\Phi_{\sf entry}$ holds at the beginning of the run then 
$\Phi_{\sf safe}$ always holds during the run. 
\begin{ex}
\label{ex-properties-global-local}
{\em 
Collision freedom can be expressed using the  
formula 
\[\Phi^g_{\sf safe}: 
 \forall i, j (i {\neq} \nil \wedge j {\neq} \nil \wedge {\sf lane}(i){=}
 {\sf lane}(j) \wedge {\sf pos}(i) {>} {\sf pos}(j) \rightarrow {\sf
   pos}(i) - {\sf pos}(j) {\geq} d_s)\]
for a suitably  
chosen constant $d_s > 0$ (global safety distance)
or  by referring only to the ``neighbors'', using 
$\Phi^l_{\sf safe} = \bigwedge_{\pointer{} \in P} \Phi^{\pointer{}}_{\sf safe}$, 
where e.g.\ $\Phi^{\sf front}_{\sf safe}$ is:  

\[\forall i  (i \neq \nil \wedge {\sf front}(i) \neq \nil  {\rightarrow}
\pos{{\sf front}(i)} - \pos{i} \geq d_s). \]
} 
\end{ex}

\

In Section~\ref{sec:safety properties} we identify a class of general safety properties with  
what we call \emph{exhaustive entry conditions} (Definition~\ref{eec}) which can be reduced to  
invariant checking for certain  
\emph{mode reachable} states (Definition~\ref{def:mode reachable}). 
In Section~\ref{sec:reduction to satisfiability} 
we then show that
for decoupled SFLHA we can reduce checking invariance for mode
reachable states of $\Phi_{\sf safe}$ to satisfiability checking in
suitable 
logical theories, which are combinations of $LI({\mathbb R})$
possibly extended with functions $x_i$ 
satisfying additional properties (boundedness, continuity, boundedness conditions 
for the slope), 
and theories of pointers for modeling the 
information provided by the sensors.

Using decidability results presented in
Section~\ref{Sec:HierarchicalReasoning}, 
in Section~\ref{inv-bmc} 
we identify situations in which the analysis of safety properties 
$\Phi_{\sf entry}  \rightarrow \Box \Phi_{\sf safe}$  
can be precisely reduced to  a neighborhood of
bounded size of the systems for which 
$\Phi_{\sf safe}$ could fail. This allows us to prove a small 
model property 
and to identify safety properties which are decidable resp.\
fixed parameter tractable. 

\

\medskip\noindent{\bf Notation.} In what follows, sequences $i_1, \dots, i_k$ of variables of sort ${\sf index}$    
are denoted with ${\overline i}$, sequences $x_1, \dots, x_n$ (resp.\ $\dot{x}_1, \dots, \dot{x}_n$)    
with ${\overline x}$ (resp.\ $\overline{\dot{x}}$).    
The sequence  $x_1(i), \dots, x_n(i)$ of all variables of $S(i)$ is   
denoted with ${\overline x}(i)$, and $\dot{x}_1(i), \dots, \dot{x}_n(i)$ with $\overline{\dot{x}}(i)$.    
To refer to the value of $x(i)$ at time $t$, we write $x(i, t)$.   
The sequence $x_1(i, t), \dots, x_n(i, t)$ of values of variables of    
system $S_i$ at a time $t$ is denoted ${\overline x}(i, t)$.

\subsection{Safety properties} 
\label{sec:safety properties}
Safety of LHA is in general undecidable; 
classes of LHA and safety properties which are decidable have been 
identified in several papers. 
In \cite{DammIS11} we discuss such approaches and propose weaker conditions 
guaranteeing decidability. The approach described here  
continues this line of research. 
The choice of the class of safety properties we consider is based on
the observation that industrial 
style guides for designing hybrid automata make sure that modes are entered in an 
``inner envelope'', chosen such that modes cannot be 
left before a fixed minimal dwelling time; this 
avoids  immediate context switching. 
In \cite{DammIS11} we showed 
that using  inner envelopes for individual LHA allows us to reduce safety checking to invariant 
checking and  the proof of bounded liveness properties to checking
bounded unfoldings. 

\subsubsection{Safety properties with exhaustive entry conditions}

In this paper we study possibilities of automatically verifying a
certain class of safety properties, namely safety properties with {\em exhaustive entry
  conditions}. 

\begin{defi}[Exhaustive Entry Conditions]
\label{eec}
A safety property with {\em ex\-haus\-tive entry conditions} has 
the form 
\[\Phi_{\sf entry} \rightarrow \Box \Phi_{\sf   safe}\]
where 
$\Phi_{\sf entry} = \forall i_1, \dots, i_m \phi_{\sf entry}({\overline x}(i_1), \dots, {\overline x}(i_m))$
is a formula in the language ${\cal L}_{{\sf index}, {\sf num}}$  such that: 
\begin{enumerate}[label=(\roman*)]
\item \label{eec:initial}If $\Phi_{\sf entry}$ holds in a state $s$, $s$ is an initial 
state of $S$;
\item \label{eec:jump}For every jump or topology update $(s, s')$, $\Phi_{\sf entry}$ holds in 
$s'$.
\end{enumerate}
\end{defi}
Condition (i) guarantees that we make minimal restrictions on initial
states: runs can start in any state satisfying $\Phi_{\sf
  entry}$.  The formula $\Phi_{\sf entry}$ can be seen as a description 
of certain ``inner envelopes'' of the modes. Condition (ii) 
expresses the fact that a jump leads into a state 
satisfying $\Phi_{\sf entry}$ (in the inner envelope of the
target mode).  

For instance,  if ${\sf Init}_{\sf top}$ describes the fact that the 
information about all variables detected by sensors in $P_S$ is 
precise, 
then condition (ii) imposes the   
restriction that sensors have to be globally updated after 
any jump or local topology update, which is clearly 
too restrictive. We can instead require that the initial 
states contain all states in which 
the positions indicated by sensors are within a given margin 
$\varepsilon$ of error (the entry condition 
$\Phi_{\sf entry}$ could describe such states). 

\begin{rem}
Conditions (i) and (ii) ensure that if we start from a state in which 
$\Phi_{\sf entry}$ holds for a given combination $a$ of the values of the
variables, then there exists at least one  tuple $q {=} (q_i)_{i \in 
  I} {\in} Q^I$ of modes of the component automata such that 
$(q, a)$ is an admissible state (i.e.\ the combination $a$ of the
values 
satisfies the invariants in mode $q$), and that any jump or 
topology update starting in a state satisfying $\Phi_{\sf entry}$
leads again to an admissible state. 
\end{rem}

\begin{ex}
\label{phi-entry-ex1}
{\em
Assume that $\Phi_{\sf entry}$ describes such a small margin of error
between the information given by sensors and the real positions 
in the running example, e.g.
\[ \Phi_{\sf entry} = \forall i\, ( i \neq
\nil \wedge {\sf front}(i) \neq \nil \rightarrow
|\lane{}_{\front{}}(i)-\lane[]{{\sf front}(i)}|<\varepsilon).\]
Since ${\sf lane}$ can be modified by a mode change (from value 1 to 2 or vice versa), 
condition (ii) is not guaranteed to hold.
For example, directly after a lane change, $\front{}$ may point to a
car which is now on a different lane, thus violating $\Phi_{\sf entry}$.

In order to guarantee (ii), we need to ensure that
\begin{itemize}
\item $\environment$ is a timed topology automaton where the interval $\Delta t$ between sensor updates is small enough and
\item after lane changes the sensors of all 
systems affected by the change are simultaneously updated.
\end{itemize}
} 
\end{ex} 

In what follows we show that checking safety properties with exhaustive
entry conditions can be reduced to checking invariance of $\Phi_{\sf
  safe}$
under all flows, and under jumps and topology updates in states which are reachable
through a flow from a state satisfying $\Phi_{\sf entry}$ (we call
such state changes GMR jumps and topology updates, cf.~Definition~\ref{def:mode reachable}). 

\subsubsection{Reduction to GMR invariant checking} 

We prove that checking safety properties with exhaustive entry 
conditions  for decoupled SFHA can be reduced to checking whether the safety property 
$\Phi_{\sf  safe}$ is invariant 
under certain jumps, flows, and topology updates. 
\begin{defi}[Globally Mode Reachable]
\label{def:mode reachable}
Let $S$ be an SFHA. 
A state $s= (q, a)$ of $S$  
is {\em globally mode reachable} (GMR, for short) if there exists a state $s_0=(q, a_0)$ of 
$S$ such that $a_0$ satisfies $\Phi_{\sf entry}$ and 
there is a flow in $S$ from $(q, a_0)$ to $(q, a)$. 

A state change $(s, s')$ of $S$ (which can be a flow, a jump, or a topology update) 
is \emph{globally mode reachable} if
$s$ is globally mode reachable.

\end{defi}

\begin{figure}[t]%
  \centering
	\begin{tikzpicture}
		\draw[fill=black!5] (1,0.5) circle (2cm and 1cm);
		\draw[rotate=-17,fill=black!20] (0,0.3) circle (1.12cm and .65cm);
	  \node[inner sep=1pt] (s0) at (-.5,.5) {$\bullet$};
	  \node[inner sep=2pt] (s) at (2,1) {$\bullet$};
		\node at (s0.east) {\ \ \ $s_0$};
		\node at (s.east) {\ $s$};
		\node at (.2,0.1) {$\Phi_{\sf Entry}\cap {\sf Inv}_q$};
		\node at (2,0) {${\sf Inv}_q$};
		\draw[jump,->,fill=none] plot [smooth,tension=1] coordinates {(s0.north) (.7,1.2) (1.7,.5) (s.south)};
  \end{tikzpicture}%
	\caption{Global mode reachability}%
  \label{fig:gmr}%
\end{figure}
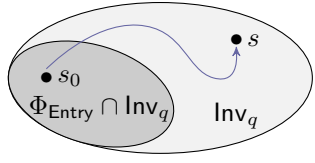

Figure~\ref{fig:gmr} visualizes the concept of global mode reachability of a state.

\begin{thm}
\label{THM:VERIFICATION CONDITIONS}
An SFHA $S = (\environment, \{S(i) \mid i\in I\})$ 
satisfies a safety property with exhaustive entry conditions 
$\Phi_{\sf entry} \rightarrow \Box \Phi_{\sf safe}$ 
if and only if the following hold: 
\begin{enumerate}
\item All states satisfying $\Phi_{\sf entry}$ satisfy $\Phi_{\sf 
    safe}$.
\item $\Phi_{\sf safe}$ is preserved under all flows starting from a state 
  satisfying  $\Phi_{\sf entry}$.  
\item $\Phi_{\sf safe}$ is preserved under all GMR jumps.  
\item $\Phi_{\sf safe}$ is preserved under all GMR topology updates. 
\end{enumerate}  
\end{thm}
\noindent {\em Proof:}  Assume $S$ satisfies the safety property
$\Phi_{\sf entry} \rightarrow \Box \Phi_{\sf safe}$. 
We prove that (1)--(4) hold. 

(1) Consider a state $s$ satisfying condition $\Phi_{\sf entry}$. 
By condition (i) from Definition~\ref{eec}, all states satisfying $\Phi_{\sf 
  entry}$ are initial. 
Since $S$ satisfies the condition $\Phi_{\sf entry} \rightarrow \Box
\Phi_{\sf safe}$, all runs consisting 
of only one state $s$ (satisfying $\Phi_{\sf entry}$) have the property
that $\Phi_{\sf safe}$ holds during the run. Hence $\Phi_{\sf safe}$
holds at state $s$.

(2) Consider now a flow $(s,s')$ starting from a state
  satisfying condition $\Phi_{\sf entry}$. Then $s$ is initial by
  condition (i) from Definition~\ref{eec}, i.e.\ $s,s'$ is a run of
  $S$. The assumption that $S$ 
satisfies the safety property implies that this flow is safe as well
(so all states during this flow are safe). 

(3) 
Consider a jump $(s, s')$, where $s$ is globally mode reachable.
Then $s$ is reachable 
using a flow in $S$ from a state $s_0$ satisfying 
condition $\Phi_{\sf entry}$ (by condition (i) from
Definition~\ref{eec}, $s_0$ is an initial state).
Because $s_0,s,s'$ is a run of $S$ and $S$ satisfies the safety property 
$\Phi_{\sf entry} \rightarrow \Box \Phi_{\sf safe}$, 
it follows that $\Phi_{\sf safe}$ holds at $s'$. 

(4) The proof for topology updates is similar to the one for jumps.
The fact that every topology update leads to an admissible state is 
a consequence of condition (ii) from
Definition~\ref{eec}. 

\medskip 
Assume now that (1)--(4) hold. We prove that $S$ satisfies the safety property 
$\Phi_{\sf entry} \rightarrow \Box \phi_{\sf safe}$. 
Let $s_0,s_1,\dots$ be a run in the composed system $S$, 
starting in an initial state satisfying condition $\Phi_{\sf entry}$. 
We prove by induction on $n$ that for every state $s_n$ in the run:
\begin{enumerate}[label=(\alph*)]
\item all states in the run up to state $s_n$ are GMR.  
\item $\Phi_{\sf safe}$ holds during the run up to state $s_n$. 
\end{enumerate}
$\Phi_{\sf entry}$ holds in state $s_0$,
hence by (1), $s_0$ is both safe and GMR. 

Assume that we have proved that for all $1 \leq i \leq n-1$, $s_i$ has
properties (a) and (b) above. 
If the change of state $(s_{n-1}, s_n)$ is due to a 
flow, then $s_{n-1}$ must be reached by a jump or topology update;
so ${\sf \Phi}_{\sf entry}$ holds at $s_{n-1}$, hence (a) $s_n$ is GMR
and (b) by (2) all the states in which the system is during the flow
from $s_{n-1}$ to $s_n$ are also safe.

Assume that the change of state $(s_{n-1},s_n)$ is due to a jump or a
topology update. 
By the induction hypothesis, $s_{n-1}$ is GMR and safe. 
Then (a) $s_n$ satisfies $\Phi_{\sf entry}$ by property \ref{eec:jump} of exhaustive entry conditions, hence is GMR and 
(b) the jump or topology update $(s_{n-1}, s_n)$ is mode reachable,  
so $s_n$ is safe by (3) if $(s_{n-1}, s_n)$ is a jump, and by (4) if it is a topology update. 
\hfill $\Box$

\subsubsection{Safety properties with GMR-exhaustive entry conditions}

Systems tend to be specified in such a way that their behavior is also
defined for situations  that cannot occur in practice. E.g.\ a car in
our running example could -- looking only at our specification -- 
be in mode \textsf{Rec} while ${\sf pos}_{\sf front}(i)=\pos{i}$. 
Jumps and updates in such a practically impossible situation may 
lead to more and more meaningless states and are nothing that we want 
to worry about when designing entry conditions. 
In this sense, condition~\ref{eec:jump} in Definition~\ref{eec} is too strong.
One way of avoiding such situations is to adapt Definition~\ref{eec}
by requiring that condition (ii) is relative to GMR jumps or topology
updates. 

\begin{defi}[GMR-Exhaustive Entry Conditions]
\label{eec-gmr}
Safety properties with {\em GMR-exhaus\-tive entry conditions} have 
the form 
\[\Phi_{\sf entry} \rightarrow \Box \Phi_{\sf   safe}\]
where 
$\Phi_{\sf entry} = \forall i_1, \dots, i_m \phi_{\sf entry}({\overline x}(i_1), \dots, {\overline x}(i_m))$
is a formula in the language ${\cal L}_{{\sf index}, {\sf num}}$  such that: 
\begin{enumerate}[label=(\roman*)]
\item If $\Phi_{\sf entry}$ holds in a state $s$, $s$ is an initial 
state of $S$;
\item For every GMR jump or GMR topology update $(s, s')$, $\Phi_{\sf entry}$ holds in 
$s'$.
\end{enumerate}
\end{defi}
The proof of Theorem~\ref{THM:VERIFICATION CONDITIONS} can easily be adapted to the case of 
safety properties with GMR-exhaustive entry conditions. 

\begin{thm}
\label{THM:VERIFICATION CONDITIONS-2}
An SFHA $S = (\environment, \{S(i) \mid i\in I\})$ 
satisfies a safety property with GMR-exhaustive entry conditions 
$\Phi_{\sf entry} \rightarrow \Box \Phi_{\sf safe}$
if and only if the following hold: 
\begin{itemize}
\item[(1)] All states satisfying $\Phi_{\sf entry}$ satisfy $\Phi_{\sf 
    safe}$.  
\item[(2)] $\Phi_{\sf safe}$ is preserved under all flows starting from a state 
  satisfying  $\Phi_{\sf entry}$.  
\item[(3)] $\Phi_{\sf safe}$ is preserved under all GMR jumps.  
\item[(4)] $\Phi_{\sf safe}$ is preserved under all GMR topology updates. 
\end{itemize}  
\end{thm}
\begin{rem}In fact, often safety cannot be guaranteed for all runs but only for 
runs with a certain structure: In the running example, we might be 
interested only in runs in which lane changes are preceded and
followed by local or global updates of the sensors. The definitions
and results presented before can be adapted without problems 
such that they are relative to classes of runs. The tests in
Section~\ref{experiments} 
show that in many cases it is not possible to guarantee safety 
for all runs, but safety can be guaranteed for runs in which jumps (corresponding e.g.\ to 
lane changes) are  preceded by local or global updates of the sensors. 
\label{rem:restrict runs}
\end{rem}
\begin{ex}
{\em Consider the running example and the safety property
\[ \Phi^g_{\sf safe}: \forall i, j (i {\neq} \nil \wedge j {\neq} \nil \wedge {\sf lane}(i){=}
 {\sf lane}(j) \wedge {\sf pos}(i) {>} {\sf pos}(j) \rightarrow {\sf
   pos}(i) - {\sf pos}(j) {\geq} d_s)\] 
We showed (using the method described in this paper) that this formula
is invariant under globally mode reachable
flows and topology updates, but not under globally mode reachable
jumps (see also the remarks in Section~\ref{sec:experiments}); the
problems with the jumps can occur because 
the information provided by sensors at the moment of a line change 
is outdated. In order to prevent this, it is necessary to ensure that 
a topology update takes place immediately before any lane change. 
We proved that {\em for all runs in which topology updates take  place
before lane changes}, formula $\Phi^g_{\sf safe}$ is invariant under 
all jumps.
}
\end{ex}

\subsection{Reducing verification tasks to satisfiability checking}
\label{sec:reduction to satisfiability} 
We consider 
safety properties $\Phi_{\sf entry} {\rightarrow} \Box
\Phi_{\sf   safe}$ 
with exhaustive entry conditions, where $\Phi_{\sf entry}$ and $\Phi_{\sf safe}$ are of the form
\begin{align*}
\Phi_{\sf entry} &= \forall i_1 \dots i_m \phi_{\sf entry}({\overline x}(i_1), \dots, {\overline x}(i_m))\\
\Phi_{\sf safe} &= \forall i_1 \dots i_n \phi_{\sf safe}({\overline x}(i_1), \dots, {\overline x}(i_n))
\end{align*}
with quantifier-free $\phi_{\sf entry}$ and $\phi_{\sf safe}$.   
We show  that for decoupled SFLHA  $S$ we can reduce checking whether
such a property holds, 
to checking
whether certain formulae $F^{\sf init}_q, F^{\sf flow}_q, F^{\sf jump}_q, F^{\sf top}_q$ 
are unsatisfiable for all combinations of modes $q = (q_i)_{i \in I}
\in Q^I$. 

\subsubsection{Sequentializing parallel jumps}

We first show that for decoupled SFLHA we do not need to consider 
parallel jumps.
\begin{lem}
\label{LEM:SEQUENTIALIZE}
	Let $S = (\environment, \{S(i) \mid i\in I\})$ be a decoupled
        SFHA. 
\begin{itemize}
\item[(1)] $\Phi_{\sf safe}$ is invariant under all (GMR) jumps in $S$ iff it is
  invariant under all  (GMR) jumps which reset the variables of a finite
  family of systems in $S$. 
\item[(2)] $\Phi_{\sf safe}$ is invariant under all (GMR) jumps
  involving  a finite
  family of systems in $S$ iff it is invariant under all (GMR) jumps in any
  component of $S$. 
\end{itemize}
\end{lem}
\noindent {\em Proof:}  
(1) The direct implication is obviously true. 
  Assume that $\Phi_{\sf safe}$ is invariant under all (GMR) jumps which
  reset the variables of a finite family of systems in $S$. 
Consider a jump in $S$ which resets the variables of an  
infinite family of systems in $S$. Assume that 
$\Phi_{\sf safe}$ is not invariant under this jump, i.e.\ 
$\Phi_{\sf safe}$ holds before the jump but there exist systems $S(i_1), \dots, S(i_n)$ 
such that after the jump $\phi_{\sf safe}({\overline x}(i_1), \dots,
{\overline x}(i_n))$ is not true. Since $S$ is decoupled, 
the value of the variables ${\overline x}(i_1), \dots,
{\overline x}(i_n)$ cannot be reset by systems not in 
 $S(i_1), \dots, S(i_n)$. This shows that already the combination 
of mode switches in the finite family $S(i_1), \dots, S(i_n)$ would 
lead from a safe to an unsafe state. Contradiction. 

(2) The direct implication is obviously true. 
We prove the converse implication. 
Let $C=\{c_1,\ldots,c_k\} \subseteq \{S(i) \mid i\in I\}$,
        let ${\sf guard}_C$ and ${\sf jump}_C$ be the formulae
        describing the guards resp. updates of a simultaneous
        (GMR) mode switch for all systems in $C$ (the other variables do not change). 
Assume that $\Phi_{\sf safe}$ is not invariant under this jump. Then the formula
\[\Phi_{\sf safe}(\overline x_0) \wedge {\sf guard}_C(\overline x_0) \wedge {\sf jump}_C(\overline x_0,\overline x_k) \wedge \neg \Phi_{\sf safe}(\overline x_k)\]
	is satisfied by some variable assignment $\beta$. 
Because of the assumptions on resets in a decoupled SFHA, 
a jump in some $S(i)$ cannot invalidate the guard 
of a simultaneous transition in another $S(j)$. 
In particular, none of $c_1,\ldots,c_k$ can invalidate the guard of a later element of this sequence. 
In other words, if ${\sf guard}_C(\overline x_0)$ is true for a
variable assignment, then -- if we sequentialize $C$ as the succession
of jumps $c_1, c_2, \dots, c_k$, sequentially changing the values of
the variables from $\overline x_0$ to $\overline x_1, \overline x_2,
\dots, \overline x_k$, ${\sf guard}_{c_i}(\overline x_{i-1})$ is also
true.\footnote{In general,  if $c_i$ is a jump in a system $S(j)$,
  ${\sf guard}_{c_i}$ is expressed using only the variables of the
  system $S(j)$, since the values of those variables are not changed
  by previous jumps, ${\sf guard}_{c_i}(\overline x_{i-1})$ is in fact
  identical with ${\sf guard}_{c_i}(\overline x_0)$.} 
Therefore, 
\[\Phi_{\sf safe}(\overline x_0)
  \wedge \bigwedge_{i\in\{1,\ldots,k\}}
    \big({\sf guard}_{c_i}(\overline x_{i-1})
         \wedge {\sf jump}_{\{c_i\}}(\overline x_{i-1},\overline x_i)\big)
  \wedge \neg \Phi_{\sf safe}(\overline x_{k})\]
is satisfiable for some extension $\beta'$ of $\beta$ to the fresh
variables $\overline x_1, \overline x_2,
\dots, \overline x_{k-1}$. 
Since for each $i$ obviously either $\Phi_{\sf safe}(\overline x_i)$
or $\neg \Phi_{\sf safe}(\overline x_i)$
is satisfied by $\beta'$, there must be at least one index
$i_0 \in \{ 1, \dots, k \}$ for which $\Phi_{\sf safe}(\overline x_{i_0-1})$
and $\neg \Phi_{\sf safe}(\overline x_{i_0})$, and thus all of
\[\Phi_{\sf safe}(\overline x_{i_0-1})
\wedge {\sf guard}_{\{c_{i_0}\}}(\overline x_{{i_0-1}}) \wedge 
{\sf jump}_{\{c_{i_0}\}}(\overline x_{{i_0-1}},\overline x_{i_0})
\wedge \neg \phi_{\sf safe}(\overline x_{i_0})\]
is satisfied by $\beta'$.
So $\Phi_{\sf safe}$ is not invariant under jumps of a single component. \hfill $\Box$

\subsubsection{Verification of safety properties and satisfiability checking}
We show that for decoupled SFLHA we can express the verification tasks
(1)--(4) in Theorem~\ref{THM:VERIFICATION CONDITIONS} as satisfiability problems.

\begin{thm}
\label{THM:INV-FORMULAE}
Let ${\cal S}$ be a decoupled SFLHA. Let $c_1, \dots, c_n$ be the Skolem constants 
obtained from the negation of $\Phi_{\sf safe}$. 
\begin{enumerate}
\item The {\em entry states} of ${\cal S}$ satisfy $\Phi_{\sf safe}$ iff 
the following formula $F^{\sf entry}$ is unsatisfiable:
\[F^{\sf entry}: \Phi_{\sf entry} \wedge  \neg \phi_{\sf safe}(\overline{x}(c_1),
\dots, \overline{x}(c_n))\]

\item
$\Phi_{\sf safe}$ is {\em invariant under flows} starting in a state satisfying $\Phi_{\sf entry}$
iff for all  $q {=} (q_i)_{i \in I} {\in} Q^I$ the following formula $F^{\sf flow}_q$ is unsatisfiable:
\begin{align*}
	F^{\sf flow}_q:\ &
t_0 < t_1 \wedge 	\Phi_{\sf entry}({\overline x}(t_0)) \wedge 
	\forall i_1, \dots, i_n \phi_{\sf safe}({\overline x}(i_1, t_0), \dots, {\overline x}(i_n, t_0))
	\\&{}\wedge \forall i \, {\sf Flow}_{q_i}({\overline x}(i, t_0), {\overline x}(i, t_1))
	\wedge \neg \phi_{\sf safe}({\overline x}(c_1, t_1), \dots, {\overline x}(c_n, t_1))  
\end{align*}
where if ${\sf flow}_q(i) = \bigwedge \big({\cal E}_f \vee \sum_{k = 1}^n 
a^q_k(i) \dot x_k(i) \leq a^q(i)\big)$ then   
\begin{align*}
  {\sf Flow}_{q_i}({\overline x}(i, t_0), {\overline x}(i,t_1)) :\ &
	\bigwedge \big({\cal E}_f \vee \sum_{k = 1}^n a^{q_i}_k(i) (x_k(i, t_1) {-} x_k(i,t_0)) {\leq} a^{q_i}(i) (t_1{-} t_0)\big) \\
  &{}\wedge  {\sf Inv}_{q_i}({\overline x}(i, t_0))
	\wedge  {\sf Inv}_{q_i}({\overline x}(i, t_1)) \ .
\end{align*}

\item $\Phi_{\sf safe}$ is {\em invariant under GMR jumps in $S$}  
iff for all  $q {=} (q_i)_{i \in I} {\in} Q^I$ the following 
formula ${F^{\sf jump}}^q_{e}(i_0)$ 
is unsatisfiable for every $i_0 \in I$ 
and $e=(q_{i_0},q'_{i_0}) \in E$, s.t.\ if $p(i_0)$ occurs 
in ${\sf guard}_e$ it is not $\nil$:  
\begin{align*}
	{F^{\sf jump}}^q_{e}(i_0):\ 	\Phi_{\sf entry}({\overline
          x}(t_0)) 
& 
{}\wedge \Bigg( \bigg(t_0 < t_1 \wedge \forall i {\sf
          Flow}_{q_i}({\overline x}(i, t_0), {\overline x}(i, t_1)) \bigg)
        \vee t_0 = t_1 \Bigg) \\
	& {}\wedge 
	\forall i_1, \dots, i_n \phi_{\sf safe}({\overline x}(i_1, t_1), \dots,{\overline x}(i_n, t_1)) \\
	& {}\wedge {\sf guard}_e({\overline x}(i_0, t_1)) \wedge {\sf jump}_{e}({\overline x}(i_0, t_1), {\overline x}'(i_0))
		{}\wedge {\sf Inv}_{q'_{i_0}}(\overline x'(i_0))\\
	& {}\wedge\forall j (j \neq i_0 \rightarrow {\overline x}'(j) = {\overline x}(j))
	 \wedge \neg \phi_{\sf safe}({\overline x}'(c_1), \dots, {\overline x}'(c_n))\ . 
\end{align*}

\item $\Phi_{\sf safe}$ is {\em invariant under GMR topology
    updates} for pointers in a set $P_1$ 
iff for all $q = (q_i)_{i \in I} \in Q^I$ the following formula $F_q^{\sf top}$ is 
unsatisfiable: 
\begin{align*}
F_q^{\sf top}:\ 	\Phi_{\sf entry}({\overline
          x}(t_0)) 
& 
{}\wedge \Bigg( \bigg(t_0 < t_1 \wedge \forall i {\sf
          Flow}_{q_i}({\overline x}(i, t_0), {\overline x}(i, t_1)) \bigg)
        \vee t_0 = t_1 \Bigg) \\
&{}\wedge 
\forall i_1, \dots, i_n \phi_{\sf safe}({\overline x}(i_1, t_1), \dots,{\overline x}(i_n, t_1)) \\
& \wedge \bigwedge_{p \in P_1} {\sf Update}(p,p') \wedge \neg \phi_{\sf safe}'({\overline x}(c_1), \dots,
{\overline x}(c_n))\ ,
\end{align*}
where $\phi_{\sf safe}'$ is obtained from $\phi_{\sf safe}$ by replacing 
every $p \in P_1$ with $p'$.
\end{enumerate}
\end{thm}

\noindent {\em Proof:}  
(1) is immediate. 

(2) Assume that $\Phi_{\sf safe}$ is not invariant 
under flows in some state $q$. Then there are functions $x(i) : {\mathbb R}
\rightarrow {\mathbb R}$ satisfying all flow conditions and such that 
$\Phi_{\sf safe}$ holds at the beginning of the flow and does not 
hold at the end of the flow. Then (using the mean value theorem) one 
can show that these functions can be used for constructing a model 
for the formula $F^{\sf flow}_q$. See \cite{DammIS11} for more
details.

Conversely, assume that formula $F^{\sf flow}_q$ is satisfiable. 
We can define the functions $x(i)$ by taking the linear interpolation 
of the functions defined at $t_0$ and $t_1$. Then ${\sf flow}_{q_i}({\overline x}(i, t_0, t_1))$
holds; it follows that the functions $x(i) : {\mathbb R}
\rightarrow {\mathbb R}$ satisfy the flow condition.
So $\Phi_{\sf safe}$ is not invariant under flows. 

 In particular, the results presented in \cite{DammIS11} 
ensure that if the numerical constraints in the mode invariants 
are conjunctions of linear inequalities (and hence convex) we do not 
need to express explicitly that the invariant needs to hold at 
all points between $t_0$ and $t_1$. (If we can construct a model of 
the formula in which the invariant holds at $t_0$ and $t_1$ we can 
construct a model in  which the invariant holds at 
all points between $t_0$ and $t_1$ using linear interpolation of the 
functions $x_i$.) 

(3) is a consequence of Lemma~\ref{LEM:SEQUENTIALIZE} using 
arguments from (2).

(4) is immediate (again, using arguments from (2)).  \hfill $\Box$

\subsubsection{Checking exhaustive  entry conditions}
\label{sec:checking exhaustive entry conditions}

We now show that for decoupled SFLHA  $S$ we can reduce checking  conditions (i) and (ii) in
Definition~\ref{eec} to satisfiability tests. 

\begin{thm}
\label{thm:check-eec}
Let $S$ be a decoupled SFHA $S$, and $\Phi_{\sf entry} \rightarrow
\Box \Phi_{\sf safe}$ be a safety condition as above. 

Then conditions (i) and (ii) in Definition~\ref{eec} hold iff: 
\begin{itemize}
\item[(i)] {\em Initial states:} 
\[ \Phi_{\sf entry}({\overline x}) \wedge \big( \neg (\bigvee_{q \in Q} {\sf 
     Init}_{q}({\overline x}(i_0))) \vee \neg {\sf Init}_{\sf     top}({\overline x}) \big) \text{ is unsatisfiable} \]


\item[(ii)] For all $q = (q_i)_{i \in I} \in Q^I$: 
\begin{itemize}
\item[(a)] {\em Topology updates:}
\[(\forall i \, {\sf Inv}_{q_i}({\overline x}_i))  \wedge {\sf Update}(p, p')
  \wedge \neg \Phi'_{\sf entry}({\overline x}) \text{ is unsatisfiable,} \]
where $\Phi'_{\sf entry}$ arises from $\Phi_{\sf entry}$ by replacing
$p$ with $p'$, and 

\item[(b)] {\em Jumps:} For all  $e = (q_{i_0}, q'_{i_0}) \in E, i_0 \in I$:
\[\begin{array}{rl}
  (\forall i \, {\sf Inv}_{q_i}({\overline x}_i)) \wedge
    {\sf guard}_e({\overline x}_{i_0}) \wedge
	  {\sf jump}_e({\overline x}_{i_0}, {\overline x}'_{i_0}) \wedge {} \\
	  \forall j (j \neq i_0 \rightarrow {\overline x}'(j) = {\overline x}(j)) \wedge \neg \Phi_{\sf entry}({\overline x}')
  & \text{ is unsatisfiable.} 
\end{array}\]
\end{itemize}
\end{itemize}
\end{thm} 

\noindent {\em Proof:} (i) Condition (i) in Definition~\ref{eec}
states that if $\Phi_{\sf entry}$ holds in a state $s$ then $s$ is
initial. This is the case if and only if  whenever $\Phi_{\sf entry}$
holds for given values of the variables, then for these values:
\begin{itemize}
\item for all $i \in I$ there exists a mode 
$q \in Q$ such that the initial condition of mode $q$ is satisfied in system
$S(i)$, and 
\item ${\sf Init}_{\sf top}$ holds.
\end{itemize}
It can be easily checked that  this is the case if and only if 
it cannot happen that $\Phi_{\sf entry}$
holds for given values of the variables and for these values ${\sf
  Init}_{\sf top}$ does not hold, or there exists a system $i_0$ 
such that for these values none of the initial conditions in $\{ {\sf
  Inv}_q(i_0) \mid q \in Q \}$ holds, i.e.\ if and only if  the following formula is unsatisfiable: 
\[ \Phi_{\sf entry}({\overline x}) \wedge \big( \neg (\bigvee_{q \in Q} {\sf 
     Init}_{q}({\overline x}(i_0))) \vee \neg {\sf Init}_{\sf
     top}({\overline x}) \big). \]

\noindent (ii) Condition (ii) in Definition~\ref{eec}
states that for every state change $(s, s')$ due to (a) a topology
update or (b) a jump,  
$\Phi_{\sf  entry}$ holds in $s'$. This happens if and only if the formulae in 
(a) and (b) are unsatisfiable (i.e.\ if and only if it cannot happen that
$S$ is in a mode $q = (q_i)_{i \in I}$ (i.e.\ the invariants of 
the systems $S(i)$ in these modes hold), and 
(a) there is an update after which $\Phi_{\sf  entry}$ does not hold
or (b) there is a jump after which $\Phi_{\sf  entry}$ does not hold). 
\hfill $\Box$

\medskip

\noindent For spatial families of {\em linear} hybrid automata, 
a similar result can be used for recognizing safety conditions with
GMR-exhaustive entry conditions. 
\begin{thm}
For a decoupled SFLHA $S$, conditions (i) and (ii') in Definition~\ref{eec-gmr} hold iff: 
\begin{itemize}
\item[(i)] 
\[ \Phi_{\sf entry}({\overline x}) \wedge \big( \neg (\bigvee_{q \in Q} {\sf 
     Init}_{q}({\overline x}(i_0))) \vee \neg {\sf Init}_{\sf
     top}({\overline x}) \big) \text{ is unsatisfiable} \]


\item[(ii')] For all $(q_i)_{i \in I} \in Q^I, e \in E, i_0 \in I$:
\begin{itemize}
\item the following conjunction is unsatisfiable: 
\[\begin{array}{l}
t_0 < t_1 ~\wedge~ \Phi_{\sf
  entry}({\overline x}(t_0)) ~\wedge~ \forall i \, {\sf Flow}_{q_i}({\overline x}(i, t_0), {\overline x}(i, t_1)) ~\wedge\\
{\sf Update}(p, p')  ~\wedge~ \neg \Phi'_{\sf entry}({\overline x}(t_1)), 
\end{array}\]
where $\Phi'_{\sf entry}$ arises from $\Phi_{\sf entry}$ by replacing
$p$ with $p'$;  and 
\item the following conjunction is unsatisfiable: 
\[\begin{array}{l@{}l}
t_0 < t_1 ~\wedge~ & \Phi_{\sf entry}({\overline x}(t_0)) ~\wedge~
\forall i {\sf Flow}({\overline x}(i, t_0), {\overline x}(i, t_1)) ~\wedge \\
& {\sf guard}_e({\overline x}(i_0, t_1)) ~\wedge~ {\sf jump}_e({\overline
   x}(i_0, t_1), {\overline x}'_{i_0}) ~\wedge {} \\
& 	  \forall j (j \neq i_0 \rightarrow {\overline x}'(j) =
          {\overline x}(j, t_1)) ~\wedge~ \neg \Phi_{\sf entry}({\overline
            x}') \\
 \end{array}\]
\end{itemize}
\end{itemize}
 where if ${\sf flow}_q(i) = \bigwedge \big({\cal E}_f \vee \sum_{k = 1}^n 
a^q_k(i) \dot x_k(i) \leq a^q(i)\big)$ then   
\begin{align*}
  {\sf Flow}_{q_i}({\overline x}(i, t_0), {\overline x}(i,t_1)) :\ &
	\bigwedge \big({\cal E}_f \vee \sum_{k = 1}^n a^{q_i}_k(i) (x_k(i, t_1) {-} x_k(i,t_0)) {\leq} a^{q_i}(i) (t_1{-} t_0)\big) \\
  &{}\wedge\forall i\, {\sf Inv}_{q_i}({\overline x}(i, t_0))
	\wedge \forall i\,{\sf Inv}_{q_i}({\overline x}(i, t_1)) \ 
\end{align*}
\end{thm} 

\noindent {\em Proof:} The proof of (ii') is similar to the proof of
Theorem~\ref{thm:check-eec}(ii), with the only difference that we need to
additionally take flows into account. \hfill $\Box$ 

\section{Automated Reasoning}
\label{Sec:HierarchicalReasoning}

We present classes of theories for which decidable fragments relevant 
for the verification tasks above exist.
We use the following complexity results 
for fragments of linear arithmetic: 
\begin{itemize}
\item 
The satisfiability over ${\mathbb R}$ 
of 
conjunctions of linear 
inequalities  can be checked in PTIME \cite{khakian}. 
\item
The problem of checking the satisfiability of sets of clauses 
in $LI({\mathbb R})$ 
is in NP \cite{sontag}.
\item
The satisfiability of any conjunction of Horn disjunctive
  linear (HDL) 
constraints\footnote{A Horn-disjunctive linear constraint 
is a disjunction $d_1 \vee \dots \vee d_n$ where each 
$d_i$ is a linear inequality or disequation, 
and the number of inequalities does not exceed one.} over ${\mathbb R}$ 
\cite{koubarakis} and the satisfiability of any
conjunction of Ord-Horn constraints\footnote{Ord-Horn constraints are  
 implications 
$\bigwedge_{i = 1}^n x_i {\leq} y_i {\rightarrow} x_0 {\leq} y_0,$ ($x_i, y_i$
are variables).}
over ${\mathbb R}$ 
\cite{nebel} 
can be decided in PTIME. 
\end{itemize}

\subsection{Local theory extensions} 
Let ${\cal T}_0$ be a base theory with signature $\Sigma_0$. 
We consider extensions ${\cal T}_1 := {\cal T}_0 \cup {\cal K}$ of ${\cal T}_0$ with new 
function symbols in a set $\Sigma_1$ of {\em extension functions}
whose properties are axiomatized with a set ${\cal K}$ of 
{\em augmented clauses}, i.e.\ of axioms of the form 
$\forall x_1 \dots x_n (\Phi(x_1, \dots, x_n) \vee C(x_1, \dots, x_n))$, 
where $\Phi(x_1, \dots, x_n)$ is a {\em first-order formula in signature $\Sigma_0$ 
and $C(x_1, \dots, x_n)$ is a {\em clause} containing extension
functions}.
In this case we refer to the (theory) extension ${\cal T}_0 \subseteq {\cal T}_0 \cup {\cal K}$. 
In \cite{Sofronie-cade-05} we introduced and studied local theory
extensions. In \cite{ihlemann-sofronie-ijcar10}, various notions of
locality of  theory extensions were introduced and studied. 

\begin{defi}[Local theory extension]
An extension ${\cal T}_0 \subseteq {\cal T}_0 \cup {\cal K}$ is a
{\em local extension} 
if for every set $G$ of 
ground $\Sigma_0 \cup \Sigma_1 \cup \Sigma_c$-clauses 
(where $\Sigma_c$ is a set of additional constants), 
if $G$ is unsatisfiable w.r.t.\ ${{\cal T}_0 {\cup} {\cal K}}$ then 
unsatisfiability can be detected using the set ${\cal K}[G]$ 
consisting of those instances 
of ${\cal K}$ in which the terms starting with 
extension functions are ground terms occurring in ${\cal K}$ or $G$.
\label{defi-loc}
\end{defi} 
\emph{Stably local} extensions are defined similarly, with the difference
that ${\cal K}[G]$ is replaced with ${\cal K}^{[G]}$, the set of 
instances of ${\cal K}$ in which the variables are instantiated 
with ground terms 
which occur in ${\cal K}$ or $G$.

\subsection{Hierarchical reasoning in local theory extensions}
For local theory extensions (or stably local theory extensions) 
hierarchical reasoning is possible. If ${\cal T}_0 \cup {\cal K}$ is a
(stably) local extension of ${\cal T}_0$ and $G$ is a set of 
ground $\Sigma_0 \cup \Sigma_1 \cup \Sigma_c$-clauses 
then, by Definition~\ref{defi-loc}, 
$ {\cal T}_0 \cup {\cal K} \cup G$ is unsatisfiable iff 
$ {\cal T}_0 \cup {\cal K}[G] \cup G$ (or resp.\ 
$ {\cal T}_0 \cup {\cal K}^{[G]} \cup G$) is unsatisfiable. 
We can reduce this last satisfiability test to 
a satisfiability test w.r.t.\ ${\cal T}_0$. 
The idea is to purify 
${\cal K}[G] \cup G$ (resp.\ ${\cal K}^{[G]} \cup G$) 
by 
\begin{itemize}
	\item introducing (bottom-up) new constants $c_t$ for subterms $t = f(g_1, \dots, g_n)$ with $f \in \Sigma$, $g_i$ ground $\Sigma_0 \cup \Sigma_c$-terms,
  \item replacing the terms $t$ with the constants $c_t$, and 
	\item adding the definitions $c_t = t$ to a set $D$. 
\end{itemize}
 We denote by ${\cal K}_0 \cup 
G_0 \cup D$ the set of formulae obtained this way. 
Then $G$ is 
satisfiable w.r.t.\ ${\cal T}_0 \cup {\cal K}$ iff 
${\cal K}_0 \cup G_0 \cup {\sf Con}_0$ is satisfiable w.r.t.\
${\cal T}_0$, where 
\[
  {\sf Con}_0  = \{ (\bigwedge_{i = 1}^n c_i {=} d_i) \rightarrow c {=} d \mid 
  f(c_1, \dots, c_n) {=} c, f(d_1, \dots, d_n){=} d \in D \}. 
\]

\begin{thm}[\cite{Sofronie-cade-05}] 
\label{lemma-rel-transl}
If ${\cal T}_0 \subseteq {\cal T}_0 \cup {\cal K}$ is a 
(stably) local extension and $G$ is a set of
(augmented) ground clauses 
then we can reduce the problem of checking whether $G$ is 
satisfiable w.r.t.\ ${\cal T}_0 \cup {\cal K}$ to checking 
the satisfiability 
w.r.t.\ ${\cal T}_0$ of the formula ${\cal K}_0 \cup G_0 \cup {\sf
  Con}_0$ constructed as explained above. 

If ${\cal K}_0 \cup G_0 \cup {\sf Con}_0$ belongs to a decidable 
fragment of ${\cal T}_0$ we can use the decision procedure for 
this fragment to decide whether $ {\cal T}_0 \cup {\cal K} \cup G$
is unsatisfiable. 
\end{thm}
As the size of  
${\cal K}_0 {\cup} G_0 {\cup} {\sf Con}_0$ is polynomial in the size of $G$
(for a given ${\cal K}$), locality allows us to express the complexity 
of the ground satisfiability problem w.r.t.\ ${\cal T}_1$  
as a function of the complexity of the satisfiability 
of ${\cal F}$-formulae w.r.t.\ ${\cal T}_0$. 

\subsection{Examples of local theories and theory extensions}

In establishing the decidability results for the verification of
safety properties of SFLHA we will use locality results for updates and for
theories of pointers.

\subsubsection{Update rules} 
We first consider update rules, in which some of the function symbols
change the way they are defined, depending on a partition of their
domain of definition. Many update rules 
define local theory extensions.  
\begin{thm}[\cite{JacobsKuncak,Sofronie-Ihlemann-Jacobs-tacas08}]
\label{thm:updates}
Let ${\cal T}_0$ be a base theory with signature $\Sigma_0$ and
$\Sigma \subseteq \Sigma_0$. 
Consider a family ${\sf Update}(\Sigma, \Sigma')$ 
of update axioms of the form: 
\begin{eqnarray}
    \forall {\overline x} (\phi^f_i({\overline x}) \rightarrow
    F^f_i(f'({\overline x}), {\overline x})) &  & i =1, \dots, m,
    \quad f \in \Sigma   
\end{eqnarray}
which describe how the values of the $\Sigma$-func\-tions change,  
depending on a partition of the state space, described by a finite set $\{ \phi^f_i \mid i \in I \}$ of
$\Sigma_0$-formulae and using $\Sigma_0$-formulae $F^f_i$ such that
\begin{enumerate}[label=(\roman*)]
\item $\phi_i({\overline x}) \wedge \phi_j({\overline x})
  \models_{{\cal T}_0} \perp $  for $i {\neq} j$ and
\item ${\cal T}_0 \models \forall {\overline x} (\phi_i({\overline x})
  \rightarrow \exists y (F_i(y, {\overline x})))$ for all $i \in I$.
\end{enumerate}
Then the extension of ${\cal T}_0$ with axioms ${\sf Update}(\Sigma, \Sigma')$
is local. 
\end{thm}


\subsubsection{A theory of pointers}
We present a fragment of the theory of pointers studied in 
 \cite{NeculaMcPeak} and later analyzed in
 \cite{Sofronie-Ihlemann-Jacobs-tacas08}. 
Consider the language ${\cal L}_{{\sf index},{\sf num}}$ 
with sorts ${\sf index}$  and 
${\sf num}$  introduced before,
with sets of unary pointer (numeric) fields $P$ ($X$), and
with a constant $\nil$ of sort ${\sf index}$.
The only predicate of sort ${\sf index}$ is equality; the signature
$\Sigma_{\sf num}$ 
of sort ${\sf num}$ depends on the theory ${\cal T}_{\sf num}$ modeling the
scalar domain. A {\em guarded {\sf index}-positive extended clause} is 
a clause of the form: 
\begin{eqnarray}
C := \forall i_1 \dots i_n ~~ {\cal E}(i_1, \dots, i_n) \vee {\cal
  C}({\overline x_i(i_1)}, \dots, {\overline x_i(i_n)}) \label{loc-ax}
\end{eqnarray}
\noindent where ${\cal C}$ is a ${\cal T}_{\sf num}$-formula over terms
of sort ${\sf num}$, 
$x_i \in X$, and ${\cal E}$ is a disjunction of equalities between
terms of sort ${\sf index}$, containing all atoms of the form 
$i = \nil,  f_n(i) = \nil, \dots, 
f_2(\dots f_n(i)) = \nil$  for all terms $f_1(f_2(\dots f_n(i)))$ occurring in 
${\cal E} \vee {\cal C}$, where $f_1 \in P \cup X, f_2,\dots, f_n \in P$. 
\begin{thm}[\cite{Sofronie-Ihlemann-Jacobs-tacas08}] 
\label{thm:pointers}
Every set ${\cal K}$ of 
guarded {\sf index}-positive extended clauses defines a stably local 
extension of ${\cal T}_{\sf num} \cup {\sf Eq}_{\sf index}$, 
where ${\sf Eq}_{\sf index}$ is the pure theory
of equality of sort ${\sf index}$.
\end{thm}


\subsection{Chains of local theory extensions}
The results we obtain in this paper will be justified by locality
properties for certain theory extensions. 
In many cases we need to perform reasoning tasks 
in an extension ${\cal T}_0 \subseteq
{\cal T}_0 \cup {\cal K}$ in which the set ${\cal
  K}$ of axioms of the extension can be written as a union 
${\cal  K} = {\cal  K}_1 \cup {\cal  K}_2$ such that 
both 
\begin{itemize}
\item[(1)] ${\cal T}_0 \subseteq
{\cal T}_0 \cup {\cal K}_1$ and 
\item[(2)] 
${\cal T}_0 \cup {\cal K}_1  \subseteq {\cal T}_0 \cup {\cal K}_1 \cup
{\cal K}_2$ 
\end{itemize}
are (stably) local theory
extensions. In this case we say that we have a chain of (stably) local 
theory extensions; the reasoning task can be hierarchically reduced 
to reasoning in ${\cal T}_0$ in two steps: 
\begin{description}
\item[Step 1:]  In a first step, we reduce checking whether 
${\cal T}_0 \cup {\cal K}_1 \cup {\cal K}_2 \cup G$ is satisfiable
to checking whether ${\cal T}_0 \cup {\cal K}_1 \cup {\cal K}_2*[G]
\cup G$
is satisfiable (where ${\cal K}_2*[G]$ is ${\cal K}_2[G]$ if the
extension is local and ${\cal K}_2^{[G]}$ if the extension is stably
local). 

We can further reduce this task to checking the satisfiability of 
${\cal T}_0 \cup {\cal K}_1 \cup ({\cal K}_2)_0 \cup G_0 \cup {\sf
  Con}_0$
as explained in Theorem~\ref{lemma-rel-transl}. 

\item[Step 2:]  if $G_1 = ({\cal K}_2)_0 \cup G_0 \cup {\sf
  Con}_0$ is a set of ground clauses, and the theory extension 
${\cal T}_0 \subseteq
{\cal T}_0 \cup {\cal K}_1$ is (stably) local, we can use again 
Theorem~\ref{lemma-rel-transl} to reduce the problem of checking the 
satisfiability of 
${\cal T}_0 \cup {\cal K}_1 \cup G_1$ to a satisfiability test w.r.t.\
${\cal T}_0$. 
\end{description}
The idea can be used also for longer chains of (stably) local theory
extensions: 
\[ {\cal T}_0 \subseteq {\cal T}_0 \cup {\cal K}_1 \subseteq {\cal
  T}_0 \cup {\cal K}_1 \cup {\cal K}_2 \subseteq \dots \subseteq  {\cal
  T}_0 \cup {\cal K}_1 \cup {\cal K}_2 \cup \dots \cup {\cal T}_n. \]
A similar reduction can be used for chains of extensions 
\[ {\cal T}_0 ~\subseteq~
{\cal T}_0 \cup {\cal K}_1  ~\subseteq~ {\cal T}_0 \cup {\cal K}_1 \cup
{\cal K}_2 \] 
in which the second  extension is (stably) local, if after using 
Step 1 above (i) the set 
of clauses obtained by instantiation 
${\cal T}_0 \cup {\cal K}_1 \cup {\cal K}_2*[G]$ or (ii) the set  of
clauses 
${\cal T}_0 \cup {\cal K}_1 \cup ({\cal K}_2)_0$ obtained 
after the hierarchical reduction described in
Theorem~\ref{lemma-rel-transl},   
define a (stably) local extension of ${\cal T}_0$. 
\begin{ex}
{\em We can for instance consider a set ${\cal K} = {\sf Update}(\Sigma, \Sigma')$ of update
rules of the form in Theorem~\ref{thm:updates}, which, by
Theorem~\ref{thm:updates}, 
defines a local extension of a base theory ${\cal T}_0$. 

Then, for every set $G$ of ground clauses, ${\cal T}_0 \cup {\cal K}
\cup G$ is satisfiable iff ${\cal T}_0 \cup {\cal K}[G] \cup G$ is
satisfiable. It can happen that ${\cal K}[G]$ (hence
also the purified set of clauses ${\cal K}_0$) is not ground, 
and that the purified set of clauses ${\cal K}_0 \cup G_0$ 
contains additional function symbols in a set $P \cup X$. 

If, for instance, ${\cal K}_0$ is a set of guarded {\sf index}-positive
extended clauses then, by Theorem~\ref{thm:pointers}, 
${\cal K}_0$ defines a stably local 
extension of ${\cal T}_{\sf num} \cup {\sf Eq}_{\sf index}$, 
where ${\sf Eq}_{\sf index}$ is the pure theory
of equality of sort ${\sf index}$.

In order to check the satisfiability of $G$ w.r.t.\ ${\cal T}_0 \cup
{\cal K}$ we need to consider the following instances of ${\cal K}$: 
$({\cal K}[G])^{[T_G]}$ where $T_G$ is the set of ground terms 
occurring in $G \cup {\cal K}[G]$.

}\end{ex}


\section{Verification: Decidability and Complexity}
\label{inv-bmc}

As mentioned in Section~\ref{verif}, we consider 
safety properties with exhaustive entry
conditions 
$\Phi_{\sf entry} \rightarrow \Box \Phi_{\sf safe}$. 
We make the following assumptions:
\newcounter{assumption}
\begin{description}
\item[Assumption 1:] \refstepcounter{assumption}\label{assumption:linear and decoupled}
  $S = (\environment, \{S(i) \mid i\in I\})$ is a decoupled
  SFLHA. 
\item[Assumption 2:] \refstepcounter{assumption}\label{assumption:entry}
  $\Phi_{\sf safe}$  is a set (conjunction) of guarded {\sf index}-positive
  extended clauses of the form $\forall i_1, \dots, i_n {\cal E} \vee
  {\cal C}$, such that ${\cal C}$ is a conjunction of 
linear inequalities, and $\Phi_{\sf entry}$  is a set (conjunction) 
containing either
\begin{enumerate}
\item only guarded {\sf index}-positive
  extended clauses of the form $\forall i_1, \dots, i_n {\cal E} \vee
  {\cal C}$,  
such that ${\cal C}$ is a conjunction of linear
  inequalities; 

\item or only ${\cal L}_{\sf index, num}$-formulae of the form $\forall i  \big(i \neq \nil
  \wedge \phi_k \rightarrow F(f(i), i)\big), k \in \{ 1, \dots, m \}$
  where $f \in \Sigma_1 \subseteq P \cup X$,
  the $\phi_k$ and $F$ are formulae satisfying the conditions in 
  Theorem~\ref{thm:updates} which do not contain any symbol in $\Sigma_1$, such that all $\phi_k$ are 
  quantifier-free; 
\item or only formulae of the form 
		$\forall i  \big(i \neq \nil \wedge \phi \rightarrow
                F_1(f'(i), i)\big) ~~~~~ \wedge$ $\forall i  \big(i \neq \nil
                \wedge \neg \phi \rightarrow
                F_2(f'(i), i)\big)$, 
where  $f \in \Sigma_1 \subseteq P \cup X$,
  the $\phi$ and $F_1, F_2$ are formulae  which do not contain any symbol in
  $\Sigma_1$,  and such that after the instantiation of the
  variable $i$, and computing the prenex normal form and Skolemization,
  the remaining formulae are either ground or 
guarded {\sf index}-positive extended clauses of the form 
${\cal E} \vee {\cal C}$, where ${\cal C}$ is a conjunction of linear inequalities.
\end{enumerate}
\item[Assumption 3:]  \refstepcounter{assumption}\label{assumption:update}
  The formulas ${\sf Update}(p, p')$ either 
	\begin{enumerate}
	\item are of the form described in Theorem~\ref{thm:updates}, with
          $\phi_k$ quantifier-free; or
	\item contain only formulae of the form 
		$\forall i  \big(i \neq \nil \wedge \phi \rightarrow
                F_1(p'(i), i)\big)~~~ \wedge$ $\forall i  \big(i \neq \nil
                \wedge \neg \phi \rightarrow
                F_2(p'(i), i)\big)$
		where for every $p \in P \cup X$,  $p'$ is a new
                function symbol
                denoting the updated value of $p$, the formulae $\phi$
                and $F$ do not contain primed function symbols and: 
		\begin{enumerate}
		\item[(i)] $\phi = \forall j_1, \dots, j_m \psi(i, j_1, \dots, j_m)$
		with $m \geq 0$ and all free variables in $F(p'(i), i)$ occur below $p'$, or 
		\item[(ii)] $\phi = \exists {\overline j} \psi(i, {\overline j})$ and $i \neq
		\nil \wedge \psi(i, {\overline j}) \rightarrow F(i', i)$ is a 
		guarded {\sf index}-positive extended clause ${\cal E} \vee {\cal C}$, where ${\cal
			C}$ is a conjunction of linear inequalities.\
		\end{enumerate}
	\end{enumerate}
\item[Assumption 4:]  \refstepcounter{assumption}\label{assumption:HDL}
  The numeric constraints in the description
  of the SFLHA $S$ (including the conditions $\phi^p_k
  {\rightarrow} F^p_k(j, i)$ obtained from $\phi^p_k
  {\rightarrow} F^p_k(p'(i), i)$ in ${\sf Update}(p, p')$ by replacing all
  occurrences of $p'(i)$ with $j$)
   and the numerical constraints in $\Phi_{\sf safe}$ and $\Phi_{\sf entry}$ 
 are all HDL constraints or all Ord-Horn constraints.  
\end{description}

\begin{ex}
{\em 
We illustrate the restrictions imposed by Assumptions 1-4 by
examples: 
\begin{itemize}
\item {\em Assumption 1:} The formulae used in the description of our
  running example (e.g.\ in  Example~\ref{run-ex}) satisfy Assumption 1. 
\item {\em Assumption 2:} The safety conditions in
  Example~\ref{ex-properties-global-local}, namely: 
\begin{itemize}
\item $\Phi^g_{\sf safe}: 
 \forall i, j (i {\neq} \nil \wedge j {\neq} \nil \wedge {\sf lane}(i){=}
 {\sf lane}(j) \wedge {\sf pos}(i) {>} {\sf pos}(j) \rightarrow {\sf
   pos}(i) - {\sf pos}(j) {\geq} d_s)$, 

\item $\Phi^l_{\sf safe} = \bigwedge_{\pointer{} \in P} \Phi^{\pointer{}}_{\sf safe}$, 
where e.g.\ $\Phi^{\sf front}_{\sf safe}$ is:  

$\forall i  (i \neq \nil \wedge {\sf front}(i) \neq \nil  {\rightarrow}
\pos{{\sf front}(i)} - \pos{i} \geq d_s)$ 
\end{itemize}
satisfy the conditions on $\Phi_{\sf safe}$ in Assumption 2. 

 \item {\em Assumption 2(1):} The entry condition in Example~\ref{phi-entry-ex1}: 
\[ \Phi_{\sf entry} = \forall i\, ( i \neq
\nil \wedge {\sf front}(i) \neq \nil \rightarrow
|\lane{}_{\front{}}(i)-\lane[]{{\sf front}(i)}|<\varepsilon)\]
satisfies the conditions in {Assumption 2}(1). 

\item {\em Assumption 2(2):} The entry condition $\Phi_{\sf entry}$: 

$\begin{array}{@{}rl} 
\forall i (i \neq \nil \wedge {\sf front}(i) = \nil & \to \forall k (k \neq \nil \wedge k \neq i \wedge  {\sf pos}(k) \geq {\sf
  pos}(i) \rightarrow {\sf lane}(k) \neq {\sf lane}(i))) \\
 \forall i (i \neq \nil \wedge {\sf front}(i) \neq \nil & \to {\sf
  pos}_{\sf front}(i) > {\sf pos}(i) + d' \wedge {\sf lane}_{\sf
  front}(i) = {\sf lane}(i) \wedge \\
 & ~~~\forall k (k \neq \nil \wedge k \neq i \wedge  {\sf pos}(k) \geq {\sf
  pos}(i) \wedge {\sf lane}(k) = {\sf lane}(i) \\
 & ~~~~~~~\to {\sf pos}(k) \geq {\sf
  pos}_{\sf front}(i)) \wedge \\
 & ~~~{\sf pos}({\sf front}(i)) = {\sf pos}_{\sf front}(i) \wedge {\sf
  lane}({\sf front}(i)) = {\sf lane}_{\sf front}(i))
\end{array}$ 

satisfies the conditions in {Assumption 2}(2). 

\item {\em Assumption 2(3):} The entry condition $\Phi_{\sf entry}$: 
 \begin{align*}
  \forall i\big(i \neq \nil \wedge {\sf Prop}(i) \wedge \neg \exists j (\textsf{ASL}(j,i)) &\to \front[]{i} = \nil\big)\\
  \forall i\big(i \neq \nil \wedge {\sf Prop}(i) \wedge \phantom{\neg}\exists j
  (\textsf{ASL}(j,i)) &\to {\sf Closest_f}(\front[]{i}, i)\big) 
\end{align*}
with the notations in Example~\ref{ex-top-updates}, namely: 
\begin{itemize}
\item  
$\textsf{ASL}(j,i){:\ }j \neq \nil \wedge \lane{j} = \lane{i} \wedge 
\pos{j} >\pos{i}$, which expresses the fact that $j$ is ahead of $i$ on the same 
lane, and 
\item ${\sf Closest_f}(j,i){:\ }\textsf{ASL}(j, i) \wedge \forall k (\textsf{ASL}(k, i) 
{\to} \pos{k} \geq \pos{j})$, which  
expresses the fact that $j$ 
is ahead of $i$ and there is no car between them
\end{itemize}
satisfies the conditions in  {Assumption 2}(3). 

\item {\em Assumption 3:} The formula ${\sf Update}({\sf front}, {\sf
    front}')$ used for the update rules in
  Example~\ref{ex-top-updates}: 
 \begin{align*}
  \forall i\big(i \neq \nil \wedge {\sf Prop}(i) \wedge \neg \exists j (\textsf{ASL}(j,i)) &\to \front[']{i} = \nil\big)\\
  \forall i\big(i \neq \nil \wedge {\sf Prop}(i) \wedge \phantom{\neg}\exists j
  (\textsf{ASL}(j,i)) &\to {\sf Closest_f}(\front[']{i}, i)\big) \\
\forall i\big(i \neq \nil \wedge \neg {\sf Prop}(i) & \to \front[']{i} = \front[]{i} \big)
\end{align*}
satisfies the conditions in  {Assumption 3}. 

\item {\em Assumption 4:} The numeric constraints in the formulae describing the invariants,
  the initial states, the flows, guards and jumps in
  Example~\ref{run-ex}  are conjunctions of 
HDL constraints, hence satisfy {Assumption 4}. 

  In the condition $\Phi^l_{\sf safe}$ above,
  the numeric constraint is ${\sf pos}({\sf front}(i)) - {\sf pos}(i)
  \geq d_s$, hence is a HDL constraint. 
\end{itemize}
}
\end{ex}

\noindent We prove that under Assumptions 1--3 the
verification problems of Theorem~\ref{THM:VERIFICATION CONDITIONS} are decidable, and analyze their complexity. 

We analyze the complexity of verifying safety properties 
with exhaustive entry conditions, by analyzing the complexity of checking 
the satisfiability of the formulae $F^{\sf entry}_q$, $F^{\sf jump}_q$, 
$F^{\sf flow}_q$, and $F^{\sf top}_q$ (cf.\ Theorem~\ref{THM:INV-FORMULAE}). 
Since the number of systems to be considered is unbounded, 
a naive approach to analyzing the satisfiability of these formulae 
for all tuples $q = (q_i)_{i \in I}  {\in} Q^I$ can be problematic. We 
identify situations which allow us to 
limit the analysis  to a ``neighborhood'' of the systems for which 
$\phi_{\sf safe}$ fails. 
For this we use the specific form of the axioms we consider.

\subsection{Verification tasks: Chains of local theory extensions}

We show that under Assumptions 1--4 the theories used for specifying
the various verification tasks in Theorem~\ref{THM:VERIFICATION
  CONDITIONS} 
and the corresponding satisfiability problems in Theorem~\ref{THM:INV-FORMULAE} 
can be structured as chains of (stably) local theory extensions. 

\begin{thm}
\label{locality-verif}
For all $(q_i)_{i \in I} {\in} Q^I$ the following hold: 
\begin{enumerate}
\item {\em Safety of entry conditions:} 
\begin{itemize}
\item[(a)] Under Assumption
~\ref{assumption:entry} (1):\\
  $ {\mathbb R} \cup {\sf Eq}_{\pointer{}} \subseteq
	{\mathbb R} \cup \Phi_{\sf entry}$ is a stably local theory
        extension. 

\medskip 
\item[(b)] Under Assumption
~\ref{assumption:entry} (2) both theory extensions below:\\
  ${\mathbb R} \cup {\sf Eq}_{\pointer{}} \subseteq {\mathbb R} \cup {\sf UIF}_{(P\cup X) \backslash \Sigma_1}  \subseteq
	{\mathbb R} \cup \Phi_{\sf entry} $ are local theory
        extensions.\footnote{If $\Sigma$ is a set of functions then
          ${\sf UIF}_{\Sigma}$ is the theory of uninterpreted function
          symbols in $\Sigma$ axiomatized only by the congruence
          axioms for the functions in $\Sigma$. Any extension of a
          theory with
          uninterpreted function symbols is local \cite{Sofronie-cade-05}.} 


\medskip
\item[(c)] Under Assumption
~\ref{assumption:entry} (3) both extensions below:\\
  ${\mathbb R} \cup {\sf Eq}_{\pointer{}} \subseteq {\mathbb R} \cup {\sf UIF}_{(P\cup X) \backslash \Sigma_1}  \subseteq
	{\mathbb R} \cup \Phi_{\sf entry}$ are local theory
        extensions. 
However, there exist sets $G$ of ground clauses  for which 
$\Phi_{\sf entry}[G]$ may not be a set of ground clauses. In this
case, the requirements in Assumption~\ref{assumption:entry} (3) ensure
that $ {\mathbb R} \cup {\sf UIF}_{(P\cup X) \backslash \Sigma_1}  \subseteq
	{\mathbb R} \cup \Phi_{\sf entry}[G]$ is a stably local theory
        extension. 
\end{itemize}
\item {\em Invariance under flows:} 

Under Assumptions~\ref{assumption:linear and decoupled}
  and~\ref{assumption:entry}(1):\\
	${\mathbb R}  \cup {\sf Eq}_{\pointer{}} 
	\subseteq {\mathbb R} \cup (\Phi_{\sf entry}({\overline
          x}(t_0)) \cup \Phi_{\sf safe}({\overline
          x}(t_0)) \cup \{ 
	\forall i \, ({\sf Flow}_{q_i}({\overline x}(i, t_0),~~
        {\overline x}(i, t_1)))\})$ is a stably local theory
        extension. 
\item {\em Invariance under GMR jumps:} 

Under Assumptions~\ref{assumption:linear and decoupled} and~\ref{assumption:entry}(1):\\
	$ \begin{array}{@{}l@{}l} 
{\mathbb R} \cup {\sf Eq}_{\pointer{}}
\subseteq {\mathbb R} \cup  (& \Phi_{\sf entry}({\overline
          x}(t_0)) \cup \{ \forall i \, ({\sf Flow}_{q_i}({\overline x}(i, t_0), {\overline
  x}(i, t_1))) \} \cup \Phi_{\sf safe}({\overline x}(t_1)) \cup  \\
&  \{ {\sf guard}_e({\overline x}(i_0, t_1)),~~ {\sf jump}_e({\overline
   x}(i_0, t_1), {\overline x}'(i_0)),~~  {\sf Inv}_{q'\!\!\raisebox{-.1em}{${}_{i_0}$}}(\overline x'(i_0,t_1))
\})
\end{array}$ \\
is a stably local theory extension 
  for every $i_0\in I$ and $e \in E$ s.t.\ if $p(i_0)$ occurs in ${\sf guard}_e$ it is not $\nil$. 

\medskip
\item {\em Invariance under topology updates:} 

Under Assumptions~\ref{assumption:linear and decoupled},
  \ref{assumption:entry}(1), and~\ref{assumption:update}, the first
  extension below is stably local:\\
  $\begin{array}{@{}l@{}l@{}l} 
{\mathbb R} \cup {\sf Eq}_{\pointer{}} & \subseteq 
  {\mathbb R} \cup (\Phi_{\sf entry}({\overline x}(t_0)) \cup 
  \Phi_{\sf safe}({\overline x}(t_0)) \cup \{ \forall
  i \, ({\sf Flow}_{q_i}({\overline x}(i, t_0), {\overline x}(i,
  t_1))) \}) \\
&	\subseteq {\mathbb R} \cup (\Phi_{\sf entry}({\overline
  x}(t_0)) \cup \Phi_{\sf safe}({\overline x}(t_0)) \cup \{ 
	 \forall i \, ({\sf Flow}_{q_i}({\overline x}(i, t_0),
         {\overline x}(i, t_1))) \}) \\
& ~~~~~	       \cup {\sf Update}(\pointer{},\pointer[']{}). 
\end{array}$ \\
and the last extension is local.
\end{enumerate}
\end{thm}

\noindent {\em Proof:} This follows immediately from the form of the
formulae and from the locality results in Theorem~\ref{thm:updates} 
and~\ref{thm:pointers}. \hfill $\Box$


\medskip
\noindent 
{\bf Notation.} 
In the following sections let $G = \neg \phi_{\sf safe}({\overline x}(c_1), \dots, {\overline x}(c_n))$.
By Assumption~\ref{assumption:entry}, $G$ consists of 
a conjunction of ground linear inequalities and a set of 
disequalities, consisting of unit clauses of the form $g \neq 
\nil$ for every ground term $g$ of sort ${\sf index}$ occurring in 
$G$ below a pointer or scalar field. We will denote by ${\sf st}(G)$
the set of all (ground) subterms of $G$. 
The results in the next subsections follow from
Theorem~\ref{locality-verif}.



\subsection{Verification of safety properties.} 
We now analyze the decidability and complexity of verifying safety properties 
with exhaustive entry conditions, by analyzing the complexity of checking 
the satisfiability of the formulae $F^{\sf entry}_q$, $F^{\sf jump}_q$, 
$F^{\sf flow}_q$, and $F^{\sf top}_q$ (cf.\
Theorem~\ref{THM:INV-FORMULAE}). 

\subsubsection{Entry conditions} 
\label{sec:entry}
We first analyze the decidability and complexity of checking whether
entry states are safe. By Theorem~\ref{THM:INV-FORMULAE}(1), this is the case iff 
$\Phi_{\sf entry} \wedge G$ is unsatisfiable, where 
$G = \neg \phi_{\sf safe}({\overline x}(c_1), \dots, {\overline
  x}(c_n))$.
In what follows we identify conditions in which the problem of 
checking the satisfiability of this formula is decidable and 
study its complexity. 
\begin{lem}
\label{lem:decidability}
Under Assumption~\ref{assumption:entry} the following hold: 
\begin{enumerate}
\item Under Assumption 2 (1), $\Phi_{\sf  entry} \wedge G$ is unsatisfiable iff 
${\Phi_{\sf entry}}^{[G]} \wedge G$ is unsatisfiable. 
\item Under Assumptions 2 (2) or (3), $\Phi_{\sf  entry} \wedge G$ is
  unsatisfiable iff $({\Phi_{\sf entry}}[G])^{[T_G]} \wedge G$ is unsatisfiable, where
$T_G$ is the set of all ground terms of sort ${\sf index}$ occurring in ${\Phi_{\sf
    entry}}[G]$. 

\item The size of the set of terms of sort ${\sf index}$ in ${\sf st}(G)$
  and hence also the number of instances in 
${\Phi_{\sf entry}}^{[G]}$ (in case (1)) resp.\ $({\Phi_{\sf
    entry}}[G])^{[T_G]}$ (in case (2)) is polynomial in the number of
terms of sort 
${\sf index}$ in $\Phi_{\sf safe}$. Therefore also the cardinality of the set
  $I^G_{\sf entry}$ of ground terms of sort ${\sf index}$ contained in
  these sets of instances is polynomial in the number of terms of sort
  ${\sf index}$ in $\Phi_{\sf safe}$.
\end{enumerate}
\end{lem}


\noindent {\em Proof:} 
(1)  Under Assumption 2 (1), by Theorem~\ref{locality-verif}(1)(a), 
$\Phi_{\sf entry} $
 defines a stably local theory extension of ${\mathbb R} \cup {\sf
  Eq}_{\sf index}$, so in order to check whether 
$\Phi_{\sf entry} \wedge G$ is satisfiable it is sufficient to check whether 
${\Phi_{\sf entry}}^{[G]} \wedge G$ is satisfiable. 

(2) Under Assumption 2 (2) or (3), by
Theorem~\ref{locality-verif}(1)(b) or (c),
 $\Phi_{\sf entry} $
 defines a  local theory extension of ${\mathbb R} \cup {\sf
   UIF}_{X}$. Therefore, in order to check whether there exists a
 model of ${\mathbb R} \cup {\Phi}_{\sf entry} $  which is a model
 for $G$ it is sufficient to check whether there is a model of 
${\mathbb R} \cup {\Phi}_{\sf entry}[G] $  which is a model
 for $G$. Note however that ${\Phi}_{\sf entry}[G]$ is in general not
 a set of ground formulae. The conditions in Assumption 2 (2) and
 (3) ensure that this set of instances is a guarded {\sf index}-positive
 extended clause. By Theorem~\ref{thm:pointers}, in order to check 
whether there is a model of 
${\mathbb R} \cup {\Phi}_{\sf entry}[G] $  which is a model
 for $G$ it is sufficient to check whether  there is a model of 
${\mathbb R} \cup {{\Phi}_{\sf entry}[G]}^{T_G}$  which is a model
 for $G$, where $T_G$ is the set of all ground terms of sort ${\sf index}$ occurring in $\Phi_{\sf
    entry}[G] \wedge G$.

(3) We show that the number of instances (and size) of ${\Phi_{\sf entry}}^{[G]}$ (resp.\ $({\Phi}_{\sf entry}[G])^{T_G}$) 
-- hence also the size of $I^G_{\sf entry}$ -- 
is polynomial in the number of terms of sort 
${\sf index}$ in $\Phi_{\sf safe}$. 

Let $np_G$ be the number of terms of sort ${\sf index}$ occurring in $G$,
and $np_{\sf entry}$ the number of terms of sort ${\sf index}$ occurring
in $\Phi_{\sf entry}$, 
and let: 
\begin{itemize}
\item $nv_{\sf entry}$ be the number of universally quantified variables in $\Phi_{\sf
    entry}$  under Assumption 2(1) or 2(2), 
\item $na_{\sf entry}$ ( $ne_{\sf entry}$ ) be the maximal number of
  universally (existentially) quantified variables in a formula in $\Phi_{\sf
    entry}$  under Assumption 2(3). 
\end{itemize}
The number $n_{\sf entry}$ of instances in ${\Phi_{\sf
    entry}}^{[G]}$ is at most ${np_G}^{nv_{\sf entry}}$; the size $s_{\sf entry}$
(number of literals) in ${\Phi_{\sf
    entry}}^{[G]}$ is at most $np_G^{nv_{\sf entry}} \cdot
{\sf size}(\Phi_{\sf entry})$. 

$I^G_{\sf entry}$ contains all terms of sort ${\sf index}$ in 
${\Phi_{\sf  entry}}^{[G]} \wedge G$. 
Under Assumption 2(1) and (2), there can be at most $np_{\sf entry} \cdot
np_G$ such terms in ${\Phi_{\sf  entry}}^{[G]}$. 
Under Assumption 2(3) we have to additionally take into account 
the Skolem constants introduced for the existentially quantified 
variables after instantiation. For each combination of values for 
the universally quantified variables, we introduce a tuple of 
Skolem functions for the existentially quantified variables. 
We have at most ${np_G}^{na_{\sf entry}}$ possible such combinations
of values, thus at most ${np_G}^{na_{\sf entry}}$ tuples of Skolem
functions. 
Since 
in Assumption 2(3), $na_{\sf entry} = 1$, we have at most ${np_G}$ 
tuples of Skolem functions for every formula in $\Phi_{\sf entry}$ 
containing existential quantifiers. 
Thus, in this case the number of terms of sort ${\sf index}$ in 
${\Phi_{\sf  entry}}^{[G]} \wedge G$ is at most 
$np_{\sf entry} \cdot (np_G + np_G)$ (the terms which can be used as 
arguments are either the $np_G$ subterms of $G$ or the newly
introduced Skolem constants). 

In all cases, the cardinality $ni_{\sf entry}$ of $I^G_{\sf entry}$
is at most $2 \cdot np_{\sf entry} \cdot np_G$, hence is linear in the number
of terms of sort 
${\sf index}$ in $\Phi_{\sf safe}$ and in the number of variables occurring
  in $\Phi_{\sf entry}$. 
\hfill $\Box$.


\begin{thm}
\label{thm:decidability}
Under Assumption~\ref{assumption:entry}  the problem of 
  checking the satisfiability of ${\sf F}^{\sf  entry}:~~ \Phi_{\sf
    entry} \wedge G$ is 
  decidable (and in NP). 
%
\end{thm} 

\noindent {\em Proof:}  The hierarchical method for reasoning in stably local theory
extensions allows us to reduce the task of checking the satisfiability
of ${\sf F}^{\sf  entry}$ to the problem of checking the
satisfiability of a formula which is a conjunction of 
guarded {\sf index}-positive extended clauses
of the form ${\cal E} \vee {\cal C}$, where ${\cal E}$ is a
disjunction of equalities between terms of sort ${\sf index}$ 
and ${\cal C}$ a constraint over
real numbers  w.r.t.\ the disjoint combination of 
the theory of real numbers ${\mathbb R}$ and the theory of
uninterpreted functions symbols in $P \cup C$.
The reduction is done in one step if Assumption 2(1) holds, 
and in two steps if Assumption 2(2) or (3) holds. 
 The problem of checking the satisfiability of such formulae is decidable. 


In both cases the variant of Assumption 2 we use guarantees that all
the clauses we obtain are 
ground or {\sf index}-positive extended clauses of the form ${\cal E} \vee {\cal C}$,
where ${\cal C}$ is a conjunction of linear inequalities.%
\footnote{The latter can happen only under Assumption~\ref{assumption:update} (2); the remaining free variables occur only as
arguments of the variables $x$; in this case we instantiate again, the
size of the set of clauses grows polynomially.} 
After the hierarchical reduction 
we obtain a set of ground clauses in the combination of $LI({\mathbb
  R})$ and ${\sf Eq}_{\sf index}$; the complexity of checking decidability of
ground clauses in such a combination is in NP. 
\hfill $\Box$


\begin{cor}
\label{cor:small-model-prop-entry}
Let $S = ({\sf Top}, \{ S(i) \mid i \in I \})$ be an SFHA. 
Under Assumption
~\ref{assumption:entry}, the following are equivalent: 
\begin{enumerate}
\item 
There exist  indices $c_1, \dots, c_n$ for which the safety condition $\Phi_{\sf
  safe}$ does not hold although $\Phi_{\sf entry}$ holds.  
\item There exists a finite set $I_{\sf entry} \subseteq I$ of
  indices, of size polynomial in the number 
of terms of sort ${\sf index}$ in $\Phi_{\sf safe}$ (assuming that the
  lengths of the formulae describing the SFHA $S$ are considered
  constants) such that the entry conditions are not safe already 
in the systems $S_{\sf entry} = ({\sf Top}_{|I_{\sf entry}}, \{ S(i)
\mid i \in I_{\sf entry} \})$. 

$I_{\sf entry}$ and the system $S_{\sf entry}$ describe a suitable
neighborhood of 
$c_1, \dots,  c_n$  which can effectively be described (the indices in
$I_{\sf entry}$ correspond to the terms in
$I^G_{\sf entry}$ in Theorem~\ref{thm:decidability}).
\end{enumerate}
\end{cor}
\noindent {\em Proof:} 
(1) $\Rightarrow$ (2) Assume that (1) holds. 
Then $\Phi_{\sf  entry} \wedge G$ is
satisfiable.  Then ${\Phi_{\sf
    entry}}^{[G]} \wedge G$ (or resp.\ ${\Phi_{\sf
    entry}[G]}^{T_G} \wedge G$) is satisfiable, i.e.\ 
there is a model ${\cal A}$ for this formula.  
Let $I^G_{\sf entry}$ be as defined in Theorem~\ref{thm:decidability}, 
and let $I_{\sf entry}$ be the set
of the values in ${\cal A}$ of the terms in $I^G_{\sf entry}$. 
The model ${\cal A}$ can easily be   transformed into 
a model of  ${\Phi}_{\sf  entry}$, describing a system referring to
the neighborhood $I_{\sf entry}$ of the indices $c_1, \dots,  c_n$ 
at which $\Phi_{\sf entry}$ holds, but $\Phi_{\sf safe}$ does not
hold. 
But then the entry conditions are not safe already 
for the system $S_{\sf entry} = ({\sf Top}_{|I_{\sf entry}}, \{ S(i)
\mid i \in I_{\sf entry} \})$. 

By Lemma~\ref{lem:decidability} (3), the size of $I^G_{\sf entry}$
(hence also the size of $I_{\sf entry}$) is  polynomial in the number 
of terms of sort ${\sf index}$ 
in $\Phi_{\sf safe}$. 

(2) $\Rightarrow$ (1) Conversely, assume that there
exists a finite set $I_{\sf entry} \subseteq I$ of indices,
corresponding to terms in $I^G_{\sf entry}$, such that in 
$S_{\sf entry}$ there are indices $c_1, \dots,  c_n$ at
which the safety property does not hold. Then 
${\Phi}_{\sf  entry} \wedge G$ is satisfiable, if quantification is
considered to be made on the finite set $I_{\sf entry}$.  
The model for this formula is a model of 
$({\Phi}_{\sf  entry})^{[G]} \wedge G$ (or resp.\ of
$({\Phi}_{\sf  entry}[G])^{T_G} \wedge G$). 
By Lemma~\ref{lem:decidability} it
follows that ${{\Phi}_{\sf  entry}} \wedge G$ is satisfiable, i.e.\
(1) holds. 
\hfill $\Box$

\

\noindent {\bf Parametric Verification.} We can consider parametric
systems, in which we assume that some of the constants used in 
the specification of the entry conditions and safety properties are
parameters. 
If we impose constraints on these parameters (in the form of constraints between
real numbers) then the results in Theorem~\ref{thm:decidability} can still be used to 
prove that the verification problems remain decidable. The complexity 
of the problems depends on the form of the constraints (for linear
constraints we still can show that the problem is in NP). 

Alternatively, we can use the method for hierarchical reasoning
combined with quantifier elimination for the theory of real numbers 
for generating constraints on the parameters which guarantee
that $\Phi_{\sf entry} \wedge G$ is unsatisfiable, as explained 
in \cite{sofronie-cade24} (the complexity is then exponential).

\

\begin{ex}
\label{ex-entry}
{\em Consider the running example, with entry states being states in
  which the information provided by the sensors is correct and every 
car is  sufficiently far away from the following car on the same
lane, described by the following formula $\Phi_{\sf entry}$: 

\medskip
\noindent $\begin{array}{@{}rl} 
\forall i (i \neq \nil \wedge {\sf front}(i) = \nil & \to \forall k (k \neq \nil \wedge k \neq i \wedge  {\sf pos}(k) \geq {\sf
  pos}(i) \rightarrow {\sf lane}(k) \neq {\sf lane}(i))) \\
\forall i (i \neq \nil \wedge {\sf front}(i) \neq \nil & \to {\sf
  pos}_{\sf front}(i) > {\sf pos}(i) + d' \wedge {\sf lane}_{\sf
  front}(i) = {\sf lane}(i) \wedge \\
& ~~~\forall k (k \neq \nil \wedge k \neq i \wedge  {\sf pos}(k) \geq {\sf
  pos}(i) \wedge {\sf lane}(k) = {\sf lane}(i) \\
& ~~~~~~~\to {\sf pos}(k) \geq {\sf
  pos}_{\sf front}(i)) \wedge \\
& ~~~{\sf pos}({\sf front}(i)) = {\sf pos}_{\sf front}(i) \wedge {\sf
  lane}({\sf front}(i)) = {\sf lane}_{\sf front}(i)). 
\end{array}$ 

\medskip 
\noindent This formula clearly satisfies
Assumption~\ref{assumption:entry}(2), as an extension of the 
theory of ${\sf front}$, ${\sf lane}_{\sf front}$ and ${\sf pos}_{\sf front}$ with the functions
${\sf pos}$ and ${\sf lane}$, satisfying the formulae above.
Consider the following safety property: 
\[
  \Phi^g_{\sf safe}= \forall i, j (i {\neq} \nil \wedge j {\neq} \nil
  \wedge i \neq j 
  \wedge {\sf pos}(i) {>} {\sf pos}(j) \wedge {\sf lane}(i) {=} {\sf lane}(j)
	\rightarrow {\sf pos}(i) {-} {\sf pos}(j) {\geq} d_s). 
\]
We check the satisfiability of $\Phi_{\sf entry} \wedge G$,
where $G = \neg \Phi^g_{\sf safe}$ is: 
\begin{align*}
  G:\ & i_0 \neq \nil  \wedge j_0 \neq \nil \wedge i_0 \neq j_0 
  \wedge {\sf lane}(i_0) = {\sf lane}(j_0)
	\\&{}\wedge {\sf pos}(i_0) > {\sf   pos}(j_0) 
  \wedge {\sf pos}(i_0) - {\sf pos}(j_0) < d_s 
\end{align*}
as follows: We compute $\Phi_{\sf entry}[G]$. For instance, by
instantiating $i$ with $j_0$ and $k$ with $i_0$ in both formulae, 
we obtain: 

\medskip
{
 $\begin{array}{@{}l} 
 (j_0 \neq \nil \wedge {\sf front}(j_0) = \nil \wedge i_0 \neq \nil \wedge i_0 \neq j_0 \wedge  {\sf pos}(i_0) \geq {\sf
  pos}(j_0) \rightarrow {\sf lane}(i_0) \neq {\sf lane}(j_0)) \\
(j_0 \neq \nil \wedge {\sf front}(j_0) \neq \nil \to {\sf
  pos}_{\sf front}(j_0) > {\sf pos}(j_0) + d'   \wedge {\sf lane}_{\sf front}(j_0) = {\sf lane}(j_0)) \\
(j_0 \neq \nil \wedge {\sf front}(j_0) \neq \nil \wedge i_0 \neq \nil \wedge i_0 \neq j_0 \wedge  {\sf pos}(i_0) \geq {\sf
  pos}(j_0) \wedge {\sf lane}(i_0) = {\sf lane}(j_0) \\
 ~~~~~~~\to {\sf pos}(i_0) \geq {\sf pos}_{\sf front}(j_0)). 
\end{array}$ } 

\medskip
\noindent After the hierarchical reduction, 
we obtain a set of clauses which is clearly unsatisfiable if $d' \geq d_s$. 
Below a short intuitive justification: From the literals in $G$ and
the first formula above we derive that ${\sf front}(j_0) \neq \nil$. 
Together with the second formula we then obtain: 
$$ {\sf pos}_{\sf front}(j_0) > {\sf pos}(j_0) + d'  \wedge {\sf lane}_{\sf front}(j_0) = {\sf lane}(j_0) ,$$
and together with the third formula we obtain: 
$$ {\sf pos}(i_0) \geq {\sf pos}_{\sf front}(j_0), \text{ hence } {\sf pos}(i_0) > {\sf pos}(j_0) + d'. $$
If $d_s$ and $d'$ are numerical values such that $d' \geq d_s$, this is
unsatisfiable. 

\medskip
\noindent {\em Parametric verification.} In this problem $d_s$ and
$d'$ can also be considered to be parameters. If we
assume that $d' \geq d_s$, we can easily see that 
${\sf pos}(i_0) > {\sf pos}(j_0) + d'  \wedge {\sf pos}(i_0) - {\sf pos}(j_0) < d_s$ is
unsatisfiable. 
Alternatively, we can use quantifier elimination after the
hierarchical reduction to prove that 
${\sf pos}(i_0) > {\sf pos}(j_0) + d'  \wedge {\sf pos}(i_0) - {\sf pos}(j_0) < d_s$ is
unsatisfiable iff $d' \geq d_s$. 

\medskip
\noindent 
{\em Small model property} The instantiation we used justifies a small model property as
explained in Corollary~\ref{cor:small-model-prop-entry}: In order to
check whether the states satisfying the entry condition $\Phi_{\sf
  entry}$ also satisfy the safety property expressed by $\Phi^g_{\sf safe}$, we
first choose two different cars for which the safety condition may not
hold, corresponding to the indices $i_0$ and $j_0$ in $G$. 
The instances of  $\Phi_{\sf entry}[G]$ contain two additional terms
of sort ${\sf index}$, namely ${\sf front}(i_0)$ and ${\sf front}(j_0)$. 
We know that $i_0$ and $j_0$ are not $\nil$ and that they are
different. We do not know however whether ${\sf front}(i_0)$ or 
${\sf front}(j_0)$ are $\nil$ (neither whether they are equal to each
other, or whether ${\sf front}(i_0) = j_0$ or ${\sf front}(j_0)
=i_0$). 
We need to consider all such combinations, i.e.\ check 
whether $\Phi_{\sf entry}$ entails $\Phi_{\sf safe}$ in all systems 
$S_{I_0} = ( {\sf Top}_{|I_0}, \{ S(i) \mid i \in I_0 \})$, where 
$I_0$ are indices corresponding to the set of terms  
$I^G_{\sf entry} = \{ \nil, i_0, j_0, {\sf front}(i_0), {\sf front}(j_0) \}$
(taking into account that one or more of the elements of  $I_0$
might be equal). 

We now analyze the complexity of checking whether in $S_{I_G}$ 
$\Phi_{\sf entry}[G] \cup G$ is satisfiable for a given $I_G$. 
Such systems describe models of 
$\Phi_{\sf entry}\cup G$ obtained by using the usual 
completion -- which sets all undefined functions 
of sort ${\sf index}$ to $\nil$ -- from 
models of $\Phi_{\sf entry}[G] \cup G$. 
Given one such system, we know precisely the equality relationships
between the terms in $I_G$. Depending on this, we have the one of the 
following situations: 
\begin{itemize}
\item some of the premises of the formulae in $\Phi_{\sf entry}[G]$
  may be false: then the corresponding instance is true in this model
\item all premises of the formulae in $\Phi_{\sf entry}[G]$ are true: 
We then only need to check the satisfiability of the conjunctions of 
linear constraints on the left-hand side, which can be done in
polynomial time.
\end{itemize}
Note that if the guards of sort ${\sf index}$ in the formulae in $\Phi_{\sf safe}$ and $\Phi_{\sf entry}$ are 
terms of the form $t = \nil$ then we do not need to take into 
account all possible equality relationships
between the terms in $I^G_{\sf entry}$, but only possible equality of 
such terms with $\nil$. The number of all possible systems which 
need to be tested is then $2^{|I^G_{\sf entry} \backslash {\sf
    st}[G]|}$, in our example $2^{|\{ {\sf front}(i_0), {\sf
    front}(j_0) \}|} = 2^2$.

}\end{ex}

\subsubsection{Flows}
\label{sec:flows}
We now analyze the decidability and complexity of checking whether
$\Phi_{\sf safe}$ is preserved under all flows starting from a state
satisfying $\Phi_{\sf entry}$. According to Theorem~\ref{THM:INV-FORMULAE}(2), this can
be expressed as the problem of checking, for all $q= (q_i)_{i \in I}
\in Q^I$, the satisfiability of the
formula: 
\begin{align*}
	F^{\sf flow}_q:\ & t_0 < t_1 \wedge 	\Phi_{\sf entry}({\overline x}(t_0)) \wedge 
	\forall i_1, \dots, i_n \phi_{\sf safe}({\overline x}(i_1, t_0), \dots, {\overline x}(i_n, t_0))
	\\&{}\wedge \forall i \, {\sf Flow}_{q_i}({\overline x}(i, t_0), {\overline x}(i, t_1))
	\wedge G 
\end{align*}
where if ${\sf flow}_q(i) = \bigwedge \big({\cal E}_f \vee \sum_{k = 1}^n 
a^q_k(i) \dot x_k(i) \leq a^q(i)\big)$ then   
\begin{align*}
  {\sf Flow}_{q_i}({\overline x}(i, t_0), {\overline x}(i,t_1)) :\ &
	\bigwedge \big({\cal E}_f \vee \sum_{k = 1}^n a^{q_i}_k(i) (x_k(i, t_1) {-} x_k(i,t_0)) {\leq} a^{q_i}(i) (t_1{-} t_0)\big) \\
  &{}\wedge  {\sf Inv}_{q_i}({\overline x}(i, t_0))
	\wedge  {\sf Inv}_{q_i}({\overline x}(i, t_1)) 
\end{align*}
and $G = \neg \phi_{\sf safe}({\overline x}(c_1, t_1), \dots,
{\overline x}(c_n, t_1))$. 

\begin{lem}[Flows]
\label{lem:decidability of flows}
Under Assumptions~\ref{assumption:linear and decoupled}
and~\ref{assumption:entry}(1) the following hold: 
\begin{enumerate}
\item For every $q = (q_i)_{i \in I} \in Q^I$, 
${\sf F}^{\sf  flow}_q$ is unsatisfiable iff ${F^{\sf
    flow}_q}^{[G]}$ is
unsatisfiable. 

\item The size of the set of terms of sort ${\sf index}$ in ${\sf st}(G)$
  and hence also the size of ${F^{\sf flow}_q}^{[G}$ is polynomial in
  the number of terms of sort ${\sf index}$ in $\Phi_{\sf safe}$. 
Therefore also the size of the set
  $I^G_{\sf flow}$ of ground terms of sort ${\sf index}$ in ${F^{\sf
      flow}_q}^{[G]}$  is polynomial in the number of terms of sort 
${\sf index}$ in $\Phi_{\sf safe}$.

The set of instances 
${F^{\sf    flow}_q}^{[G]}$ contain formulae ${\sf Inv}_{q_i}$ and
${\sf Flow}_{q_i}$ for indices $i$ corresponding to
  terms in $I^G_{\sf flow}$.  
\end{enumerate} 
\end{lem}

\noindent \noindent {\em Proof:} 
(1) If $\Phi_{\sf entry}$ satisfies Assumption 2(1) then, by
Theorem~\ref{locality-verif}(2), 
for every $q = (q_i)_{i \in I}
\in Q^I$ the set of axioms: 
\[ \begin{array}{ll} 
{\cal K}_{\sf flow} = & \Phi_{\sf entry}({\overline x}_0) \wedge \Phi_{\sf
  safe}({\overline x}_0)  \wedge \forall i \, {\sf Flow}_{q_i}({\overline x}_0(i),
{\overline x}_1(i))
\end{array} \]
 defines a stably local theory extension of ${\mathbb R} \cup {\sf
  Eq}_{\sf index}$, so in order to check whether 
${\sf F}^{\sf flow}_q$ is satisfiable it is sufficient to check whether 
${{\cal K}_{\sf flow}}^{[G]} \wedge G$ is satisfiable.

(2) Clearly, the size of ${{\cal K}_{\sf flow}}^{[G]}$ (hence also the size
of $I^G_{\sf flow}$)
is polynomial in the number of terms of sort 
${\sf index}$ in $\Phi_{\sf safe}$. 
Because of Assumption~\ref{assumption:entry}, this set of instances 
contains only the instances of $\forall i {\sf Inv}_{q_i}$ in which
$i$ is replaced by a term in $I^G_{\sf flow}$. But this means that 
only the states $q_i$, where $i \in I^G_{\sf flow}$ need to be
considered. 
(This also means that in order to check invariance of the safety condition
under all flows, we only need to consider combinations of states 
of systems corresponding to the indices in $ I^G_{\sf flow}$). 

With the notation used in the proof of Lemma~\ref{lem:decidability} (3) we
have the following upper bounds for the size of ${{\cal K}_{\sf
    flow}}^{[G]}$ and of $I^G_{\sf flow}$: 
\begin{itemize}
\item the number $n_{\sf flow}$ of clauses in ${{\cal K}_{\sf   flow}}^{[G]}$ is
  $n_{\sf flow} = n_{\sf entry} + n_{\sf safe} + n_{\sf Flow} \leq
  {np_G}^{nv_{\sf entry}} + {np_G}^{nv_{\sf safe}} + c \cdot np_G$,  

where $n_{\sf entry}$ is the number of instances in ${\Phi_{\sf
    entry}}^{[G]}$ (thus at most ${np_G}^{nv_{\sf entry}}$); $n_{\sf safe}$ is the number of instances in ${\Phi_{\sf
    safe}}^{[G]}$ (thus at most ${np_G}^{nv_{\sf safe}}$, proof
analogous to the proof of Lemma~\ref{lem:decidability}(3)), and $n_{\sf Flow}$ is the number
of instances of $\forall i {\sf Flow}_{q_i}({\overline x}_0(i),
{\overline x}_1(i))$. Since ${\sf Flow}$ is a conjunction of $c$ formulae,
each having only one universally quantified variable, the number of
instances is at most $c \cdot np_G$. 

\item the number $ni_{\sf flow}$ of elements in $I^G_{\sf flow}$ is 
$ni_{\sf flow} = ni_{\sf entry} + ni_{\sf safe}+ ni_{\sf Flow} \leq
(np_{\sf entry} + np_{\sf safe} + np_{\sf Flow}) \cdot np_G$ (the
justification is the same as that used in the proof of
Lemma~\ref{lem:decidability}(3)). \hfill $\Box$
\end{itemize}

\begin{thm}
\label{thm:decidability of flows} 
For every  $q \in Q^I$, the satisfiability of the formulae ${F^{\sf
    flow}_{q}}$ is decidable (and in NP). 
\end{thm} 

\noindent {\em Proof:} The hierarchical method for reasoning in stably local theory
extensions allows us to reduce the task of checking the satisfiability
of ${\sf F}^{\sf  flow}_q$ to the problem of checking the
satisfiability of a formula which is a conjunction of guarded {\sf index}-positive extended clauses
of the form ${\cal E} \vee {\cal C}$, where ${\cal E}$ is a
disjunction of equalities of sort ${\sf index}$ and ${\cal C}$ a constraint over
real numbers  w.r.t.\ the disjoint combination of 
the theory of real numbers ${\mathbb R}$ and the theory of
uninterpreted function symbols in $P \cup X$. 

Due to Assumption~\ref{assumption:linear and decoupled}, 
all the clauses in ${\sf F}^{\sf  flow}_q$ are ground or {\sf index}-positive
 extended clauses of the form ${\cal E} \vee {\cal C}$,
where ${\cal C}$ is a conjunction of linear inequalities.%
We obtain a set of ground clauses in the combination of $LI({\mathbb
  R})$ and ${\sf Eq}_{\sf index}$. 
\hfill $\Box$

\medskip

The locality result mentioned above shows that in order to check invariance of the safety condition 
under all flows, we only need to consider combinations of states 
of systems corresponding to the indices in $ I^G_{\sf 
  flow}$. Therefore checking invariance under all flows is decidable.  

\begin{cor}
\label{cor:fin-st-flows}
Under Assumptions~\ref{assumption:linear and decoupled} and~\ref{assumption:entry}(1), 
there exists a finite set $I_{\sf flow} \subseteq I$ of indices,
such that the following are equivalent:
\begin{enumerate}
	\item $F^{\sf flow}_q$ is satisfiable for some $q  \in  Q^I$
	\item $F^{\sf flow}_{q_0}$ is satisfiable for some $q_0 \in Q^{I_{\sf flow}}$.
\end{enumerate}
Therefore checking invariance under all flows is decidable (and in NP). 
\end{cor}

\noindent {\em Proof:} 
(1) $\Rightarrow$ (2) Assume that for some $q \in Q^I$,  ${\sf F}^{\sf  flow}_q$ is
satisfiable. By Theorem~\ref{thm:decidability}, ${F^{\sf flow}_q}^{[G]}$ is
satisfiable.
Then there is a model ${\cal A}$ for this formula.  
Let $I^G_{\sf entry}$ be the set of ground terms of sort ${\sf index}$ in 
${F^{\sf    flow}_q}^{[G]}$, and let $I_{\sf flow}$ be the set
of the values of the terms in the model ${\cal A}$. The model ${\cal A}$ can
easily be   transformed into 
a model of  ${\sf F}^{\sf  flow}_{q_0}$, where $q_0$ is the
restriction of $q$ to $I_{\sf flow}$. 

(2) $\Rightarrow$ (1) Conversely, assume that there
exists a finite set $I_{\sf flow} \subseteq I$ of indices,
corresponding to terms in $I^G_{\sf flow}$ (and
thus to a neighborhood of the indices of cars that may violate the
safety condition) and a tuple of modes  $q_0 \in Q^{I_{\sf flow}}$ 
such that ${\sf F}^{\sf  flow}_{q_0}$ is satisfiable. This model is 
a model of ${{\sf F}^{\sf  flow}_{q}}^{[G]}$. By Theorem~\ref{thm:decidability} it
follows that ${{\sf F}^{\sf  flow}_{q}}$ is satisfiable. \hfill $\Box$

\medskip
The results in Lemma~\ref{lem:decidability of flows},
Theorem~\ref{thm:decidability of flows} 
and Corollary~\ref{cor:fin-st-flows}
immediately imply the following small model property. 

\begin{cor}
\label{cor:small-model-prop-flow}
Let $S = ({\sf Top}, \{ S(i) \mid i \in I \})$ be an SFHA. 
Under Assumption~\ref{assumption:linear and decoupled}
and~\ref{assumption:entry}(1), 
the following are equivalent: 
\begin{enumerate}
\item 
There exist  indices $c_1, \dots, c_n$ for which the safety condition 
$\Phi_{\sf  safe}$ is not preserved under flows starting in a state in 
which $\Phi_{\sf entry}$ holds.  
\item There exists a finite set $I_{\sf flow} \subseteq I$ of
  indices, of size polynomial in the size of $n$ (assuming that the
  lengths of the formulae describing the SFHA $S$ are considered
  constants) describing a suitable neighborhood of $c_1, \dots,
  c_n$  which can effectively be described (they correspond to the terms in
$I^G_{\sf flow}$ in Theorem~\ref{thm:decidability}) 
such that  already 
in the systems $S_{\sf flow} = ({\sf Top}_{|I_{\sf flow}}, \{ S(i)
\mid i \in I_{\sf flow} \})$ the safety condition 
$\Phi_{\sf  safe}$ is not preserved under flows starting in a state in which $\Phi_{\sf entry}$ holds.  
\end{enumerate}
\end{cor}
\noindent {\em Proof:} 
(1) $\Rightarrow$ (2) Assume that (1) holds. Then for some 
$q = (q_i)_{i \in I} \in Q^I$, ${\cal K}_{\sf  flow}
\wedge G$ is satisfiable (with the notation in the proof of
Lemma~\ref{lem:decidability of flows}). 
By Theorem~\ref{thm:decidability}, ${{\cal K}_{\sf
    flow}}^{[G]} \wedge G$ is satisfiable.
Then there is a model ${\cal A}$ for this formula.  
Let $I^G_{\sf flow}$ be as defined in Theorem~\ref{thm:decidability of
flows}, and let $I_{\sf flow}$ be the set
of the values in ${\cal A}$ of the terms in $I^G_{\sf flow}$. 
The model ${\cal A}$ can easily be   transformed into 
a model of  ${\cal K}_{\sf flow} \wedge G$, describing a system referring to
the neighborhood $I_{\sf flow}$ of the indices $c_1, \dots,  c_n$ 
at which $\Phi_{\sf safe}$ does not
hold, although $\Phi_{\sf safe}$ and $\Phi_{\sf entry}$ hold at the
beginning of the flow. 
But then for $q = (q_i)_{i \in I} \in Q^{I_{\sf flow}}$, 
$\Phi_{\sf safe}$ is not invariant under flows starting in a 
state in which $\Phi_{\sf entry}$ holds already 
for the system $S_{\sf entry} = ({\sf Top}_{|I_{\sf flow}}, \{ S(i)
\mid i \in I_{\sf flow} \})$. 

(2) $\Rightarrow$ (1) Conversely, assume that there
exists a finite set $I_{\sf flow} \subseteq I$ of indices,
corresponding to terms in $I^G_{\sf flow}$, a tuple $q = (q_i)_{i \in
  I} \in Q^{I_{\sf flow}}$, 
and that in $S_{\sf flow}$ there are indices $c_1, \dots,  c_n$ at 
which the safety property does not hold at the end of a flow starting
in a state in which $\Phi_{\sf safe}$ and $\Phi_{\sf entry}$
hold. Then ${{\cal K}_{\sf flow}} \wedge G$ (with instantiation over
$I_{\sf flow}$ is satisfiable, i.e.\ it has a model. As $I_{\sf flow}$
corresponds to $I^G_{\sf flow}$, we can obtain 
a model of ${{\cal K}_{\sf flow}}^{[G]} \wedge G$. 
By Theorem~\ref{thm:decidability} it
follows that ${{\cal K}_{\sf flow}}^{[G]} \wedge G$ is satisfiable, i.e.\
(1) holds. 
\hfill $\Box$

\

\noindent {\bf Parametric Verification.} If we consider parametric
systems, in which some of the constants used in 
the specification of the entry conditions, flows, and safety properties are
parameters, we have again the following options: 
If we impose constraints on these parameters (in the form of constraints between
real numbers) then the results in Theorem~\ref{thm:decidability of
  flows} and Corollary~\ref{cor:fin-st-flows} can still be used to 
prove that the verification problems remain decidable. The complexity 
of the problems depends on the form of the constraints (for linear
constraints, in particular when Assumptions 1-3 hold and parameters
are not allowed as coefficients and do not appear as bounds in the
flow conditions we still can show that the problem is in NP). 
For systems in which parameters are allowed as coefficients 
or appear in the flow conditions, the 
complexity is exponential. 

We can use the method for hierarchical reasoning
combined with quantifier elimination for the theory of real numbers 
for generating constraints on the parameters which guarantee
that $F^{\sf flow}_{q_0}$ is unsatisfiable for all 
$q_0 \in Q^{I_{\sf flow}}$  (the complexity is exponential).

\

\begin{ex}
{\em We consider the following safety property: 
$$ \Phi^l_{\sf safe}: \forall i (i \neq \nil \wedge {\sf front}(i) \neq \nil  
\rightarrow {\sf pos}({\sf front}(i)) - {\sf pos}(i) \geq d_s).$$
Consider the tuple
$(q_i)_{i \in I}$ consisting of the acceleration modes for all
systems 
\[ {\sf Inv}_{q_i}(i)  := i \neq \nil \wedge {\sf front}(i) \neq \nil \rightarrow {\sf
  pos}({\sf front}(i),t_0) - {\sf pos}(i,t_0) \geq d.\]
%
$\Phi^l_{\sf safe}$ is invariant under flows in mode $(q_i)_{i \in
  I}$ if and only if the following formula is unsatisfiable: 

\vspace{-2mm}
{\[ \begin{array}{@{}rlr}
0 \leq t_0 < t_1 \leq \Delta t\ \wedge & 
  \forall i (i \neq \nil \wedge {\sf front}(i) \neq \nil  
\rightarrow {\sf pos}({\sf front}(i),t_0) - {\sf pos}(i,t_0) > d_s) & ~~~\Phi^l_{\sf safe}(t_0)
 \\
\wedge&  \forall i (i \neq \nil \wedge {\sf front}(i) \neq \nil  
\rightarrow {\sf pos}({\sf front}(i),t_0) - {\sf pos}(i,t_0) \geq d) &
\forall i \, {\sf Inv}_{q_i}(t_0)  
\\
\wedge & \forall i (i \neq \nil \wedge {\sf front}(i) \neq \nil  
\rightarrow {\sf pos}({\sf front}(i),t_1) - {\sf pos}(i,t_1) \geq d) &
\forall i \, {\sf Inv}_{q_i}(t_1)  
 \\
\wedge & \forall i( i \neq \nil \rightarrow {\sf pos}(i, t_1) - {\sf pos}(i, t_0) \leq v_{\sf
  max}(t_1 - t_0)) & {\sf Flow}(t_0, t_1) \\ 
\wedge & i_0 \neq \nil \wedge {\sf front}(i_0) \neq \nil \wedge {\sf
  pos}({\sf front}(i_0), t_1) - {\sf pos}(i_0, t_1) \leq d_s & G
\end{array} \]}%

\medskip

The universally quantified conjuncts in the formula are guarded {\sf index}-positive
clauses.  After instantiation and purification, 
we obtain: 

\smallskip
{\small \quad \quad  $\begin{array}{@{}l|ll}
D &  & {{\cal K}_{\sf flow}^{[G]}}_0 \wedge {G}_0 \\
\hline 
f = {\sf front}(i_0) & & 0 \leq t_0 < t_1 \leq \Delta t \\
f'= {\sf front}(f) &  { {\Phi^l_{\sf safe}}^{[G]}}_0 & i_0 \neq \nil \wedge f \neq \nil  
\rightarrow p_{10} - p_{00} > d_s \\ 
p_{00} = {\sf pos}(i_0,t_0) & & f \neq \nil \wedge f' \neq \nil  
\rightarrow p_{20} - p_{10} > d_s \\ 
p_{01} = {\sf pos}(i_0,t_1) & {{{\sf Inv}_a(t_0)}^{[G]}}_0  & i_0 \neq \nil \wedge f \neq \nil  
\rightarrow p_{10} - p_{00} > d \\ 
p_{10} = {\sf pos}(f,t_0) & & f \neq \nil \wedge f' \neq \nil  
\rightarrow p_{20} - p_{10} > d \\ 
p_{11} = {\sf pos}(f, t_1) & {{{\sf Inv}_a(t_1)}^{[G]}}_0 & i_0 \neq \nil \wedge f \neq \nil  
\rightarrow p_{11} - p_{01} > d\\  
p_{20} = {\sf pos}(f',t_0) & & f \neq \nil \wedge f' \neq \nil  
\rightarrow p_{21} - p_{11} > d\\  
p_{21} = {\sf pos}(f',t_1) & {{\sf Flow}_q^{[G]}}_0 &  i_0 \neq \nil
\wedge f \neq \nil  \rightarrow p_{01} - p_{00} \leq v_{\sf max} (t_1 - t_0) \\
& &  f \neq \nil
\wedge f' \neq \nil  \rightarrow p_{11} - p_{10} \leq v_{\sf max} (t_1 - t_0) \\
& G_0 &  i_0 \neq \nil \wedge f \neq \nil 
\wedge p_{11} - p_{01} \leq d_s \\
& N_0 & \!\!\!\!\!\!\!\!\!\!\!\!\!
\text{(instances of the congruence axioms)}\\
\end{array}$} 

\

\noindent It is easy to check unsatisfiability if $d > d_s$.
This proves that if $d > d_s$ then $\Phi^l_{\textsf{Safe}}$ is invariant under flows. 

\

The modularity/small model property result in Corollary~\ref{cor:small-model-prop-flow} can be 
used as follows: From the safety property, we can determine 
the index set $I_{\sf flow}$ which we need to consider
(which describes the instances of the universally quantified formulae 
which we need to take into account). 
For the example described above, $I^G_{\sf flow} = \{ i_0, {\sf
  front}(i_0), {\sf front}({\sf front}(i_0)) \}$. 
Since we know that $i_0 \neq 0$ and ${\sf front}(i_0) \neq 0$, 
we have two situations to consider: one in which ${\sf front}({\sf
  front}(i_0)) = \nil$ and one in which  ${\sf front}({\sf
  front}(i_0)) \neq \nil$ (equalities between $i_0,  {\sf
  front}(i_0)$ and ${\sf front}({\sf front}(i_0))$ are ruled out by 
the conditions on ${\sf pos}$). 

By Corollary~\ref{cor:small-model-prop-flow}, in order to check whether 
all initial states are safe, it is sufficient to restrict to 
families of systems $(\environment_{I_{\sf flow}}, \{ S(i) \mid i \in 
  I_{\sf flow} \} )$ for the two situations: 
\begin{itemize}
\item $I_{\sf flow} = \{ c_0,  c_2 \}$ 
where ${\sf front}(c_0) = c_1$ and ${\sf front}(c_1) = \nil$, 
and 
\item $I_{\sf flow} = \{ c_0,  c_2, c_3 \}$, where 
 ${\sf front}(c_0) = c_1$ and ${\sf front}(c_2) = c_3$,  ${\sf
   front}(c_3) = \nil$. 
\end{itemize}

We will need to consider combinations of modes 
(${\sf Appr}$/${\sf Rec}$) only for the systems in this family, 
thus we need to try only $2^2 + 2^3$  possible 
combinations of modes.


\smallskip
The  global safety condition: 
$\forall i, j (i \neq \nil \wedge j \neq \nil \wedge {\sf lane}(i) =
{\sf lane}(j) \wedge {\sf pos}(i) > {\sf pos}(j) 
\rightarrow {\sf pos}(i) - {\sf pos}(j) \geq d)$ 
can be checked only together with properties which guarantee that 
the imprecise information of the sensors does not impact on safety. 
For proving such properties, we use timed 
topologies and timed topology updates. 
} 
\end{ex}


\subsubsection{Jumps}
\label{sec:jumps}
We now analyze the decidability and complexity of checking whether
$\Phi_{\sf safe}$ is preserved under all jumps starting from a state
reachable by a flow from a state satisfying $\Phi_{\sf entry}$. 
According to Theorem~\ref{THM:INV-FORMULAE}(3), this can
be expressed as the problem of checking whether 
for all  $q {=} (q_i)_{i \in I} {\in} Q^I$ the following 
formula ${F^{\sf jump}}^q_{e}(i_0)$ 
is unsatisfiable for every $i_0 \in I$ 
and $e=(q_{i_0},q'_{i_0}) \in E$, s.t.\ if $p(i_0)$ occurs 
in ${\sf guard}_e$ it is not $\nil$:  
\begin{align*}
	{F^{\sf jump}}^q_{e}(i_0):\ 	\Phi_{\sf entry}({\overline
          x}(t_0)) 
& 
{}\wedge \Bigg( \bigg(t_0 < t_1 \wedge \forall i {\sf
          Flow}_{q_i}({\overline x}(i, t_0), {\overline x}(i, t_1)) \bigg)
        \vee t_0 = t_1 \Bigg) \\
	& {}\wedge 
	\forall i_1, \dots, i_n \phi_{\sf safe}({\overline x}(j_1, t_1), \dots,{\overline x}(i_n, t_1)) \\
	& {}\wedge {\sf guard}_e({\overline x}(i_0, t_1)) \wedge {\sf jump}_{e}({\overline x}(i_0, t_1), {\overline x}'(i_0))
		{}\wedge {\sf Inv}_{q'_{i_0}}(\overline x'(i_0))\\
	& {}\wedge\forall j (j \neq i_0 \rightarrow {\overline x}'(j) = {\overline x}(j))
	 \wedge G, 
\end{align*}
where $G = \neg \phi_{\sf safe}({\overline x}(c_1, t_1), \dots,
{\overline x}(c_n, t_1))$. 

\begin{lem}[Jumps]
\label{lem:decidability of jumps}
Under Assumptions~\ref{assumption:linear and decoupled} 
and~\ref{assumption:entry}(1) the following hold: 
\begin{enumerate}
\item For every $q = \in Q^I$, 
${\sf F}^{\sf  jump}_q$ is unsatisfiable iff ${F^{\sf jump}_q}^{[G]}$ is
unsatisfiable. 

\item The size of the set of terms of sort ${\sf index}$ in ${\sf st}(G)$
  and hence also the size of ${F^{\sf jump}_q}^{[G]}$ is polynomial in
  the number of terms of sort 
${\sf index}$ in $\Phi_{\sf safe}$. Therefore also the size of the set
  $I^G_{\sf jump}$ of ground terms of sort ${\sf index}$ in ${F^{\sf
      jump}_q}^{[G]}$  is polynomial in the number of terms of sort
 ${\sf index}$ in $\Phi_{\sf safe}$.

The set of instances 
${F^{\sf    jump}_q}^{[G]}$ contain formulae ${\sf Inv}_{q_i}$ and
${\sf Flow}_{q_i}$ corresponding to
  terms $i \in  I^G_{\sf jump}$. 
\end{enumerate} 
\end{lem}

\noindent {\em Proof:} 
The proof is similar to the one of Lemma~\ref{lem:decidability} 
and  Lemma~\ref{lem:decidability of flows} 
using Theorem~\ref{locality-verif}(3).
The set of terms $I^G_{\sf jump}$ corresponding to $i_0$ 
is the set of all ground terms of sort ${\sf index}$ in ${{\cal K}_{\sf
    jump}}^{[G]}$.

The estimation of the number $n_{\sf jump}$ of instances in ${{\cal K}_{\sf
    jump}}^{[G]}$ and on the number of terms $ni_{\sf jump}$ in $I^G_{\sf jump}$ is
similar to that made in the proofs of Lemma~\ref{lem:decidability}(3)
and Lemma~\ref{lem:decidability of flows}(2).  With the notations used
in the proofs of these Lemmata we have: 
\begin{itemize}
\item $n_{\sf jump} = n_{\sf entry} + n_{\sf safe} + n_{\sf Flow} +
  {np}_G \leq  {np_G}^{nv_{\sf entry}} + {np_G}^{nv_{\sf safe}} +
  (c +1) \cdot np_G$; 

\item $ni_{\sf jump} =  ni_{\sf entry} + ni_{\sf safe}+ ni_{\sf Flow}
  + ni_{\sf Jump} \leq (np_{sf entry} + np_{\sf safe} + np_{\sf Flow})
  \cdot np_G +   ni_{\sf Jump}$, 

where $ni_{Jump}$ is the number of terms
  of sort ${\sf index}$ occurring in 
$${\sf guard}_e({\overline x}(i_0, t_1)) \wedge {\sf jump}_{e}({\overline x}(i_0, t_1), {\overline x}'(i_0))
		{}\wedge {\sf Inv}_{q'_{i_0}}(\overline x'(i_0)).$$
\end{itemize}
\hfill $\Box$

\begin{thm}[Jumps]
\label{thm:decidability of jumps}
 For every $q \in Q^I$, the satisfiability of 
${F^{\sf jump}_{q}}$ is decidable (and in NP). 
\end{thm}

\noindent {\em Proof:} 
Follows from Lemma~\ref{lem:decidability of jumps} and the fact that 
for every $q_0 \in Q^{I_{\sf jump}}$, the satisfiability of 
${F^{\sf jump}_{q}}^{[G]}$ is decidable (and it is in NP). \hfill $\Box$

\

\noindent The following two results can be proved as in the case of flows.

\begin{cor}
\label{cor:fin-st-jumps}
Let $S = ({\sf Top}, \{ S(i) \mid i \in I \})$ be an SFHA. 
Under Assumptions~\ref{assumption:linear and decoupled}
and~\ref{assumption:entry}(1), 
there exists a finite set $I_{\sf jump} \subseteq I$ of indices,
such that the following are equivalent:
\begin{enumerate}
	\item $F^{\sf jump}_q$ is satisfiable for some $q  \in  Q^I$
	\item $F^{\sf jump}_{q_0}$ is satisfiable for some $q_0 \in Q^{I_{\sf jump}}$.
\end{enumerate}
Therefore checking invariance under all GMR jumps is decidable (and in
NP). 
\end{cor}

\begin{cor}
\label{cor:small-model-prop-jumps}
Under Assumptions~\ref{assumption:linear and decoupled} and \ref{assumption:entry}(1), the following are equivalent: 
\begin{enumerate}
\item 
There exist  indices $c_1, \dots, c_n$ for which the safety condition $\Phi_{\sf
  safe}$ does not hold after a jump following a flow starting in a state satisfying
$\Phi_{\sf entry}$. 
\item There exists a finite set $I_{\sf jump} \subseteq I$ of
  indices, of size polynomial in the size of $n$ (assuming that the
  length of the formulae describing the SFHA $S$ are considered
  constants) such that already 
in the system $S_{\sf jump} = ({\sf Top}_{|I_{\sf jump}}, \{ S(i)
\mid i \in I_{\sf jump} \})$ the safety condition $\Phi_{\sf
  safe}$ does not hold after a jump following a flow starting in a state satisfying
$\Phi_{\sf entry}$. 

The set of indices $I_{\sf jump}$ and the system $S_{\sf jump} = ({\sf Top}_{|I_{\sf jump}}, \{ S(i)
\mid i \in I_{\sf jump} \})$ describe a suitable neighborhood of the
systems $c_1, \dots,  c_n$  at which the safety property is not
preserved under jumps, which can effectively be described (they correspond to the terms in
$I^G_{\sf jump}$ in Lemma~\ref{lem:decidability of jumps}).
\end{enumerate}
\end{cor}

\

\noindent {\bf Parametric Verification.} 
If we impose constraints on these parameters (in the form of constraints between
real numbers) then the results in Theorem~\ref{thm:decidability of
  jumps} and Corollary~\ref{cor:fin-st-jumps} can be used to 
prove that the verification problems remain decidable. For linear
constraints, in particular when Assumptions 1-3 hold and parameters
are not allowed as coefficients and do not appear as bounds in the
flow conditions, the problem is in NP. For systems in which parameters 
are allowed as coefficients or appear in the flow conditions, the 
complexity is exponential. 

We can use the method for hierarchical reasoning
combined with quantifier elimination for the theory of real numbers 
for generating constraints on the parameters which guarantee
$\Phi_{\sf safe}$ is preserved under GMR jumps  (the complexity is exponential).

\

\begin{ex}
{\em 
We consider the following safety property
$\Phi_{\sf safe}$: 
\[\Phi_{\sf safe}{:\ } \forall i,j( i\not=\nil\wedge j \neq \nil
\wedge {\sf lane}(i) =
{\sf lane}(j) \wedge \pos{i}=\pos{j}\to i=j).\]
Because jumps are instantaneous and $\pos{}$ is a continuous variable, 
$\Phi_{\sf safe}$ is obviously invariant under jumps where the lane is not
changed, i.e.\ where no variables are updated. To verify a jump where
an update of the lane occurs, we look at a transition from the first to the second
lane. We assume that car $i_0$ is in mode {\sf Appr};
the modes of other cars will not affect the verification.

Verifying the safety condition in general for such a jump will require
the  afore-mentioned interplay with other components of a global 
safety condition, because $\front{i}$ may not actually be the car in 
front of $i$ if another car cut in in front of $i$ after the last
topology  update. To keep the presentation simple, we instead assume 
for this example that the lane change follows directly on an update, 
so that the sensors show correct information (i.e.~the state of 
$\environment$ is an initial state).
This is a special case of global mode reachability that is much easier to follow by hand than the general case.
 In particular, we use that 
there is no car between $\sidefront{}$ and $\sideback{}$.
Invariance under lane-changing jumps can then be reduced to checking
whether the following set is unsatisfiable:

\smallskip

{\footnotesize 
\begin{align*}
  \Phi_{\sf safe}\\
  \guard{:\ }
	  & k_0\not=\nil \wedge \front{k_0}\not=\nil \wedge \lane{k_0}
          = 1 \wedge \pos{\front{k_0}}-\pos{k_0}\leq D'\\[-1ex]
    & \back{k_0}\not=\nil\to \pos{k_0}-\pos{\back{k_0}}\geq d'\\[-1ex]
    & \sideback{k_0}\not=\nil\to \pos{k_0}-\pos{\sideback{k_0}}\geq d'\\[-1ex]
    & \sidefront{k_0}\not=\nil\to \pos{\sidefront{k_0}}-\pos{k_0}\geq d'\\
  \invariant_{\sf before}{:\ }
    & \forall i ((\lane{i}=1\vee \lane{i}=2) \wedge \\[-1ex]
		& ~~~~ i \neq \nil \wedge \front{i}\not=\nil\to\pos{\front{i}}-\pos{i}\geq d)\\
  \invariant_{\sf after}{:\ }
    & \forall i ((\lane[']{i}=1\vee\lane[']{i}=2) \wedge \\[-1ex]
		& ~~~~ \front{i}\not=\nil\to\pos[']{\front{i}}-\pos[']{i}\geq d)\\
  \jump{:\ }
    & \lane[']{k_0}=2 \wedge \forall i(i\not=k_0 \to \lane[']{i}=\lane{i})\\[-1ex]
    & \forall i(\pos[']{i}=\pos{i})\\
  {\sf Init}_{\environment}{:\ }
	  & \forall i, j (\sideback{i},\sidefront{i},j\not=\nil\wedge\lane{j}=2 \to
		\pos{j}\leq \pos{\sideback{i}} \vee \pos{j}\geq \pos{\sidefront{i}}\\[-1ex]
	  & ~~~~ \sideback{i},j\not=\nil\wedge\sidefront{i}=\nil\wedge\lane{j}=2 \to
		\pos{j}\leq \pos{\sideback{i}}\\[-1ex]
	  & ~~~~ \sidefront{i},j\not=\nil\wedge\sideback{i}=\nil\wedge\lane{j}=2 \to
		\pos{j}\geq \pos{\sidefront{i}})\\
	\neg\Phi_{\sf safe}'{:\ }
	  & i_0\not=\nil \wedge  j_0\not=\nil \wedge i_0\not=j_0
          \wedge \lane[']{i_0}=\lane[']{j_0} \wedge \pos[']{i_0}=\pos[']{j_0}\\[-4ex]
\end{align*}}%

\

\noindent These axioms define a chain of local theory extensions: 
{\small $$ {\mathbb R} \cup {\sf Eq}_{\sf index} \subseteq {\mathbb R} \cup {\sf
  Inv}_{\sf before} \cup {\sf Init}_{\sf top} \subseteq  {\mathbb R} \cup {\sf
  Inv}_{\sf before} \cup {\sf Init}_{\sf top} \cup {\sf jump} \cup {\sf Inv}_{\sf after}$$}%
After instantiation and
purification the problem
is reduced to a satisfiability test in the combination of 
linear arithmetic with pure equality (for the {\sf index} sort). 
Below, we explain intuitively why the set of clauses above is 
unsatisfiable. 

Due to the implication in the jump condition, the verification will be
a case distinction on whether or not $i_0$ or $j_0$ equals
$k_0$. Since the case $k_0\not\in\{i_0,j_0\}$ is trivial, we
concentrate the manual analysis on $k_0=i_0\not=j_0$. 
From the jump condition, we obtain: 
\[ \lane[']{j_0}=\lane{j_0} \quad  \pos[']{k_0} = \pos{k_0} \quad 
\pos[']{i_0} = \pos{i_0} \quad  \pos[']{j_0} = \pos{j_0} \]
From the information from ${\sf Top}$, we obtain: 

\medskip

{\footnotesize \begin{align*}
	  & \sideback{k_0},\sidefront{k_0},j_0\not=\nil\wedge\lane{j_0}=2 \to
		\pos{j_0}\leq \pos{\sideback{k_0}}\vee \pos{j_0}\geq \pos{\sidefront{k_0}}\\
	  & \sideback{k_0},j_0\not=\nil\wedge\sidefront{k_0}=\nil\wedge\lane{j_0}=2 \to
		\pos{j_0}\leq \pos{\sideback{k_0}}\\
	  & \sidefront{k_0},j_0\not=\nil\wedge\sideback{k_0}=\nil\wedge\lane{j_0}=2 \to
		\pos{j_0}\geq \pos{\sidefront{k_0}}
\end{align*}}%

\medskip

We know that $j_0\not=\nil$ and $\lane{j_0}=2$
 (because $\lane{j_0}=\lane[']{j_0}=\lane[']{i_0}=\lane[']{k_0}$). 
If either of $\sideback{k_0}$ or $\sidefront{k_0}$ is defined, 
then the guard condition states that they are at least $d'$ 
away from $k_0$, and the instances that we just derived state 
that then the same must hold for $j_0$. 
In particular, $j_0\not=k_0$ if $d'>0$. 
This means that the derived set of ground instances is unsatisfiable if $d'>0$.

}
\end{ex}

\subsubsection{Topology updates}
\label{sec:updates}
We now analyze the decidability and complexity of checking whether 
$\Phi_{\sf safe}$ is preserved under all GMR topology updates. 
By Theorem~\ref{THM:INV-FORMULAE}, this can be reduced to 
checking whether for all $q = (q_i)_{i \in I} \in Q^I$ the following formula $F_q^{\sf top}$ is 
unsatisfiable: 
\begin{align*}
F_q^{\sf top}:\ 	\Phi_{\sf entry}({\overline
          x}(t_0)) 
& 
{}\wedge \Bigg( \bigg(t_0 < t_1 \wedge \forall i {\sf
          Flow}_{q_i}({\overline x}(i, t_0), {\overline x}(i, t_1)) \bigg)
        \vee t_0 = t_1 \Bigg) \\
&{}\wedge 
\forall i_1, \dots, i_n \phi_{\sf safe}({\overline x}(j_1, t_1), \dots,{\overline x}(i_n, t_1)) \wedge \bigwedge_{p \in P_1} {\sf Update}(p,p') \wedge G, 
\end{align*}
where $G = \neg \phi_{\sf safe}'({\overline x}(c_1), \dots,{\overline
  x}(c_n))$ and $\phi_{\sf safe}'$ is obtained from $\phi_{\sf safe}$ by replacing 
every $p \in P_1$ with $p'$.

\begin{lem}[Topology updates]
\label{lem:decidability of updates}
Under Assumptions~\ref{assumption:linear and decoupled},
~\ref{assumption:entry}(1) and~\ref{assumption:update}
the following hold: 
\begin{enumerate}
\item For every $q = (q_i)_{i \in I} \in Q^I$, 
${\sf F}^{\sf  top}_q$ is unsatisfiable iff ${F^{\sf top}_q}^{[T_G]}$ is
unsatisfiable for a suitable set of ground terms $T_G$. 

\item The size of the set of terms of sort ${\sf index}$ in ${\sf st}(G)$
  and hence also the size of ${F^{\sf top}_q}^{[T_G]}$ is polynomial
  in the number of terms of sort 
${\sf index}$ in $\Phi_{\sf safe}$. Therefore also the size of the set
  $I^G_{\sf top}$ of ground terms of sort ${\sf index}$ in ${F^{\sf
      top}_q}^{[T_G]}$  is polynomial in the number of terms of sort 
${\sf index}$ in $\Phi_{\sf safe}$.

The set of instances 
${F^{\sf    top}_q}^{[T_G]}$ contains only formulae  corresponding
states $q_i$ where $i$ are indices 
corresponding to  terms in $I^G_{\sf top}$. 
\end{enumerate}
\end{lem}

\noindent {\em Proof:} 
The proof is similar to the one of Lemma~\ref{lem:decidability of  
  flows}, using Theorem~\ref{locality-verif}(4) and is only sketched
here. 
Let ${\cal K}_{\sf top} = {\cal K}_1 \cup \bigwedge_{p \in P_1} {\sf
  Update}(p,p'),$ where ${\cal K}_1$ is the following formula: 
\[ {\cal K}_1 =  \Phi_{\sf entry}({\overline
          x}(t_0))  \wedge \Bigg( \bigg(t_0 < t_1 \wedge \forall i {\sf
          Flow}_{q_i}({\overline x}(i, t_0), {\overline x}(i, t_1)) \bigg)
        \vee t_0 = t_1 \Bigg) \wedge \Phi_{\sf safe}({\overline x}(t_1)) \]  
By Theorem~\ref{locality-verif}(4), the extension  of the theory 
${\mathbb R} \cup {\cal K}_1$ with the additional function symbols $\{ p' \mid p \in
  P_1 \}$ axiomatzed by 
 $\bigwedge_{p \in P_1} {\sf Update}(p,p')$ 
is local. Thus, ${\mathbb R} \cup  {\cal K}_1 \cup \bigwedge_{p \in P_1} {\sf
  Update}(p,p') \cup G$ is satisfiable iff ${\mathbb R} \cup  {\cal K}_1 \cup \bigwedge_{p \in P_1} {\sf
  Update}(p,p')[G] \cup G$ is satisfiable. 

\medskip
\noindent We can distinguish two cases: 

\noindent {\bf Case 1: $\displaystyle{\bigwedge_{p \in P_1} {\sf
  Update}(p,p')[G] \cup G}$ is a ground formula $G'$.} Then we can
proceed as in the proof of Lemma~\ref{lem:decidability of flows}, with
the difference that $G$ is replaced by $G'$. ${\mathbb R} \cup  {\cal
  K}_1 \cup G'$ is satisfiable iff ${\mathbb R} \cup  {\cal
  K}_1^{[G']}$ is satisfiable. The set $T_G$ consists of all the
ground terms of sort ${\sf index}$ in ${\sf st}(G')$, and depends not
only of $G$ but also on the form of the update rules. 

\noindent {\bf Case 2: $\displaystyle{\bigwedge_{p \in P_1} {\sf
  Update}(p,p')[G] \cup G}$ contains free variables.} 
Then the proof 
proceeds as the proof of Lemma~\ref{lem:decidability}. 
The conditions in Assumption 2(1) and 3 ensure also in this case that 
after at most two instantiation steps we can reduce the satisfiability
test to testing the satisfiability of ground clauses. 
Under Assumption 3(1), the set $T_G$  contains the ground terms of 
sort ${\sf index}$ in $\displaystyle{\bigwedge_{p \in P_1} {\sf
  Update}(p,p')[G] \cup G}$. 
Under Assumption 3(2) it contains additional Skolem constants 
which need to be introduced because of the existential quantifiers 
in some of the updates.

\noindent $I^G_{\sf top}$ consists of 
the set of all ground terms of sort ${\sf index}$ in ${{\cal K}_{\sf
    top}}^{[T^{\sf top}_G]}$
together with all terms obtained by replacing the variables with Skolem constants 
$c_p$, $p \in P$ which occur from Skolemization in the instances of ${\sf Update}(p,p')$. 

The estimation of the number $n_{\sf update}$ of instances in
${F_q^{\sf top}}^{[T^{\sf top}_G]}$ and on the number of terms $ni_{\sf update}$ in $I^G_{\sf top}$ is
similar to that made in the proofs of Lemma~\ref{lem:decidability}(3), 
Lemma~\ref{lem:decidability of flows}(2) and
Lemma~\ref{lem:decidability of jumps}(2).  
With the notations used in the proofs of these Lemmata we have: 
\begin{itemize}
\item $n_{\sf update} = n_{\sf entry} + n_{\sf safe} + n_{\sf Flow} +
  n_{\sf Update} \leq  {np_G}^{nv_{\sf entry}} + {np_G}^{nv_{\sf
      safe}} +  {np_G}^{nv_{\sf update}} + c \cdot np_G$; 

\item $ni_{\sf update} =  ni_{\sf entry} + ni_{\sf safe}+ ni_{\sf Flow}
  + ni_{\sf update} \leq (np_{\sf entry} + np_{\sf safe} + np_{\sf
    Flow} + 2 np_{\sf update})
  \cdot np_G$, 

where $ni_{\sf update}$ is the number of terms
  of sort ${\sf index}$ occurring in 
$\bigwedge_{p \in P_1} {\sf Update}(p, p')$. \hfill $\Box$
\end{itemize}

\begin{thm}
\label{thm:decidability of updates} 
For every $q \in Q^I$, the satisfiability of the formulae ${F^{\sf
    top}_{q}}$ is decidable (and in NP). 
\end{thm} 

\begin{cor}
\label{cor:fin-st-updates}
Under Assumptions~\ref{assumption:linear and decoupled},
~\ref{assumption:entry}(1) and~\ref{assumption:update}
there exists a finite set $I_{\sf top} \subseteq I$ of indices,
such that the following are equivalent:
\begin{enumerate}
	\item $F^{\sf top}_q$ is satisfiable for some $q  \in  Q^I$
	\item $F^{\sf top}_{q_0}$ is satisfiable for some $q_0 \in Q^{I_{\sf flow}}$.
\end{enumerate}
Therefore checking invariance under all topology updates is decidable
(and in NP). 
\end{cor}

\begin{cor}
\label{cor:small-model-prop-update}
Let $S = ({\sf Top}, \{ S(i) \mid i \in I \})$ be an SFHA. 
Under Assumption~\ref{assumption:linear and decoupled}, ~\ref{assumption:entry}(1)
and~\ref{assumption:update}, 
the following are equivalent: 
\begin{enumerate}
\item 
There exist  indices $c_1, \dots, c_n$ for which the safety condition $\Phi_{\sf
  safe}$ is not preserved under updates reachable from a state in which $\Phi_{\sf entry}$ holds.  
\item There exists a finite set $I_{\sf update} \subseteq I$ of
  indices, of size polynomial in the size of $n$ (assuming that the
  lengths of the formulae describing the SFHA $S$ are considered
  constants) describing a suitable neighborhood of $c_1, \dots,
  c_n$  which can effectively be described (they correspond to the terms in
$I^G_{\sf update}$ in Theorem~\ref{thm:decidability}) 
such that  already 
in the systems $S_{\sf update} = ({\sf Top}_{|I_{\sf update}}, \{ S(i)
\mid i \in I_{\sf update} \})$ the safety condition $\Phi_{\sf
  safe}$ is not preserved under updates in states reachable from a state in which $\Phi_{\sf entry}$ holds.  
\end{enumerate}
\end{cor}

\noindent The proofs are in all cases analogous to the proofs for the case of
flows and jumps (Corollaries~\ref{cor:small-model-prop-jumps} and~\ref{cor:small-model-prop-jumps}). 




\

\noindent {\bf Parametric Verification.} 
Also in this case, if we impose constraints on these parameters (in the form of constraints between
real numbers) then the results in 
Lemma~\ref{lem:decidability of updates} and Corollary~\ref{cor:fin-st-updates} can be used to 
prove that the verification problems remain decidable. The complexity
of the problems is similar to that for jumps. 
We can also use hierarchical reasoning
combined with quantifier elimination for the theory of real numbers 
for generating constraints on the parameters which guarantee
$\Phi_{\sf safe}$ is preserved under GMR updates, as 
in \cite{sofronie-cade24} (the complexity is exponential).

\begin{ex}
\label{ex:top-update}
{\em 
Consider the topology updates in Example~\ref{ex-top-updates}. 
Invariance of $\Phi^g_{\sf safe}$ under these updates can be proved
(cf. Section~\ref{experiments}). 
$\Phi^l_{\textsf{safe}}$ is not invariant. We now consider a variant
${\overline \Phi^l}_{\textsf{safe}}$ of
$\Phi^l_{\textsf{safe}}$ where: 
\[ \overline{\Phi}^{\sf front}_{\textsf{safe}}{:\ } \forall i\big(i\not=\nil\wedge  
\front{i}\not=\nil \wedge {\sf lane}(i) = {\sf lane}({\sf front}(i))
\to \pos{\front{i}}-\pos{i} > d_s\big)\] 

\noindent In order to prove that  ${\overline \Phi}^{\sf front}_{\textsf{safe}}$ is
 preserved by topology updates,  
we prove that the formula 
\[ {\overline \Phi}^{\sf front}_{\sf safe} \wedge 
{\sf Update}({\sf  front},{\sf front}') \wedge G\]  
is unsatisfiable, where 
$G = \neg {\overline \Phi}^{\sf front'}_{\sf safe}$ is the ground clause 
\[ i_0 {\neq} \nil \wedge {\sf front}'(i_0) {\neq} \nil \wedge {\sf
  lane}(i_0) {=} {\sf lane}({\sf front}'(i_0)) \wedge {\sf
  pos}({\sf front}'(i_0)) - {\sf pos}(i_0) {\leq} d_s.\]  
The extension:
$\mathbb{R}\cup {\overline \Phi}^{\sf front}_{\sf safe}
	\ \subseteq\ \mathbb{R}\cup {\overline \Phi}^{\sf front}_{\sf safe} \cup {\textsf{Update}(\front{},\front[']{})}$
 is local. We 
determine the conjuncts of ${\textsf{Update}(\front{},\front[']{})}[G]$, 
where ${\sf st}(K,G)=\{\front[']{i_0}\}$. 
After instantiation and purification (replacing ${\sf front}'(i_0)$
with  $f'$) we obtain: 

{\small \[ 
\begin{array}{l}
   i_0{\not=}\nil\wedge \neg \exists j ({\sf ASL}(j,i_0)) \to f' {=}
   \nil\\
  i_0{\not=}\nil\wedge \phantom{\neg}\exists j ({\sf ASL}(j,i_0)) \to {\sf Closest}_f(f', i_0)
\end{array}
\]}%
with the notations in Example~\ref{ex-top-updates}. 
Transforming these formulae into prenex form and skolemizing the
existential quantifier, we obtain (with Skolem constant
$c_{0}$): 
\begin{align*}
 C_1: ~& i_0{\not=}\nil\wedge \neg {\sf ASL}(c_{0},i_0) \to f' {=} \nil\\
 C_2: ~& i_0{\not=}\nil\wedge {\sf ASL}(j,i_0) \to
 {\sf Closest}(f', i_0). 
\end{align*}
The formula $C_1$ is ground.
To check the satisfiability of $\Phi_{\sf safe} \cup C_2 \cup G_1$
where 
$G_1 = C_1 \wedge G_0$ 
(where $G_0$ is $i_0 {\neq} \nil \wedge f' {\neq} \nil \wedge {\sf
  lane}(i_0) {=} {\sf lane}(f') \wedge {\sf 
  pos}(f') {-} {\sf pos}(i_0) {\leq} d_s$), it is sufficient to check the satisfiability
of  $\Phi_{\sf safe}[G_1] \cup C_2[G_1] \cup G_1$.

} 
\end{ex}

\subsection{Checking exhaustive entry conditions}

In Theorem~\ref{thm:check-eec} we showed that for decoupled 
SFLHA  $S$ we can reduce checking  conditions (i) and (ii) in
Definition~\ref{eec} (exhaustive entry conditions) to checking the
satisfiability of the following formulae: 

\begin{itemize}
\item[(i)] 
$\Phi_{\sf entry}({\overline x}) \wedge \big( \neg (\bigvee_{q \in Q} {\sf 
    Init}_{q}({\overline x}(i_0))) \vee \neg {\sf Init}_{\sf 
    top}({\overline x}) \big)$ is unsatisfiable. 
\item[(ii)] for all $(q_i)_{i \in I} \in Q^I$: 
\begin{itemize}
\item[(a)] {\em Topology updates:}
\[(\forall i \, {\sf Inv}_{q_i}({\overline x}_i))  \wedge {\sf Update}(p, p')
  \wedge \neg \Phi'_{\sf entry}({\overline x})\text{  is unsatisfiable},\]
where $\Phi'_{\sf entry}$ arises from $\Phi_{\sf entry}$ by replacing
$p$ with $p'$, and 

\item[(b)] {\em Jumps:} For all  $e \in E, i_0 \in I$:
\[\begin{array}{rl}
  (\forall i \, {\sf Inv}_{q_i}({\overline x}_i)) \wedge
    {\sf guard}_e({\overline x}_{i_0}) \wedge
	  {\sf jump}_e({\overline x}_{i_0}, {\overline x}'_{i_0}) \wedge {} \\
	  \forall j (j \neq i_0 \rightarrow {\overline x}'(j) = {\overline x}(j)) \wedge \neg \Phi_{\sf entry}({\overline x}')
  &\text{  is unsatisfiable}.
\end{array}\]
\end{itemize}
\end{itemize}
We now identify conditions under which these tasks are decidable and
analyze their complexity. 
\begin{thm}
Under Assumption~\ref{assumption:linear and decoupled}, and if both $\Phi_{\sf
  entry}$ and ${\sf Init}_{\sf top}$ satisfy the conditions on
$\Phi_{\sf entry}$ in Assumption~\ref{assumption:entry}(1), then the following hold: 
\begin{itemize}
\item[(i)] The following are equivalent: 
\begin{itemize}
\item[(1)] $\Phi_{\sf entry}({\overline x}) \wedge \big( \neg (\bigvee_{q \in Q} {\sf 
    Init}_{q}({\overline x}(i_0))) \vee \neg {\sf Init}_{\sf 
    top}({\overline x}) \big)$ is unsatisfiable. 
\item[(2)] $\Phi_{\sf entry}({\overline x}) \wedge G_1$ is
  unsatisfiable, where $G_1 = \bigwedge_{q \in Q}
  \neg  {\sf     Init}_{q}({\overline x}(i_0))$ and 

 $\Phi_{\sf entry}({\overline x}) \wedge G_2$ is unsatisfiable, where
 $G_2 = \neg {\sf Init}_{\sf  top}({\overline x})$. 

\item[(3)] $\Phi_{\sf entry}({\overline x})^{[G_1]} \wedge G_1$ is
  unsatisfiable, where $G_1 = \bigwedge_{q \in Q}
  \neg  {\sf     Init}_{q}({\overline x}(i_0))$ and 

 $\Phi_{\sf entry}({\overline x})^{[G_2]} \wedge G_2$ is unsatisfiable, where
 $G_2 = \neg {\sf Init}_{\sf  top}({\overline x})$. 
\end{itemize}

The size of the set of terms of sort ${\sf index}$ in ${\sf st}(G_1), {\sf st}(G_2)$
  and hence also the size of the sets of instances in (3) is
  polynomial in the number 
of terms of sort ${\sf index}$ in $G_1, G_2$. 

\item[(ii)] (a) For every $q = (q_i)_{i \in I} \in Q^I$ the following
  are equivalent: 
\begin{itemize}
\item[(a1)] $(\forall i \, {\sf Inv}_{q_i}({\overline x}(i)))  \wedge {\sf Update}(p, p')
  \wedge G_3$   is unsatisfiable, where $G_3 = \neg \Phi'_{\sf
    entry}({\overline x})$. 
\item[(a2)] $(\forall i \, {\sf Inv}_{q_i}({\overline x}(i)))  \wedge {\sf Update}(p, p')[G_3]
  \wedge G_3$ is unsatisfiable.
\item[(a3)] $[(\forall i \, {\sf Inv}_{q_i}({\overline x}(i)))  \wedge
  {\sf Update}(p, p')[G_3]]^{[T_{G_3}]}   \wedge G_3$ is
  unsatisfiable, where $T_{G_3}$ is the set of all  ground terms of sort
  ${\sf index}$ in the formula in (2). 
\end{itemize}

(b) For every $q = (q_i)_{i \in I} \in Q^I$ the following
  are equivalent: 

\begin{itemize}
\item[(b1)] $
  (\forall i \, {\sf Inv}_{q_i}({\overline x}_i)) \wedge
    {\sf guard}_e({\overline x}_{i_0}) \wedge
	  {\sf jump}_e({\overline x}_{i_0}, {\overline x}'_{i_0}) \wedge 
	  \forall j (j \neq i_0 \rightarrow {\overline x}'(j) =
          {\overline x}(j)) \wedge G_4,$ is unsatisfiable, where $G_4 = \neg \Phi_{\sf
    entry}({\overline x}')$. 
\item[(b2)] $((\forall i \, {\sf Inv}_{q_i}({\overline x}_i)) \wedge
    {\sf guard}_e({\overline x}_{i_0}) \wedge
	  {\sf jump}_e({\overline x}_{i_0}, {\overline x}'_{i_0}) \wedge {} 
	  \forall j (j \neq i_0 \rightarrow {\overline x}'(j) =
          {\overline x}(j)))^{[G_4]} \wedge G_4$ is unsatisfiable.
\end{itemize}
\end{itemize}
\end{thm}

 \begin{thm}[Decidability and complexity]
The problem of checking the satisfiability of the formula in (i)(3) is
decidable (and in NP). For every $q = (q_i)_{i \in I} \in Q^I$, the
problem of checking the satisfiability of the formulae in (ii)(a3) and
(ii)(b2) is decidable (and in NP). 
\label{thm:compl-eec}
\end{thm}

\begin{cor}
Under Assumption~\ref{assumption:linear and decoupled}, and if $\Phi_{\sf
  entry}$ and ${\sf Init}_{\sf top}$ satisfy the conditions in 
Assumption~\ref{assumption:entry}(1), there exists a finite set $I_0 \subseteq I$ of indices,
such that the following are equivalent:
\begin{enumerate}
	\item The formula in (i)(a) is satisfiable for some $q \in Q^I$
	\item The formula in (i)(a) is satisfiable for some $q \in Q^{I_0}$.
\end{enumerate}
Therefore checking invariance under all GMR jumps is decidable (and in
NP). 
\end{cor} 

\

\noindent {\bf Parametric Verification.} These results can be used
also for parametric systems, either for {\em checking} whether a
safety property has exhaustive entry conditions (assuming that 
certain constraints on the parameters are known) or for generating 
constraints on parameters used in the specification of the system, 
and of $\Phi_{\sf entry}$ under which Definition~\ref{eec}  holds. 

\

\begin{ex}
{\em 
\label{ex:checking entry conditions}
Consider the running example. Assume that the initial conditions for
the topology automaton are expressed by the formulae ${\sf Init}_{\sf
  Top}$, stating that all sensor pointers have the correct value, as if they had 
just been updated. For ${\sf front}$ this can be expressed by
the following set of formulae:

\medskip

\noindent $\begin{array}{@{}rl} 
\forall i (i \neq \nil \wedge {\sf front}(i) = \nil & \to \forall k (k \neq \nil \wedge k \neq i \wedge  {\sf pos}(k) \geq {\sf
  pos}(i) \rightarrow {\sf lane}(k) \neq {\sf lane}(i))) \\
\forall i (i \neq \nil \wedge {\sf front}(i) \neq \nil & \to {\sf
  pos}_{\sf front}(i) > {\sf pos}(i)  \wedge {\sf lane}_{\sf
  front}(i) = {\sf lane}(i) \wedge \\
& ~~~\forall k (k \neq \nil \wedge k \neq i \wedge  {\sf pos}(k) \geq {\sf
  pos}(i) \wedge {\sf lane}(k) = {\sf lane}(i) \\
& ~~~~~~~\to {\sf pos}(k) \geq {\sf
  pos}_{\sf front}(i)) \wedge \\
& ~~~{\sf pos}({\sf front}(i)) = {\sf pos}_{\sf front}(i) \wedge {\sf
  lane}({\sf front}(i)) = {\sf lane}_{\sf front}(i)). 
\end{array}$ 

\medskip
\noindent In Example~\ref{run-ex}, the initial conditions of the two modes {\sf
  Appr} and {\sf Rec} are: 
\[ {\sf Init}_{\sf Appr} = {\sf Init}_{\sf Rec} = \forall i ( i \not= \nil \wedge
\front{i}\not=\nil \rightarrow {\sf pos}_{\sf front}(i)-\pos{i}\geq
d'). \] 

\medskip
\noindent Consider a safety property $\Phi_{\sf entry} \rightarrow \Box
\Phi_{\sf safe}$, with entry states being states in
  which the information provided by the sensors is correct and every 
car is  sufficiently far away from the following car on the same
lane, described by the following formula $\Phi_{\sf entry}$ (again
stated only for ${\sf front}$):

\medskip
\noindent $\begin{array}{@{}rl} 
\forall i (i \neq \nil \wedge {\sf front}(i) = \nil & \to \forall k (k \neq \nil \wedge k \neq i \wedge  {\sf pos}(k) \geq {\sf
  pos}(i) \rightarrow {\sf lane}(k) \neq {\sf lane}(i))) \\
\forall i (i \neq \nil \wedge {\sf front}(i) \neq \nil & \to {\sf
  pos}_{\sf front}(i) > {\sf pos}(i) + d' \wedge {\sf lane}_{\sf
  front}(i) = {\sf lane}(i) \wedge \\
& ~~~\forall k (k \neq \nil \wedge k \neq i \wedge  {\sf pos}(k) \geq {\sf
  pos}(i) \wedge {\sf lane}(k) = {\sf lane}(i) \\
& ~~~~~~~\to {\sf pos}(k) \geq {\sf
  pos}_{\sf front}(i)) \wedge \\
& ~~~{\sf pos}({\sf front}(i)) = {\sf pos}_{\sf front}(i) \wedge {\sf
  lane}({\sf front}(i)) = {\sf lane}_{\sf front}(i)). 
\end{array}$ 

\medskip
\noindent 
It can be easily checked that $\Phi_{\sf entry} \wedge \neg {\sf
  Init}_{\sf top}$ is unsatisfiable and that $\Phi_{\sf entry} \wedge G_1$, where 
\[ G_1 = \neg {\sf Init}_a \wedge \neg {\sf Init}_r = (c \neq \nil
\wedge {\sf front}(c) \neq \nil \wedge  {\sf pos}_{\sf
  front}(c)-\pos{c} < d' )\]
is unsatisfiable. 

\medskip
\noindent In general, we can only guarantee that $\forall i {\sf
  Inv}_{q_i}({\overline x}(i)) \wedge {\sf Update}(p, p') \wedge \neg
\Phi'_{\sf entry}$ is unsatisfiable if the invariants and the update rules are designed
such that after an update each car is sufficiently far away from the
following car on the same lane. 

Similarly, we can only guarantee that $\forall i {\sf
  Inv}_{q_i}({\overline x}(i)) \wedge {\sf guard}_e({\overline x}) \wedge
{\sf jump}_e({\overline x}, {\overline x}') \wedge \neg
\Phi_{\sf entry}({\overline x}')$ is unsatisfiable if the jump rules are designed
such that after a jump that resets some of the variables (e.g.\ after a lane change) 
each car is sufficiently far away from the
following car on the same lane. 
}
\end{ex}

\section{Consequences of Locality}
\label{sec:consequences of locality}
In what follows we present two applications of the previous results: a
small model property and a complexity result which refines the
NP-complexity results established in Section~\ref{inv-bmc}. 

\subsection{A small model property}

From Corollaries~\ref{cor:small-model-prop-entry},~\ref{cor:small-model-prop-flow},~\ref{cor:small-model-prop-jumps}
and~\ref{cor:small-model-prop-update}
we obtain the following  small model property for the verification of
safety properties with exhaustive entry conditions. 

\begin{thm}[Small model property]
\label{thm: small model property}
Under Assumptions~\ref{assumption:linear and
  decoupled},~\ref{assumption:entry}(1) and~\ref{assumption:update}, 
a decoupled SFLHA $S$ satisfies a safety property with 
exhaustive entry conditions iff 
 the property holds in all systems of the form 
$S_0 = (\environment, \{ S(i) \mid i \in I_0 \} )$, where 
$I_0$  is a set of indices corresponding to ground terms in 
$G = \neg \Phi_{\sf safe}$ occurring in the instances of the 
formulae ${F^{\sf entry}}^{[G]}, {F^{\sf flow}_q}^{[G]}, {F^{\sf
    jump}_q}^{[G]}$, or ${F^{\sf  top}_q}^{[G]}$.

The size $|I_0|$ of $I_0$ is polynomial in the number of terms of 
sort ${\sf index}$ occurring in $\Phi_{\sf safe}$, and can be precisely 
determined from the form of the formulae $\Phi_{\sf safe}, 
{F^{\sf entry}}, {F^{\sf flow}_q},  {F^{\sf 
    jump}_q}$, or ${F^{\sf  top}_q}$. 

\end{thm}
\noindent {\em Proof:} Direct consequence of  
Corollaries~\ref{cor:small-model-prop-entry},~\ref{cor:small-model-prop-flow},~\ref{cor:small-model-prop-jumps}
and~\ref{cor:small-model-prop-update}. 
From the proofs of Lemma~\ref{lem:decidability},
~\ref{lem:decidability of flows},
~\ref{lem:decidability of jumps} 
and~\ref{lem:decidability of updates}, we know that for checking the
safety of entry conditions and invariance under flows and GMR jumps
and topology update we only need to analyze systems with  set of indices 
of cardinality at most 
 $(np_{\sf entry} + np_{\sf safe} + np_{\sf Flow} + 2 np_{\sf
   Update}) \cdot np_G + np_{\sf jump}$, 
where $np_{\sf entry}, np_{sf safe}, np_{\sf Flow}, 
np_{\sf Update}$ is the number of all terms of sort 
${\sf index}$ occurring in the corresponding formulae 
($\Phi_{\sf entry}, \Phi_{\sf safe}, {\sf Flow}, {\sf Update}(p, p')$)
and $np_G$ is the set of ground terms of sort ${\sf index}$ occurring in 
$G$. \hfill $\Box$


\subsection{Decidability, Complexity} 

From Theorems~\ref{thm:decidability},
~\ref{thm:decidability of flows},
~\ref{thm:decidability of jumps} 
and~\ref{thm:decidability of updates} 
and from Theorem~\ref{thm:compl-eec} and  
Corollaries~\ref{cor:small-model-prop-entry},~\ref{cor:small-model-prop-flow},~\ref{cor:small-model-prop-jumps}
and~\ref{cor:small-model-prop-update}
we obtain the following decidability and complexity results: 

\begin{thm}
\label{thm:complexity-all1} 
Under Assumptions 1, 2(1) and 3, 
the problem of checking invariance of a safety condition in an SFLHA
$S$ is decidable (and in NP). 
\end{thm}

\noindent {\em Proof:} Direct consequence of Theorems~\ref{thm:decidability},
~\ref{thm:decidability of flows},
~\ref{thm:decidability of jumps} 
and~\ref{thm:decidability of updates} . \hfill $\Box$. 

\begin{thm}
\label{thm:complexity-all2} 
Under Assumptions 1, 2(1), and 3, and if ${\sf Init}_{\sf top}$
consists of guarded {\sf index}-positive extended clauses where the scalar
constraint is a conjunction of linear inequalities, 
the problem of checking whether a safety property $\Phi_{\sf entry}
\rightarrow \Box \Phi_{\sf safe}$ has extended entry condition in an SFLHA
$S$ is decidable (and in NP). 
\end{thm}

\noindent {\em Proof:} 
Direct consequence of Theorem~\ref{thm:compl-eec}. \hfill $\Box$. 

\

\noindent Under Assumption 4, some of the verification problems can be solved in
PTIME: 

\begin{thm}
\label{thm-tractability} 
With the notation introduced in Theorem~\ref{THM:INV-FORMULAE} and
used in Sections~\ref{sec:entry}--\ref{sec:updates}, and
under Assumptions~\ref{assumption:linear and
  decoupled}, 2(1), 3 and \ref{assumption:HDL}, 
the following hold for every conjunction 
${\sf Def}: \bigwedge_{p(t) \in T_1} p(t) 
{=} \nil \wedge \bigwedge_{p(t) \in T_2} p(t) \neq \nil$, where $T_1 \cup  
T_2 =  \{ p(t) \mid t \text{ subterm  of sort {\sf index} of } G, p \in P, 
p(t) \text{ not in } G \}$  
and every $q \in  Q^{I_{\sf entry}}$ (resp. $Q^{I_{\sf flow}}$ or $Q^{I_{\sf update}}$): 
\begin{itemize}
\item[(1)] 
The satisfiability of ${\sf F}^{\sf  
  entry}_q \wedge {\sf Def}$ can be checked  in PTIME. 
\item[(2)] 
The satisfiability of ${\sf F}^{\sf flow}_q \wedge {\sf  
    Def}$ can be checked  in PTIME. 
\item[(2)] 
The satisfiability of ${{\sf F}_q^{\sf jump}} \wedge {\sf  
    Def}$ can be checked  in PTIME. 
\item[(4)] 
Assuming that  
either (a) $P_S$ is empty, or else (b)  ${\sf Update}(p,p')$ has the  
form in Theorem~\ref{thm:updates},
 the satisfiability of ${\sf F}_q^{\sf  update} \wedge {\sf Def}$ can be checked  in PTIME. 
\end{itemize}
If we consider $|Q|, |E|$ and $|P|$ to be constant and the number 
of terms of sort ${\sf index}$ 
in $\Phi_{\sf safe}$, and the maximal number of variables in the update 
axioms as a parameter, these problems can be considered to be 
fixed parameter tractable. 
\end{thm}
 
\noindent {\em Proof:} 
All transformations in the hierarchical reduction 
increase the size of the ground formulae to be 
checked polynomially. 
If the constraints over ${\mathbb R}$ we 
obtain after this reduction lie in a tractable fragment of
linear arithmetic,  and if the ground 
constraints involving terms of sort ${\sf index}$ are unit and contain definedness  or undefinedness 
conditions\footnote{A definedness condition for a term $t$ of sort
  ${\sf index}$ is a
  literal $t \neq \nil$; an undefinedness condition for $t$ is a
  literal of the form $t = \nil$.}  for all ground terms of sort 
${\sf index}$, 
then checking satisfiability can be done in PTIME. 
The number of possible choices for ${\sf Def}$ is $2^{(T_1 \cup T_2)
  \backslash {\sf st}(G)}$.  
Since each of the verification tasks for a fixed ${\sf Def}$ can be
solved in PTIME, this yields the fixed parameter tractability
result. \hfill $\Box$

\begin{thm}[Parametric systems]
\label{thm:parametric systems}
The complexity results in
Theorems~\ref{thm:decidability}--\ref{thm:decidability of updates} and
\ref{thm-tractability}, as well as the small model property,  
also hold for parametric SFLHA in which only 
the bounds in $\Phi_{\sf entry}, \Phi_{\sf safe}$, ${\sf Inv}_q,$
${\sf Init}_q,$, 
${\sf  guard}_e, {\sf jump}_e,$ and ${\sf Update}$ are parameters. 
For systems in which parameters are allowed as coefficients 
or appear in the flow conditions, the 
complexity is exponential. 
\end{thm} 

\noindent {\em Proof:} This follows from the fact that all 
verification problems can be reduced to checking satisfiability for 
quantifier-free formulae (i.e.\ validity of existentially quantified
formulae). If the parameters occur only in the bounds in 
$\Phi_{\sf entry}, \Phi_{\sf safe}$, ${\sf Inv}_q,$ ${\sf Init}_q,$, 
${\sf  guard}_e, {\sf jump}_e,$ and ${\sf Update}$ then the numerical
constraints are still linear hence the complexity
is as in the non-parametric case, and the satisfiability of 
quantifier-free formulae over the theory of real-closed fields
(${\mathbb R}$) can be checked in EXPTIME \cite{Ben-OrKR:86}. \hfill $\Box$.

\begin{thm}[Parametric synthesis]
\label{thm: synthesis}
Under Assumptions 1, 2(1) and 3, the complexity of synthesizing 
constraints on parameters which guarantee that a parametric SFLHA satisfies 
a safety condition with exhaustive entries (using quantifier
elimination) is exponential. 
\end{thm}

\noindent {\em Proof:} The proof is similar to the proof of
Theorem~\ref{thm:parametric systems}, taking into account that 
the complexity of quantifier  elimination for formulae 
without alternation quantifiers (hence also for existential 
formulae) is EXPTIME \cite{Collins:75,Ben-OrKR:86}. \hfill $\Box$.

\

\noindent 
Similar methods can be used for showing that under
Assumptions~\ref{assumption:linear and decoupled},~\ref{assumption:entry}(1) and \ref{assumption:update} the problem
of checking conditions (i) and (ii) in the definition of exhaustive entry conditions 
is in NP.  
We can also express $\Phi_{\sf entry}$ and $S$ 
parametrically and infer 
constraints on parameters under which conditions (i) and (ii) hold.

\begin{rem}
Similar results can also be obtained under Assumption 2(2) or 2(3), but 
because in those cases we need to instantiate in two steps the 
description of the instances needed is a bit more complicated (the 
number of instances and the size of $I_0$ is still polynomial 
in these situations. 

In fact, all decidability results directly translate to
situations where the involved formulas do not satisfy Assumptions 2 or 3
but belong to other fragments for which the theory extensions in
Theorem~\ref{locality-verif} are local or stably local; the complexity
depends on the complexity of checking satisfiability for formulae
obtained after instantiation. 
\end{rem} 

\section{Tool Support}
\label{experiments}
In order to perform the verification tasks automatically, we 
implemented our approach in the tool HAHA (Hierarchic Analysis of Hybrid Automata)\footnote{\url{http://userp.uni-koblenz.de/~sofronie/horbach/haha.html}}.
HAHA employs
H-PILoT\footnote{\url{http://userp.uni-koblenz.de/~sofronie/hpilot/}}, 
a program for hierarchical reasoning in extensions of logical theories
\cite{Sofronie-Ihlemann-hpilot08}, to perform
reductions of the verification proof tasks to satisfiability problems in a combination of 
linear arithmetic over $\mathbb{R}$ and pure equality. These 
are then solved using the theorem prover Z3~\cite{MouraB08}.

\subsection{Input syntax}

\begin{figure}[t]%
\small
\begin{lstlisting}
<mode id="approach">
    <invariant>OR(lane(i) = 1,lane(i) = 2)</invariant>
    <invariant>pos(front(i))-pos(i) >= mindistance</invariant>
    <flow derivatives=".lane(i)" value="0"/>
    <flow derivatives=".pos(i)" upperbound="100"/>
    <flow derivatives=".pos(i)" lowerbound="0"/>
    <flow derivatives=".pos(i)-.pos(front(i))" lowerbound="0"/>
</mode>
<jump source="__any__" target="approach">
    <guard>mindistance >= pos(front(i))-pos(i)</guard>
    <guard>pos(sidefront(i))-pos(i) >= mindistance</guard>
    <guard>pos(i)-pos(sideback(i)) >= mindistance</guard>
    <guard>pos(i)-pos(back(i)) >= mindistance</guard>
    <reset variable="lane" value="3-lane(i)"/>
</jump>
\end{lstlisting}%
\caption{XML specification of the \textsf{Appr} mode and of the lane change jump}%
\label{fig:xml example}%
\end{figure}

We specify spatial families of linear hybrid automata in XML files, whose structure directly mirrors the constituent structure of such a family. For example, the specifications of the approach mode and the lane-changing jump for our running example are presented in Figure~\ref{fig:xml example}. Note that we do not explicitly specify the definedness guards $\mathcal E$. Instead, they are added automatically by \hpilot.

\subsection{System architecture}
\begin{figure}[t]%
\centering\includegraphics{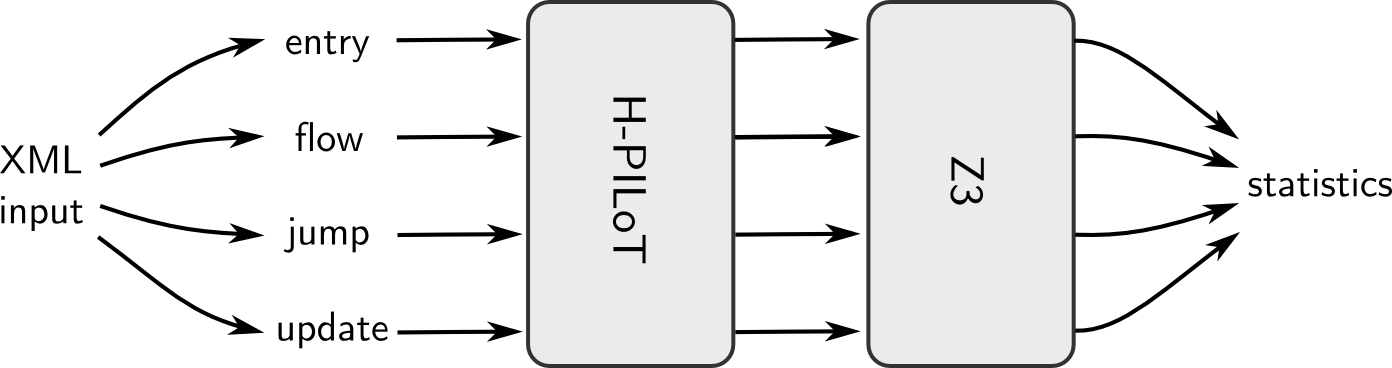}%
\caption{Implementation Data Flow Overview. }%
\label{fig:haha}%
\end{figure}
An overview of our implementation is depicted in Figure~\ref{fig:haha}.
In a first step, HAHA parses the problem from the XML specification and creates internal representations of the four verification tasks explained in Theorem~\ref{THM:INV-FORMULAE}.

Each of them is then translated into H-PILoT syntax, and
H-PILoT performs the reduction to quantifier-free problems as in the proofs of Theorems~\ref{thm:decidability}--\ref{thm:decidability of updates}. 
\mbox{H-PILoT}'s output consists of problems in linear real and integer arithmetic, whose satisfiability is checked by Z3.

If Z3 detects unsatisfiability, the proof task was successful.
For satisfiable formulae, H-PILoT returns a model which can be
used to visualize the counterexample to the invariance properties \cite{Krawez2012}.
Finally, HAHA collects statistics on run times, satisfiability, and model sizes for the individual verification problems.

The check whether a given entry condition satisfies the properties in Definition~\ref{eec} or~\ref{eec-gmr} works similarly.

The use of GMR constraints is not always necessary to prove safety, because some safety properties are maintained by \emph{all} jumps and updates, not just by globally mode reachable ones. Because the inclusion of GMR constraints affects the performance of the approach, HAHA can also run in a mode that does not create them (c.f.\ our experimental results below).


\subsection{Experiments}
\label{sec:experiments}

We evaluated HAHA on variations of  our running problem and on examples from the Passel benchmark suite \cite{JohnsonMitra12}. In the following sections, we describe the results of the verification of some of the safety conditions presented throughout the paper. The list is not exhaustive, but includes safety properties that demonstrates a variety of features of our approach.
On the HAHA homepage, we provide all source data for these examples, including an xml description of the automaton, the verification problems that are handed over to H-PILoT, and finally the SMT problems handled by Z3.
We also provide formalizations of several of the examples from the Passel benchmark suite.

\subsubsection{Decision Problems}
We considered our running example with the entry condition $\Phi_{\sf entry}$ from Example~\ref{ex:checking entry conditions}:
\[\begin{array}{@{}r@{\ }l} 
\forall i (i \neq \nil \wedge {\sf front}(i) = \nil & \to \forall k (k \neq \nil \wedge k \neq i \wedge  {\sf pos}(k) \geq {\sf
	pos}(i) \rightarrow {\sf lane}(k) \neq {\sf lane}(i))) \\
\forall i (i \neq \nil \wedge {\sf front}(i) \neq \nil & \to {\sf
	pos}_{\sf front}(i) > {\sf pos}(i) + d' \wedge {\sf lane}_{\sf
	front}(i) = {\sf lane}(i) \wedge \\
& ~~~\forall k (k \neq \nil \wedge k \neq i \wedge  {\sf pos}(k) \geq {\sf
	pos}(i) \wedge {\sf lane}(k) = {\sf lane}(i) \\
& ~~~~~~~\to {\sf pos}(k) \geq {\sf
	pos}_{\sf front}(i)) \wedge \\
& ~~~{\sf pos}({\sf front}(i)) = {\sf pos}_{\sf front}(i) \wedge {\sf
	lane}({\sf front}(i)) = {\sf lane}_{\sf front}(i))
\end{array} \]
As safety conditions, we chose the following:
\begin{align*}
\Phi^{\sf top}_{\sf safe}:\ 
 & \forall i(i\not=\nil \to \front{i}\not=i)\\
\Phi^g_{\sf safe}:\ 
 & \forall i, j (i {\neq} \nil \wedge j {\neq} \nil \wedge {\sf lane}(i){=}
{\sf lane}(j) \wedge {\sf pos}(i) {>} {\sf pos}(j) \rightarrow {\sf	pos}(i) - {\sf pos}(j) {\geq} d_s)\\
\Phi^{\sf front}_{\sf safe}:\ 
 & \forall i  (i \neq \nil \wedge {\sf front}(i) \neq \nil  \rightarrow \pos{{\sf front}(i)} - \pos{i} \geq d_s)
\end{align*}
The first condition states a basic consistency property of the sensor information; the next two are the ones first introduced in Example~\ref{ex-properties-global-local}.
We provided constraints for all parameters, stating e.g.~that the minimal distance between cars in mode \textsf{Appr} does not exceed the maximal distance between cars in mode \textsf{Rec} ($d\leq D$), and both are nonnegative ($d\geq 0$, $D\geq 0$).

\begin{figure}[t]%
	\centering
	\includegraphics[width=.6\columnwidth]{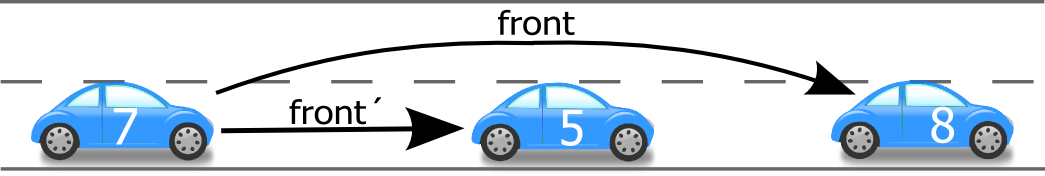}%
	\caption{The property $\Phi^{\sf front}_{\sf safe}$ is violated by the depicted update if the distance between cars 7 and 5 is below the minimal safe distance $d_s$. Restriction to globally mode reachable jumps avoids this situation.}%
	\label{fig:local safety violation}%
\end{figure}

\medskip
Results of experiments with our running example are summarized in
Figure~\ref{fig:times}.
The left half of the diagram shows the results and run times as well as the maximal model sizes (cf.\ Theorem~\ref{thm: small model property}) of verification attempts that ignore the entry condition and global mode reachability.
A result of \textsf{unsat} means that HAHA could prove the respective verification task, \textsf{sat} means that it found a counter example.
As can be seen, the analysis without regard to global mode reachability is faster but not always powerful enough.
For example, $\Phi^{\sf front}_{\sf safe}$ is not invariant under all updates;
Figure~\ref{fig:local safety violation} shows an example of such an update that violates $\Phi^{\sf front}_{\sf safe}$.

\begin{figure}[t]%
	\centering 
	\includegraphics[width=.7\columnwidth]{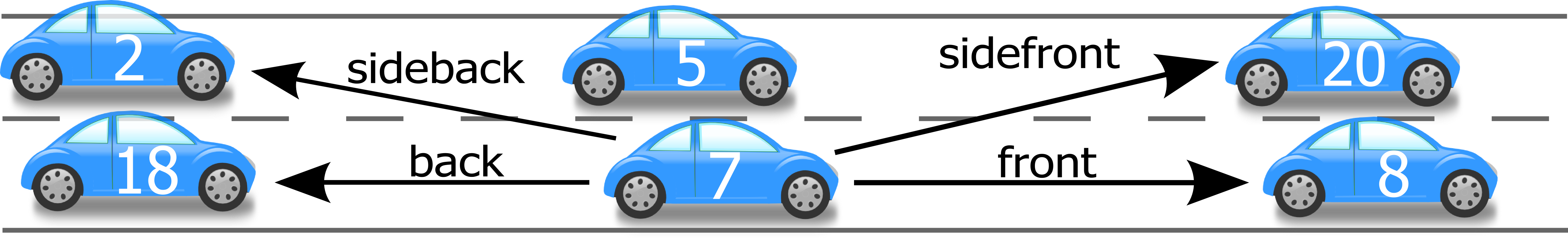}%
	\caption{The property $\Phi_{\sf safe}^g$ is violated by by a lane change if there is another car between \textsf{sidefront} and \textsf{sideback}. This can happen even for globally mode reachable jumps.}%
	\label{fig:global safety violation}%
\end{figure}

The right half of the diagram shows the results of verification
including global mode reachability.
In this mode, we could prove that $\Phi^{\sf front}_{\sf safe}$ holds in all runs.

\begin{figure}[t]
\centering\small
\begin{tabular}{l@{\quad}cccc@{\qquad}ccccc}
	\toprule
	& \multicolumn{4}{c@{\qquad}}{without mode reachability}
	& \multicolumn{4}{c}{with mode reachability}\\
	& init & flow  & jump & update & entry & flow  & jump & update\\ \midrule
	$\Phi^{\sf top}_{\sf safe}$ & \text{unsat} & \text{unsat} & \text{unsat} & \text{unsat} & \text{unsat} & \text{unsat} & \text{unsat} & \text{unsat} & \text{verified} \\
	constants       & 11& 43& 141& 19& 11& 43& 206& 51 \\
	reduction       & 0.028& 0.072& 0.460& 0.024& 0.028& 0.060& 2.224& 0.108 \\
	SMT             & 0.008& 0.020& 0.050& 0.012& 0.004& 0.000& 0.072& 0.020 \\
	total time      & 0.036& 0.092& 0.510& 0.036& 0.032& 0.060& 2.296& 0.128 \\ \addlinespace
	$\Phi^{\sf front}_{\sf safe}$ & \text{unsat} & \text{unsat} & \text{unsat} & \textbf{sat} & \text{unsat} & \text{unsat} & \text{unsat} & \text{unsat} & \text{verified} \\
	constants       & 11& 43& 141& 19& 11& 43& 206& 51 \\
	reduction       & 0.020& 0.048& 0.420& 0.024& 0.020& 0.060& 2.260& 0.140 \\
	SMT             & 0.000& 0.008& 0.060& 0.008& 0.000& 0.000& 0.080& 0.020 \\
	total time      & 0.020& 0.056& 0.480& 0.032& 0.020& 0.060& 2.340& 0.160 \\ \addlinespace
	$\Phi_{\sf safe}^g$ & \text{unsat} & \textbf{sat} & \textbf{sat} & \text{unsat} & \text{unsat} & \text{unsat} & \textbf{sat} & \text{unsat} & \text{not verified} \\
	constants       & 9& 33& 131& 15& 9& 33& 191& 39 \\
	reduction       & 0.012& 0.028& 0.692& 0.020& 0.020& 0.044& 2.372& 0.100 \\
	SMT             & 0.012& 0.004& 0.048& 0.000& 0.000& 0.012& 0.292& 0.000 \\
	total time      & 0.024& 0.032& 0.740& 0.020& 0.020& 0.056& 2.664& 0.100 \\ \addlinespace
	\toprule
	&\multicolumn{8}{c}{including forced topology updates before every jump}\\\midrule
	$\Phi_{\sf safe}^g$ & \text{unsat} & \textbf{sat} & \text{unsat} & \text{unsat} & \text{unsat} & \text{unsat} & \text{unsat} & \text{unsat} & \text{verified} \\
	constants       & 9& 33& 155& 15& 9& 33& 215& 39 \\
	reduction       & 0.012& 0.032& 2.240& 0.016& 0.016& 0.040& 4.784& 0.072 \\
	SMT             & 0.008& 0.012& 0.070& 0.012& 0.012& 0.010& 0.144& 0.020 \\
	total time      & 0.020& 0.044& 2.310& 0.028& 0.028& 0.050& 4.928& 0.092 \\ \addlinespace
\bottomrule\end{tabular}

\caption{Verification times (in seconds) for the given safety
  properties and number of constants of {\sf index} type in the reduced satisfiability problem}%
\label{fig:times}%
\end{figure}%

\medskip
From the tests presented in Figure~\ref{fig:times}, we observe the following facts: 
\begin{itemize}
\item The formula $\Phi^{\sf top}_{\sf safe}$ is an invariant of the system, and is also invariant under 
globally mode reachable flows, jumps and topology updates. 
\item The formula $\Phi^{\sf front}_{\sf safe}$ is true in the initial states and is invariant
under jumps and flows, but not under all topology updates. It is however 
invariant under all globally mode reachable topology updates. 
\item The formula $\Phi^g_{\sf safe}$ is true in the initial states and is invariant under topology updates. 
However, the formula is not invariant under jumps and flows. 
We could show that it is invariant under globally mode reachable
flows and topology updates, but not under globally mode reachable
jumps.
\end{itemize} 
\subsubsection{Model generation} 
The fact that we could show that $\Phi^g_{\sf
  safe}$ is not invariant under globally mode reachable jumps
contradicted our intuition, because a lane change (and no other jump could be the culprit) can only take place if the adjacent cars \textsf{front}, \textsf{back}, \textsf{sidefront} and \textsf{sideback} are sufficiently far away.
In order to understand the problem, we used the model returned by 
H-PILoT to construct a counterexample to safety. After simplifying this
model, we obtained a model describing the situation 
presented in Figure~\ref{fig:global safety violation}: Because we do not specify in $\Phi_{\sf entry}$ that sensors have to be set correctly, there may be another car between \textsf{sidefront} and \textsf{sideback} which will cause a lane change to lead to a collision.

A jump in the situation described in Figure~\ref{fig:global safety violation} can only occur because 
the information provided by sensors at the moment of a line change 
is outdated. One way to avoid this is to ensure that 
a topology update takes place immediately before any lane change.
This is exactly what a human driver would do: to recheck the surroundings immediately before a lane change.
We proved that for all runs in which topology updates take place
before lane changes, formula $\Phi^g_{\sf safe}$ is invariant under 
all jumps.
The detailed results are presented in the bottom rows of
Figure~\ref{fig:times}.


\subsubsection{Complexity}

From the detailed run times in Figure~\ref{fig:times}, one can see that the locality-based reduction of the problem usually dominates the overall run time. The final satisfiability check with Z3 is much faster, especially when the problem size increases. 
We could partially reduce the gap by adding several optimizations to \hpilot. The results reported in the table are thus an order of magnitude faster than the ones we reported in~\cite{DammHS15}. 

Comparing runs with and without consideration of entry states, we can see that the analysis of entry conditions and flows starting in an entry state is only marginally slower than the analysis of initial conditions and general flows. For jumps and topology updates, on the other hand, the additional flow formulae lead to larger ground problems, corresponding to larger potential counter models (cf.\ Theorem~\ref{thm: small model property}). Of course, a similar effect also occurs when every jump is preceded by an update.

\section{Conclusions}
\label{sec:conclusions}

\subsection{Summary of results} 
\label{sec:summary}

We proved that safety properties with exhaustive
entry conditions 
for spatial families of similar linear hybrid
automata can be verified efficiently: 
We reduced 
the proof task
to invariant checking for certain mode reachable states 
and analyzed the complexity of such problems. 
As a by-product, we obtained a 
modularity result for checking safety properties. 
The results can also be used for invariant checking 
(for this the information about mode reachability in the formulae is 
ignored).
The results we obtained are summarized in Figure~\ref{fig:summary}.


\noindent 
\begin{figure}[t]
\centering
{\small \begin{tabular}{@{}l@{\quad}cc@{\quad}cc@{}}\toprule 
  & \multicolumn{2}{c@{\quad}}{\!\!\!\!\!\textbf{Safety} of $\Phi_{\sf Entry}\to\Box\Phi_{\sf Safe}$} &
    \multicolumn{2}{c@{\quad}}{\textbf{Exh. entry conds} $\Phi_{\sf Entry}$}\\
  & {\footnotesize Assumptions 1--3} & {\footnotesize Assumptions 1--4} & {\footnotesize Assumptions 1--3} & {\footnotesize Assumptions 1--4}\\\midrule 
  \textbf{Verification}      & decidable & decidable & decidable & decidable \\
  (Thm.~\ref{thm:decidability}--\ref{thm:decidability of updates},\ref{thm:complexity-all1},\ref{thm:complexity-all2},\ref{thm-tractability}) 
                           & NP & {\footnotesize fixed parameter}  & NP &   {\footnotesize fixed parameter} \\[-1ex]
                            &      & {\footnotesize tractable} & & {\footnotesize tractable}\\
  \textbf{Small model property}     & yes & yes & yes & yes \\
  (Thm.~\ref{thm: small model property})\\
  \textbf{Parametric verification}  
& decidable & decidable & decidable & decidable \\
  (Thm.~\ref{thm:parametric systems})  & & & & \\
~~~ non-param. coefficients/  & NP & {\footnotesize fixed parameter}  & NP & {\footnotesize fixed parameter}  \\[-1ex]
~~~~~~ bounds flows:  & & {\footnotesize tractable} & & {\footnotesize
  tractable} \\[0.5ex]
~~~ parametric coefficients   & EXPTIME & EXPTIME & EXPTIME & EXPTIME \\
~~~ parametric bounds flows & EXPTIME & EXPTIME & EXPTIME & EXPTIME \\
  \addlinespace 
  \textbf{Parameter Synthesis}  & EXPTIME  & EXPTIME  & EXPTIME  &  EXPTIME \\
  (Thm.~\ref{thm: synthesis}) \\
  
\bottomrule 
\end{tabular}
}\caption{Summary of Results}%
\label{fig:summary}%
\end{figure}%

The decidability and complexity results and the small 
model property were established under Assumptions 1, 2(1), 3 
(and possibly 4 for tractability). 
Similar results can also be obtained under Assumption 2(2) or (3) (we
did not present these situations explicitly in this paper because the 
instances obtained due to the locality results are more complicated to
describe (the instantiation takes place in several steps); however it 
can be proved that the 
number of instances and the size of $T_0$ is still polynomial. 

All decidability results directly translate to
situations where the involved formulas do not satisfy Assumptions 2 or 3
but belong to other fragments for which the theory extensions in
Theorem~\ref{locality-verif} are local or stably local; the complexity
depends on the complexity of checking satisfiability for formulae
obtained after instantiation.

\

We would like to point out that although in this paper we refer to a
countable set  $I$ of car identities, due to the verification method
we use the concrete identities of the cars are not important. 
If we prove safety, then we prove it for any model (and thus for any
possible index set);  if we cannot prove it then a counterexample
gives us a possible index set for which the safety propery fails 
(thus a set of possible identities of the cars for which we can
construct a counterexample to safety). On the other hand, 
fixing a set of car identities is not a restriction. 
In all the models that can be obtained in case the formulae we
consider 
are satisfiable, the index sets are quotients (finite or countably infinite) 
of a countable set (which can for instance be chosen to be $I$ or 
the set of natural numbers); all countable models are isomorphic to
this set 
($I$ or the set of natural numbers). In the paper this is handled by 
introducing Skolem 
constants for the indexes of the cars at which the safety condition might not 
hold. A model gives values for these constants (in $I$ or in ${\mathbb
  N}$).  

\subsection{Plans for further work} 

\noindent Another important class of properties, related to timely completion of
maneuvers, are bounded reachability properties.  
They state that for every run starting in a suitable initial
configuration 
$\Phi_{\sf entry}$, a maneuver completion condition $\Phi_{\sf complete}$ becomes true in a given bounded time frame. 
Similar methods can be used for efficiently checking
also this type of properties if we guarantee that the number of 
jumps and topology updates in any fixed interval is bounded. 
We did not include such considerations here in order 
to keep the presentation and the required logics simpler.

\

\noindent {\bf Acknowledgments.} 
This work was partly supported by the German Research Council (DFG) as part of the Trans\-re\-gional Collaborative Research Center ``Automatic Verification and Analysis of Complex Systems'' {\small (SFB/TR 14 AVACS)} \texttt{www.avacs.org}.

\bibliographystyle{apalike}

\end{document}